\documentclass[preprint,12pt]{elsarticle}

\usepackage{amsmath}
\usepackage{amssymb}
\usepackage{url}

\journal{Computer Physics Communications}

\begin{document}

\newcommand{\apj}{Astrophs. J.}
\newcommand{\apjl}{Astrophs. J. Lett.}
\newcommand{\apjs}{Astrophs. J. Suppl.}
\newcommand{\jgr}{J. Geophys. Res.}
\newcommand{\aap}{Astron. Astrophys.}
\newcommand{\nat}{Nature}
\newcommand{\solphys}{Sol. Phys.}
\newcommand{\pasj}{Publ. Astron. Soc. Japan}
\newcommand{\figures}{.}

\begin{frontmatter}



\title{A new MHD code with adaptive mesh refinement and parallelization for
astrophysics}


\author{R. L. Jiang\corref{author}}
\author{C. Fang and P. F. Chen}

\cortext[author]{Corresponding author.\\\textit{E-mail address:} rljiang@nju.edu.cn}
\address{School of Astronomy and Space Science, Nanjing University,
Nanjing 210093, China}
\address{Key Laboratory of Modern Astronomy and Astrophysics (Nanjing
        University), Ministry of Education, China}

\begin{abstract}
A new code, named MAP, is written in FORTRAN language for magnetohydrodynamics
(MHD) calculation with the adaptive mesh refinement (AMR) and Message Passing
Interface (MPI) parallelization. There are several optional numerical schemes for
computing the MHD part, namely, modified Mac Cormack Scheme (MMC), Lax-Fridrichs
scheme (LF) and weighted essentially non-oscillatory (WENO) scheme. All of them
are second order, two-step, component-wise schemes for hyperbolic conservative
equations. The total variation diminishing (TVD) limiters and approximate
Riemann solvers are also equipped. A high resolution can be achieved by the
hierarchical block-structured AMR mesh. We use the extended generalized Lagrange
multiplier (EGLM) MHD equations to reduce the non-divergence free error produced
by the scheme in the magnetic induction equation. The numerical algorithms for
the non-ideal terms, e.g., the resistivity and the thermal conduction, are also
equipped in the MAP code. The details of the AMR and MPI algorithms are
described in the paper.

\end{abstract}

\begin{keyword}

Magnetohydrodynamics \sep Numerical methods \sep Adaptive mesh refinement

\end{keyword}

\end{frontmatter}


\section{Introduction}
\label{introduction}
The adaptive mesh refinement (AMR) algorithm was firstly proposed by 
\citet{Berger1984} and \citet{Berger1989}, which can recursively create finer
overlapping meshes to a given accuracy based on the Richardson error estimation.
Using AMR, one can resolve a very small region in a very large scale
simulation, for instance, the simulation of the propagation of solar
wind from the Sun to the Earth, the protostellar collapse in the
processes of star formation, the very thin current sheets in
magnetic reconnection, and so on. There are two advantages comparing
to uniform grid: AMR is very fast and effective to get a global high
resolution; it can also refine any part of the computational box.

The preliminary block structured AMR method~\citep{Berger1984, Berger1989} 
is not so easy to extend
to multiprocessor calculation because the sizes of the blocks in
the AMR hierarchical grid are different which generate
difficulty in load balance between different processors, although the
arbitrary shapes of the blocks can get a flexible and efficient
memory usage. The data blocks, which are the subgrids of the base
grid, allow to have arbitrary shapes and can merge with other blocks
at the same level. In order to overcome this drawback, a simple approach
was developed by \citet{DeZeeuw1993}. The basic idea is to build
a hierachical binarytree for one dimension (1D), quadtree for two
dimension (2D) and octree for three dimension (3D). These tree
structures contain the necessary information between the parent
blocks, child blocks and neighbor blocks, which can be used to
update the block data from the finer blocks or exchange boundary
data from the neighbor blocks. All blocks in all refinement levels
have the exact same shape and can be easily parallelled using the
Morton space filling curve (Z-curve \citep{Morton1966}), Hilbert
space filling curve (H-curve \citep{Hilbert1891}), or other
curves.

There are many existing AMR codes including 
FLASH~\citep{Fryxell2000}, AMRVAC~\citep{Keppens2003},
PLUTO~\citep{Mignone2007}, SFUMATO~\citep{Matsumoto2007},
NIRVANA~\citep{Ziegler2005, Ziegler2008},
RAMSES~\citep{Fromang2006}, RIEMANN~\citep{Balsara2001}, CRASH~\citep{vanderHolst2011}, 
CHARM~\citep{Miniati2011}, CASTRO~\citep{Almgren2010, Zhang2011} and so on.
Some of them, e.g. FLASH and PLUTO, are using the AMR library like
PARAMESH~\citep{MacNeice2000} toolkit, CHOMBO library \cite{Chombo} or 
BoxLib library \cite{Boxlib}. Other codes, however, developed 
their own AMR algorithms for better
performance. Our code, named as MAP (MHD code with adaptive mesh
refinement and parallelization for astrophysics), was also designed with the same
considerations. Besides, being easy to use and modify is another purpose.
In the developing process, we learn much experience from these
pioneers' papers, for instance, the block structured idea
from \citet{Berger1989} and \citet{DeZeeuw1993}, the error estimation
method from \citet{Ziegler2008}, the file management method from the
code AMRVAC developed by \citet{Keppens2003} and CANS (Coordinated
Astronomical Numerical Softwares).

The major difference between our code and most of the existing
codes is that we use the extended generalized Lagrange multiplier
(EGLM) MHD equations which have one more variable and one more equation
representing the damping and transfer of the non-divergence free
error as described by \citet{Dedner2002}. The method of constrained
transport (CT) \citep{Evans1988} is not included in our MAP code,
although it is almost perfect to control the divergence free
condition to the machine round-off error. The reasons are: (1) 
it is complicated in an AMR code as it requires a staggered mesh; (2) additional
variables have to be allocated for the cell center value of magnetic
field, and the boundary data exchanging process is complex which
takes much longer time than the unstraggered grids; (3) it requires
a strict divergence free boundary condition, because CT can only
guarantee the zero divergence condition from the old time to new
time. When we use a rapidly-changing boundary, for instance, an
emerging flux boundary, it will inevitably produce non-divergence
free magnetic field. In this case, the CT method may not be a good
choice. There are several numerical schemes optionally for computing the
EGLM-MHD equations, namely, modified Mac Cormack Scheme 
(MMC)~\citep{Yu2001}, Lax-Fridrichs (LF)~\citep{Toth1996} and
weighted essentially non-oscillatory (WENO)~\citep{Jiang1999}. Since all
of them are second-order, two-step, component-wise schemes for
hyperbolic conservation laws, the code is fast, effective and lightweight. 
One can easily understand what is going on in the FORTRAN codes of the
three schemes according to the formulae described in Section~\ref{scheme}. 
It is also convenient to modify the code or even add a new MHD solver.
This is another feature of our code. The three schemes have their own
special implements. The first two schemes, i.e. the MMC and LF schemes, 
are equipped with the total variation diminishing \citep[TVD][]{Harten1997}
limiters, while the latter two, i.e. the LF and WENO, are
implemented with the approximate Riemann solvers. In MHD, the exact
solver for Riemann problem is too complex and time consuming, thus
only the simple solver like nonlinear Harten-Lax-van Leer Contact
(HLLC)~\citep{Gurski2001}, Harten-Lax-van Leer Discontinuities
(HLLD)~\citep{Miyoshi2005} and Roe linear solver~\citep{Roe1981,
Brio1988} are adopted in the code MAP. There are several reasons for us not
using higher-resolution numerical schemes (for instance, the piecewise parabolic
method (PPM) by~\citet{Colella1984, Dai1994} or corner transport
upwind PPM (CTU-PPM) by~\citet{Gardiner2005, Gardiner2008} or 5th order
WENO by~\citet{Jiang1996, Jiang1999}): (1) the lack
in accuracy can be compensated by the hierarchical block-structured
AMR algorithm; (2) the accuracy of many parts of the equations
like the thermal conduction and the resistivity, divergence cleanance
method, as well as AMR reflux and interpolation method, can not always
guarantee a high-order precision, which would lead to a poor global
accuracy; (3) the high-order method is complex and time consuming.

Our current MAP code is based only on the Cartesian coordinates, and the
extension to cylindrical and spherical grids will be conducted in
the next version. The other improvements like finite difference (FD)
and finite volume (FC) schemes, the radiation cooling for both
optically thin and thick situations, and the relativistic MHD module are
in consideration too. The AMR strategy for high-order FD and FV schemes
are almost the same except the algorithm of boundary reflux between
coarse and fine meshes, interpolation from parent blocks to child
blocks and update parent blocks from child blocks at the new time.
For a high-order FD scheme, if we use the same procedure as in FV, the
scheme may lose some conservation and accuracy according to our practical
experience. That is our additional reason for using only second-order schemes.

The organization of this paper is as follows: Section~\ref{equations}
introduces the basic EGLM-MHD equations for the divergence clean 
in our code; Section~\ref{scheme} describes the
numerical schemes for solving the EGLM-MHD equations including the
resistivity and thermal conduction; The detailed AMR algorithm with
MPI is given in Section~\ref{AMR}; Some numerical tests in 1D, 2D
and 3D are presented in Section~\ref{test}; Finally, we give a
summary in Section~\ref{summary}.

\section{EGLM-MHD equation}
\label{equations}

The dimensionless MHD equations with gravity,
resistivity and thermal conduction included are given in
conservative form as follows:

\begin{equation}
\frac{\partial \rho}{\partial t} + \nabla \cdot \left(\rho \mathbf{v}\right)
 = 0 \,\, , \label{MHD-01}
\end{equation}

\begin{equation}
\frac{\partial \left(\rho \mathbf{v} \right)} {\partial t} + \nabla \cdot
\left(\left( p + \frac{1}{2}{B}^2 \right) \mathbf{I} + \rho \mathbf{v}
\mathbf{v} - \mathbf{B} \mathbf{B} \right) = \rho \mathbf{g} \,\, ,
\label{MHD-02}
\end{equation}

\begin{equation}
 \frac{\partial \mathbf{B}}{\partial t} + \nabla \cdot \left(\mathbf{v}
 \mathbf{B} - \mathbf{B} \mathbf{v} \right) = - \nabla \times \left(\eta
 \nabla \times \mathbf{B} \right) \,\, , \label{MHD-03}
\end{equation}

\begin{equation}
\frac{\partial e}{\partial t} + \nabla \cdot \left(\mathbf{v} \left(e +
\frac{1}{2}{B}^2 + p \right) - \mathbf{B} \left (\mathbf{B} \cdot \mathbf{v}
\right) \right) = -\nabla \cdot \left( \left(\eta \nabla \times \mathbf{B}
\right) \times \mathbf{B} \right) + \nabla \cdot \left (\kappa
\nabla T \right) + \rho \mathbf{g} \cdot \mathbf{v} \,\, ,
\label{MHD-04}
\end{equation}

\noindent 
here, eight independent conserved variables are the
density ($\rho$), momentum ($\rho v_x$, $\rho v_y$, $\rho v_z$),
magnetic field ($B_{x}$, $B_{y}$, $B_{z}$), and total energy density
($e$). The expression of the total energy density is $e =
p/(\gamma-1) + \rho v^2/2 + B^2/2$. The pressure $p$ and temperature
$T$ are dependent on the eight conserved variables, $\mathbf{g}$ is
the gravity vector, and $\eta$ the magnetic resistivity coefficient, and
$\kappa$ is the thermal conductivity coefficient.
It is noted that $\kappa$ is not always a constant.
It may relate to the local
temperature value with the form $\kappa_0 T^{5 / 2}$ (here
$\kappa_0$ is another parameter for describing the conductivity).
This thermal flux may associate with the direction of the magnetic
field, the direction of this thermal flux may change to $(\mathbf{B} \cdot \nabla T)
\mathbf{B} / B^2$ in some cases. Finally, a unity matrix
$\mathbf{I}$ is involved for matrix operations. The vacuum
permeability is omitted in this dimensionless MHD equation.

Almost all modern MHD codes have to face the problem of how to
guarantee the divergence free requirement. Because of the discretization and
numerical errors, the performance of the MHD code can be unphysical
\citep{Brackbill1980}. There are several ways to maintain $\nabla
\cdot \mathbf{B} = 0$ for MHD equations (\ref{MHD-01}) -
(\ref{MHD-04}): (1) 8-wave formulation \citep{Powell1999}, (2) the
CT method \citep{Evans1988}, (3) the
projection scheme \citep{Brackbill1980}. The 8-wave formulation can
be easily implemented in a code by so-called ``divergence source
terms" without modifying the MHD solver. The numerical value of $\nabla
\cdot \mathbf{B}$ can be controlled to a truncation error. The CT
method is much robuster which can maintain the zero divergence
condition to the machine round off error, but this method needs the
staggered mesh which increase the difficulty in coding especially
for AMR algorithm. The projection scheme
introduces the Poisson equation to clean the numerical error of
$\nabla \cdot \mathbf{B}$ which can preserve the conservative
properties and the efficiency of the base scheme \citep{Toth2000}.
Another way to keep divergence-free condition is to modify the MHD
equations (\ref{MHD-01}) - (\ref{MHD-04}) by using \itshape vector
potential \upshape $\mathbf{A}$ instead of the magnetic field
$\mathbf{B}$ as it was used in \citet{Chen1999}, or by using the
vector magnetic potential or Euler poential. The advantage is that the
divergence free condition is always satisfied, however, the MHD
equation should be rewritten.

Recently, \citet{Dedner2002} proposed the extended generalized
Lagrange multiplier (EGLM) formulation of the MHD equations. This 
method uses two additional waves to transfer the numerical error of
$\nabla \cdot \mathbf{B}$. Thus the local divergence error can be
damped and passed out of the computational domain. It can also
tansfer and damp the numerical error coming from the boundaries. As
we described above, the rapidly-changing boundary will inevitably
produce non-divergence free magnetic field. 
Actually, another method, the GLM-MHD system (i.e. mixed GLM scheme in Dedner's paper), 
is also proposed in Dedner's paper and 
\citet{Dedner2002} themselves recommend GLM-MHD system rather than the EGLM-MHD system.
The EGLM-MHD equations (\ref{MHD-1}) - (\ref{MHD-5}) is the GLM-MHD 
system extended by additional terms which may lead 
to some conservation loss. However, in some of our applications, 
we found that the GLM-MHD system 
works not so well as the EGLM-MHD system. Another reason, once we 
implement the EGLM-MHD, it is very easy to roll back to the GLM-MHD by 
disabling the additional source terms. Hence we adopted the
EGLM-MHD equations of \citet{Dedner2002} in our simulations. The EGLM-MHD
equations with resistivity, thermal conduction and gravity included 
are given as follows:

\begin{equation}
\frac{\partial \rho}{\partial t} + \nabla \cdot \left(\rho \mathbf{v}\right)
= 0 \,\, , \label{MHD-1}
\end{equation}

\begin{equation}
\frac{\partial \left(\rho \mathbf{v}\right)}{\partial t} + \nabla \cdot
\left( \left(p + \frac{1}{2}{B}^2 \right) \mathbf{I} + \rho \mathbf{v}
\mathbf{v} - \mathbf{B} \mathbf{B}\right) =  - \left(\nabla \cdot \mathbf{B}
\right) \mathbf{B} + \rho \mathbf{g} \,\, , \label{MHD-2}
\end{equation}

\begin{equation}
 \frac{\partial \mathbf{B}}{\partial t} + \nabla \cdot \left(\mathbf{v}
 \mathbf{B} - \mathbf{B} \mathbf{v} + \psi \mathbf{I} \right) = - \nabla \times
 \left(\eta \nabla \times \mathbf{B}\right) \,\, , \label{MHD-3}
\end{equation}

\begin{equation}
\frac{\partial e}{\partial t} + \nabla \cdot \left(\mathbf{v}
\left(e + \frac{1}{2}{B}^2 + p\right) - \mathbf{B} \left(\mathbf{B}
\cdot \mathbf{v} \right)\right) = - \mathbf{B} \cdot \left(\nabla
\psi \right) - \nabla \cdot \left( \left(\eta \nabla \times
\mathbf{B} \right) \times \mathbf{B} \right) + \nabla \cdot \left
(\kappa \nabla T \right) + \rho \mathbf{g} \cdot \mathbf{v} \,\, ,
\label{MHD-4}
\end{equation}

\begin{equation}
\frac{\partial \psi}{\partial t} + c_h^2 \nabla \cdot \mathbf{B} =
-\frac{c_h^2}{c_p^2} \psi \,\, , \label{MHD-5}
\end{equation}

\noindent
where $\psi$ is a scalar potential propagating 
divergence error, $c_h$ the wave speed, and $c_p$ the damping rate
of the wave \citep{Matsumoto2007, Dedner2002}. The other symbols
have their normal meanings as in Eq. (\ref{MHD-01}) - (\ref{MHD-04}). 
As suggested in \citet{Dedner2002}, the
expressions for $c_h$ and $c_p$ are

\begin{equation}
c_h = \frac{c_{cfl}} {\Delta t} \min (\Delta x, \Delta y, \Delta z) \,\, ,
\end{equation}

\begin{equation}
c_p = \sqrt{-\Delta t \frac{c_h^2} {\ln{c_d}}} \,\, .
\end{equation}

\noindent
where, $\Delta t$ is the time step, $\Delta x$, $\Delta y$ and $\Delta
z$ are the space steps, $c_{cfl}$ is a safety coefficient less than
1. $c_d \in (0, 1)$ is a problem dependent coefficient to decide the
damping rate for the waves of divergence errors. The value of $c_d$ is 0.18 in
most of our tests. We can see that
$c_h$ and $c_p$ is not independent of the grid resolution and the
scheme used. Hence we may have to adjust their values for different
situations.

\section{Numerical schemes}
\label{scheme}
We use three different optional numerical schemes to solve the MHD part, i.e.
MMC, LF and WENO as we mentioned above. In our calculations presented here,
we mainly use the WENO scheme, the other two are used for comparison
(see the accuracy test of these schemes in Section~\ref{test}).
Although the accuracy of all three schemes is not so high in both
space and time, the advantage is that the coding is simple and the
running speed is fast. In this section,
we briefly review the three schemes by using the following 1D equation:

\begin{equation}
\frac{\partial u(x, t)}{\partial t} + \frac{\partial f\left(u(x,
t)\right)}{\partial x} = 0 \,\, , \label{scalar_law}
\end{equation}

\noindent
here $u(x, t)$ represents the eight conserved variables ($\rho$, $\rho v_x$, $\rho v_y$, 
$\rho v_z$, $B_x$, $B_y$, $B_z$, $e$) and
we assume the grid in domain $[a, b]$ is uniform and
it is discretized into $n$ points (for FD) or cells (for FV):

\begin{equation}
a < x_i < b \, , i = 1, 2, 3, 4,..., n \,\, ,
\end{equation}
\begin{equation}
\Delta x = x_i - x_{i - 1} \, , i = 2 ,3, 4,..., n \,\, ,
\end{equation}

\noindent the discretization of Eq.  (\ref{scalar_law}) in
finite difference form at the point $i$ is as follows (using $u_{i}
= u(x_i, t_n)$ and $f_{i} = f\left(u(x_i, t_n)\right)$ for simplicity):

\begin{equation}
\frac{1}{\Delta t} \left(u_{i} ^ {n + 1} - u_{i}\right) + \frac{1}{\Delta x}
\left( \hat{f}_{i + \frac{1}{2}} - \hat{f}_{i - \frac{1}{2}}\right) = 0 \,\, ,
\label{fd}
\end{equation}

\noindent
where $\hat{f}_{i \pm \frac{1}{2}}$ is the approximate numerical flux. 
The notation $n + 1$ means the values of $u$ at the
new time. Other terms without the superscript $n + 1$ are old values. The
time step is $\Delta t$, $i \pm \frac{1}{2}$ means the half grid
points. The discretization of Eq. (\ref{scalar_law}) in the finite volume
form at the cell $i$ is:

\begin{equation}
\frac{1}{\Delta t} \left(\bar{u}_{i} ^ {n + 1} - \bar{u}_{i}\right) +
\frac{1}{\Delta x} \left( \hat{f}_{i + \frac{1} {2}} - \hat{f}_{i -
\frac{1} {2}}\right) = 0 \,\, ,
\end{equation}

\noindent
where $i \pm \frac{1}{2}$ are the cell boundaries and the
cell average is:

\begin{equation}
\bar{u}_{i} = \frac{1}{\Delta x} \int^{x_{i + \frac{1}{2}}}_{x_{i -
\frac{1}{2}}} u(\xi, t)\, \mathrm{d} \xi \,\, ,
\label{fv}
\end{equation}

\noindent
where we use the cell average values to get $\hat{f}_{i
\pm \frac{1}{2}}$ for this equation.

Whether FD schemes or FV schemes, the new value can be obtained by 
the following formula (if we omit the average bar on the variable $u$):

\begin{equation}
u_{i} ^ {n + 1} = u_{i} - \frac{\Delta t}{\Delta x}\left(\hat{f}_{i +
\frac{1} {2}} - \hat{f}_{i - \frac{1} {2}}\right) \,\, .
\label{Equ:basic}
\end{equation}

As we mentioned above, the AMR algorithms for high-order FD schemes
and high-order FV schemes are not exactly the same. The main
differences include boundary reflux between coarse and fine meshes,
interpolation from parent blocks to child blocks and updating parent
blocks from child blocks at the new time. Note that the initial
definitions of the cell average value and the point value are
different. However, under our second-order approximation, there is
no difference in the mathematic expressions between FV and FD and
the point values are regarded as the cell average values in our MAP
code.

\subsection{MMC scheme}
\label{mmc}
We briefly give the FD formulae here,
for more information we refer the readers to the paper by~\citet{Yu2001}. 
The numerical flux ($\hat{f}_{i+\frac{1}{2}}$) needed in the Eq. (\ref{Equ:basic}) 
is formed by using the Lax-Friedrichs splitting method with an 
additional higher order modified term:

\begin{equation}
\hat{f}_{i + \frac{1} {2}} = \hat{h}_{i + \frac{1}{2}} +
\frac{1}{2} \left(\phi\left(r_{i + \frac{1}{2}} ^ +\right) \hat{g}_{i +
\frac{1}{2}} ^ + - \phi\left(r_{i + \frac{1}{2}} ^ -\right) \hat{g}_{i +
\frac{1}{2}} ^ - \right) \,\, ,
\end{equation}

\noindent
where

\begin{equation}
\hat{h}_{i + \frac{1}{2}} = f_i^+ + f_{i+1}^{-} \,\, ,
\end{equation}

\begin{equation}
r_{i + \frac{1}{2}} ^ + = \frac{\hat{g}_{i - \frac{1}{2}} ^
{+}}{\hat{g}_{i + \frac{1}{2}} ^ {+}} \, , \,\,\, r_{i +
\frac{1}{2}} ^ - = \frac{\hat{g}_{i + \frac{3}{2}} ^ {-}}{\hat{g}_{i
+ \frac{1}{2}} ^ {-}} \,\, ,
\end{equation}

\begin{equation}
\hat{g}_{i + \frac{1}{2}} ^ {+} = \tilde f_{i + 1}^{+} - f_{i}^{+}
\, , \,\,\, \hat{g}_{i + \frac{1}{2}} ^ {-} = f_{i + 1}^{-} - \tilde
f_{i}^{-} \,\, ,
\end{equation}

\begin{equation}
\tilde{u}_{i} = u_{i} - \frac{\Delta t}{\Delta x}\left(\hat{h}_{i +
\frac{1} {2}} - \hat{h}_{i - \frac{1}{2}}\right) \,\, ,
\end{equation}

\begin{equation}
f_i^\pm = \frac{1}{2} \left(f_i \pm \alpha u_i\right) \, , \,\,\,
\tilde f_i^\pm = \frac{1}{2} \left(\tilde f_i \pm \tilde{\alpha} \tilde u_i\right) \,\, .
\end{equation}

The values of $\tilde f$ and $\tilde{\alpha}$ are obtained from the value
$\tilde u$. $\alpha$ and $\tilde{\alpha}$ is the local maximum
eigenvalue of $f$ and $\tilde f$ (namely the maximum wave speed
taken from the relevant range of $u$ and $\tilde{u}$), respectively.
The TVD flux limiter is taken as the form~\citep{Roe1989}:

\begin{equation}
\phi(r)=\left\{
\begin{array}{ll}
r  &, \ \ |r| \le 1  \\
1  &, \ \ |r| > 1
\end{array}
. %
\right.
\end{equation}


\subsection{LF scheme}
\label{lf}
The FV formulae of LF scheme is given here, see~\citet{Toth1996} for more
information. The numerical flux ($\hat{f}_{i+\frac{1}{2}}$) 
needed by Eq. (\ref{Equ:basic}) is obtained by the following flux ($f_m$):

\begin{equation}
\hat{f}_{i + \frac{1}{2}} = f_m\left(\tilde{u}_{i + \frac{1}{2}} ^ {-}, 
\tilde{u}_{i + \frac{1}{2}} ^ {+} \right) \,\, .
\end{equation}

The simplest $f_m$ flux is Lax-Friedrichs flux but our MAP code also 
implemented approximate Riemann solvers like 
HLLC~\citep{Gurski2001}, HLLD~\citep{Miyoshi2005} and Roe
solver~\citep{Roe1981, Brio1988}. Since the formulae of these Riemann solvers 
are too complicated to be given here, thus we only list the Lax-Friedrichs flux as

\begin{equation}
f_m(a, b) = \frac{1}{2} \left(f(a) + f(b) - \alpha (b - a) \right) \,\, ,
\label{Equ:LFFlux}
\end{equation}

\noindent
where $\alpha$ is the local maximum speed. Other
variables are given as:

\begin{equation}
\tilde{u}_{i} = u_{i} - \frac{\Delta t}{2 \Delta x}\left(f_{i + \frac{1}
{2}} ^ {-} - f_{i - \frac{1} {2}} ^ {+}\right) \,\, ,
\end{equation}

\begin{equation}
u_{i + \frac{1}{2}}^{-} = u_i + \frac{1}{2} \phi\left(\Delta u_{i -
\frac{1}{2}}, \Delta u_{i + \frac{1}{2}}\right) \, , \,\,\, u_{i -
\frac{1}{2}}^{+} = u_i - \frac{1}{2} \phi\left(\Delta u_{i -
\frac{1}{2}}, \Delta u_{i + \frac{1}{2}}\right) \,\, , \label{limiter1}
\end{equation}

\begin{equation}
\tilde u_{i + \frac{1}{2}}^{-} = \tilde u_i + \frac{1}{2}
\phi\left(\Delta u_{i - \frac{1}{2}}, \Delta u_{i +
\frac{1}{2}}\right) \, , \,\,\, \tilde u_{i - \frac{1}{2}}^{+} =
\tilde u_i - \frac{1}{2} \phi\left(\Delta u_{i - \frac{1}{2}}, \Delta
u_{i + \frac{1}{2}}\right) \,\, , \label{limiter2}
\end{equation}

\begin{equation}
\Delta u_{i + \frac{1}{2}} = u_{i + 1} - u_{i} \,\, ,
\end{equation}

\noindent
where $\phi$ is a TVD slope limiter. Note that the slopes in equations
(\ref{limiter1}) and (\ref{limiter2}) are calculated from the same
variable differences $\Delta u_{i \pm \frac{1}{2}}$ for better
performance~\citep{Toth1996}. The form of the limiter used in our
code is

\begin{equation}
\phi\left(a, b \right) = \mathrm{minmod}\left(a, b\right) =
\mathrm{sgn}\left(a\right) \max \left(0, \min\left(|a|,
\mathrm{sgn}\left(a\right) b \right) \right) \,\, , \label{minmod}
\end{equation}

\noindent
where $\mathrm{sgn}\left(a\right)$ stands for the sign of value $a$.


\subsection{WENO scheme}
As for the component-wise FV WENO scheme, in order to form Eq. (\ref{Equ:basic}),
we can also use the same Lax-Friedrichs flux Eq. (\ref{Equ:LFFlux}) 
or Riemann solvers like HLLC, HLLD and Roe solvers in 


\begin{equation}
\hat{f}_{i + \frac{1}{2}} = f_m\left(u_{i + \frac{1}{2}} ^ {-}, 
u_{i + \frac{1}{2}} ^ {+} \right) \,\, ,
\end{equation}

\noindent
where the reconstructions of values $u_{i + \frac{1}{2}} ^ {\pm}$ are

\begin{equation}
u_{i + \frac{1}{2}} ^ {-} = \frac{2 a_{i + \frac{1}{2}} w_{i +
\frac{1}{2}} ^ {-} + a_{i - \frac{1}{2}} v_{i - \frac{1}{2}} ^ {-}}
{2 a_{i + \frac{1}{2}} + a_{i - \frac{1}{2}}} \, , \,\,\, u_{i -
\frac{1}{2}} ^ {+} = \frac{a_{i + \frac{1}{2}} w_{i + \frac{1}{2}} ^
{+} + 2 a_{i - \frac{1}{2}} v_{i - \frac{1}{2}} ^ {+}} {a_{i -
\frac{1}{2}} + 2 a_{i - \frac{1}{2}}} \,\, ,
\end{equation}

\noindent
where

\begin{equation}
w_{i + \frac{1}{2}} ^ {-} = \frac{1}{2} \left(u_i + u_{i + 1}\right)
\, , \,\,\, v_{i - \frac{1}{2}} ^ {-} = \frac{1}{2} \left(3 u_i -
u_{i - 1}\right) \,\, ,
\end{equation}

\begin{equation}
w_{i + \frac{1}{2}} ^ {+} = \frac{1}{2} \left(3 u_i - u_{i +
1}\right) \, , \,\,\, v_{i - \frac{1}{2}} ^ {+} = \frac{1}{2}
\left(u_i + u_{i - 1}\right) \,\, ,
\end{equation}

\begin{equation}
a_{i + \frac{1}{2}} = \frac{1}{\left(\epsilon + \left(u_{i + 1} -
u_{i}\right) ^ 2 \right) ^ 2} \,\, .
\end{equation}

\noindent
where the constant $\epsilon = 10^{-12}$ in our code, which can avoid the
denominator becoming zero. Following the formulae listed above, we can
obtain the value of $u$ at the new time. However, the time accuracy
is only the first-order. It is recommended to use the optimal second-order
TVD Runge-Kutta method~\citep{Butcher2005} instead of formula
(\ref{Equ:basic}), i.e.,

\begin{equation}
u_i^1 = u_i - \frac{\Delta t}{\Delta x}(\hat{f}_{i + \frac{1} {2}} -
\hat{f}_{i - \frac{1} {2}}) \,\, ,
\end{equation}

\begin{equation}
u_i^{n + 1} = \frac{1}{2} \left(u_i + u_i^1 -
\frac{\Delta t}{\Delta x} \left(\hat{f}_{i + \frac{1} {2}}^1 -
\hat{f}_{i - \frac{1} {2}}^1 \right) \right).
\end{equation}

Note that $u_i^1$ is an intermediate variable and the numerical
flux $\hat{f}_{i \pm \frac{1} {2}}^1$ is taken from $u_i^1$.

\subsection{Thermal conduction}
\begin{figure}[htbp]
\centering
\includegraphics[height=500pt]{\figures/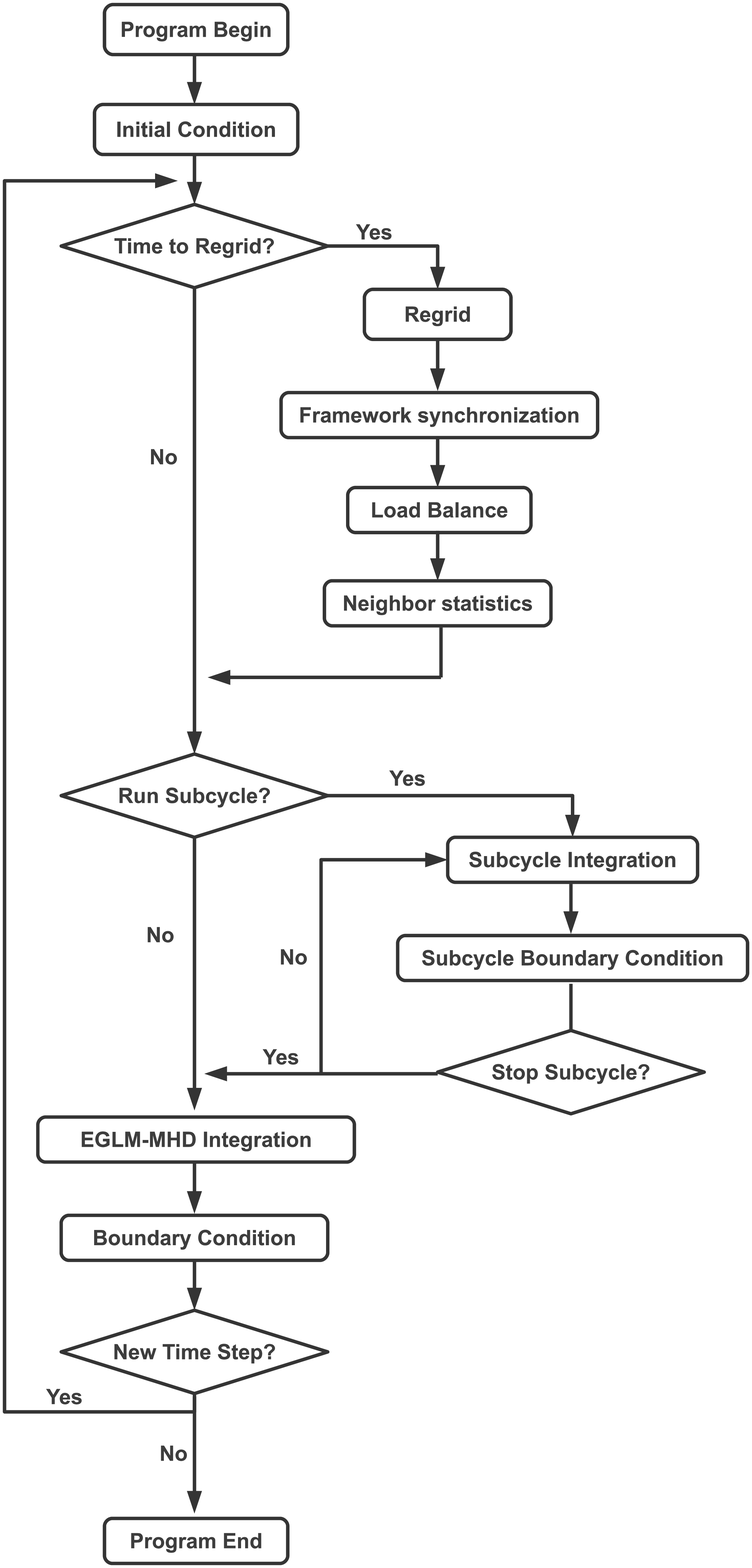}
\caption{Schematic chart of our MAP code. The AMR algorithm includes the
regriding, framework synchronization, load balance and neighbor
statistics procedures. The time step in the subcycle is controlled by
actual physical problem, e.g., the thermal conduction. The boundary
condition of the subcycle is treated only for the specific physical
quantity, i.e., the total energy. The boundary conditions contain
the refluxing, data exchange and boundary fixing procedures.}
\label{fig01}
\end{figure}

Thermal conduction is added into the MAP code by an explicit scheme. 
Generally speaking, once we consider the thermal conduction, 
the safety time step become very small. 
It may take a long wall time to get the result if the code evolves the 
whole EGLM-MHD part with such a small time step. 
Thus we set a subcycle for
treating the thermal conduction. The subcycle means that the code
only computes the thermal conduction part of energy equation many
times with the time step determined by the speed of thermal
conduction within an EGLM-MHD time step, as shown in the schematic chart
(Fig.~\ref{fig01}). The equation of the thermal conduction for
integration of subcycle is written as

\begin{equation}
\frac{\partial e}{\partial t} = \nabla \cdot \left (\kappa \nabla T
\right) \,\, .
\label{eqn_thermal}
\end{equation}

Assuming the grid is uniform, the one dimensional numerical scheme for
this equation is given as

\begin{equation}
e_i^{n + 1} = e_i + \frac{\Delta t}{{\Delta x}^2} \left(
\sqrt{\kappa_i \kappa_{i + 1}}\left(T_{i + 1} - T_i\right) -
\sqrt{\kappa_i \kappa_{i - 1}}\left(T_{i} - T_{i - 1}\right)
\right) \,\, .
\end{equation}

The time step used in the subcycle is given by

\begin{equation}
\Delta t = c_{cfl} \frac{\left(\min(\Delta x, \Delta y,
\Delta z)\right)^2}{n_{dim}\max(\kappa T / p)} \,\, ,
\end{equation}

\noindent where $n_{dim}$ is the number of dimensions. The
subcycle is simple and fast but one should be careful to deal with
it. Since the other variables do not change with time during the
subcycle, in our experience it is better to reduce the number of
cycles to dozens or less to get the reliable results. Moreover,
there should be a threshold for the temperature gradient in 
Eq. (\ref{eqn_thermal}), because for
any matter the speed of thermal conduction can not be infinite. That
is a problem dependent parameter. So far, only the thermal
conduction is calculated by the subcycle. If necessary, it is easy to
add other physical process into the subcycle.

\subsection{Resistivity}
As for the resistivity terms $\mathbf{R}(R_x, R_y, R_z) = - \nabla
\times \left(\eta \nabla \times \mathbf{B} \right)$ and $ R_e =
-\nabla \cdot \left( \left(\eta \nabla \times \mathbf{B} \right)
\times \mathbf{B} \right)$ in the induction equations (\ref{MHD-3})
and (\ref{MHD-4}), we use the simple central finite difference
scheme to compute the current density first ($\mathbf{J}(J_x, J_y,
J_z) = \nabla \times \mathbf{B}$), and then the resistivity terms,
i.e.:

\begin{eqnarray}
(J_x)_{i, j, k} = \frac{(B_z)_{i, j + 1, k} - (B_z)_{i, j - 1, k}}{2
\Delta y} - \frac{(B_y)_{i, j, k + 1} - (B_y)_{i, j, k - 1}}{2 \Delta z} \,\,\, \notag \\
(J_y)_{i, j, k} = \frac{(B_x)_{i, j, k + 1} - (B_x)_{i, j, k - 1}}{2 \Delta z} -
                  \frac{(B_z)_{i + 1, j, k} - (B_z)_{i - 1, j, k}}{2 \Delta x} \,\,\, \\
(J_z)_{i, j, k} = \frac{(B_y)_{i + 1, j, k} - (B_y)_{i - 1, j, k}}{2 \Delta x} -
                  \frac{(B_x)_{i, j + 1, k} - (B_x)_{i, j - 1, k}}{2 \Delta y} \,\, , \notag
\end{eqnarray}

\begin{eqnarray}
(R_x)_{i, j, k} = \frac{(\eta J_y)_{i, j, k + 1} - (\eta J_y)_{i, j,
k - 1}}{2 \Delta z} -
                  \frac{(\eta J_z)_{i, j + 1, k} - (\eta J_z)_{i, j - 1, k}}{2 \Delta y} \,\,\, \notag \\
(R_y)_{i, j, k} = \frac{(\eta J_z)_{i + 1, j, k} - (\eta J_z)_{i - 1, j, k}}{2 \Delta x} -
                  \frac{(\eta J_x)_{i, j, k + 1} - (\eta J_x)_{i, j, k - 1}}{2 \Delta z} \,\,\, \\
(R_z)_{i, j, k} = \frac{(\eta J_x)_{i, j + 1, k} - (\eta J_x)_{i, j - 1, k}}{2 \Delta y} -
                  \frac{(\eta J_y)_{i + 1, j, k} - (\eta J_y)_{i - 1, j, k}}{2 \Delta x} \,\, , \notag
\end{eqnarray}

\begin{eqnarray}
R_e = \frac{\left(\eta(J_z B_y - J_y B_z)\right)_{i + 1, j, k} - 
\left(\eta(J_z B_y - J_y B_z)\right)_{i - 1, j, k}}{2 \Delta x} \,\,\, \notag \\
      \frac{\left(\eta(J_x B_z - J_z B_x)\right)_{i, j + 1, k} - 
\left(\eta(J_x B_z - J_z B_x)\right)_{i, j - 1, k}}{2 \Delta y} \,\,\, \notag \\
      \frac{\left(\eta(J_y B_x - J_x B_y)\right)_{i, j, k + 1} - 
\left(\eta(J_y B_x - J_x B_y)\right)_{i, j, k - 1}}{2 \Delta z} \,\, ,
\end{eqnarray}

\noindent where $R_x$, $R_y$ and $R_z$ are for the induction
equation (\ref{MHD-3}) and $R_e$ for the energy Eq. 
(\ref{MHD-4}). The resistivity model can be modified to any form
according to the user's need.

\subsection{Damping zone}
\begin{figure}[htbp]
\centering
\includegraphics[height=180pt]{\figures/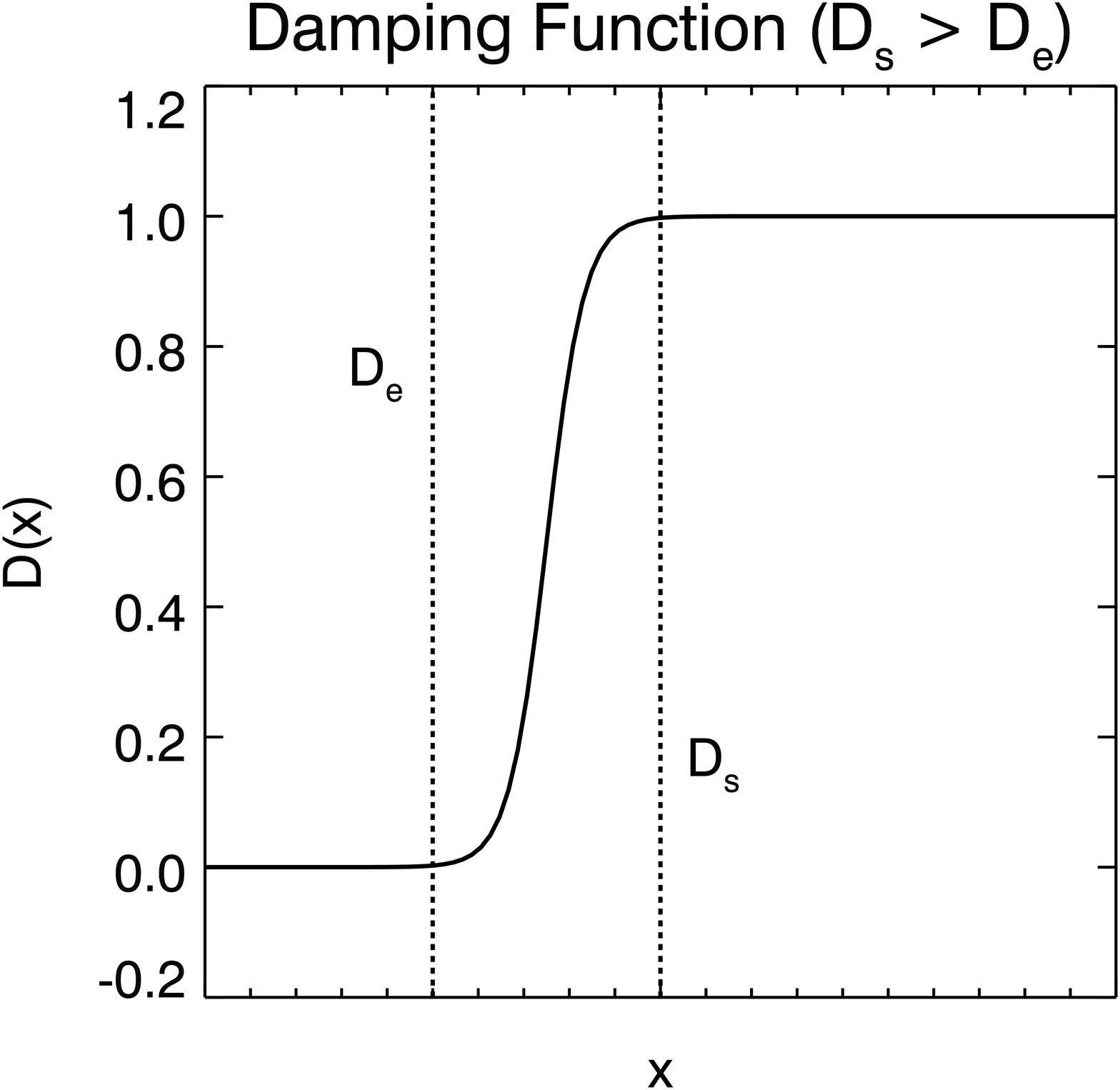}
\includegraphics[height=180pt]{\figures/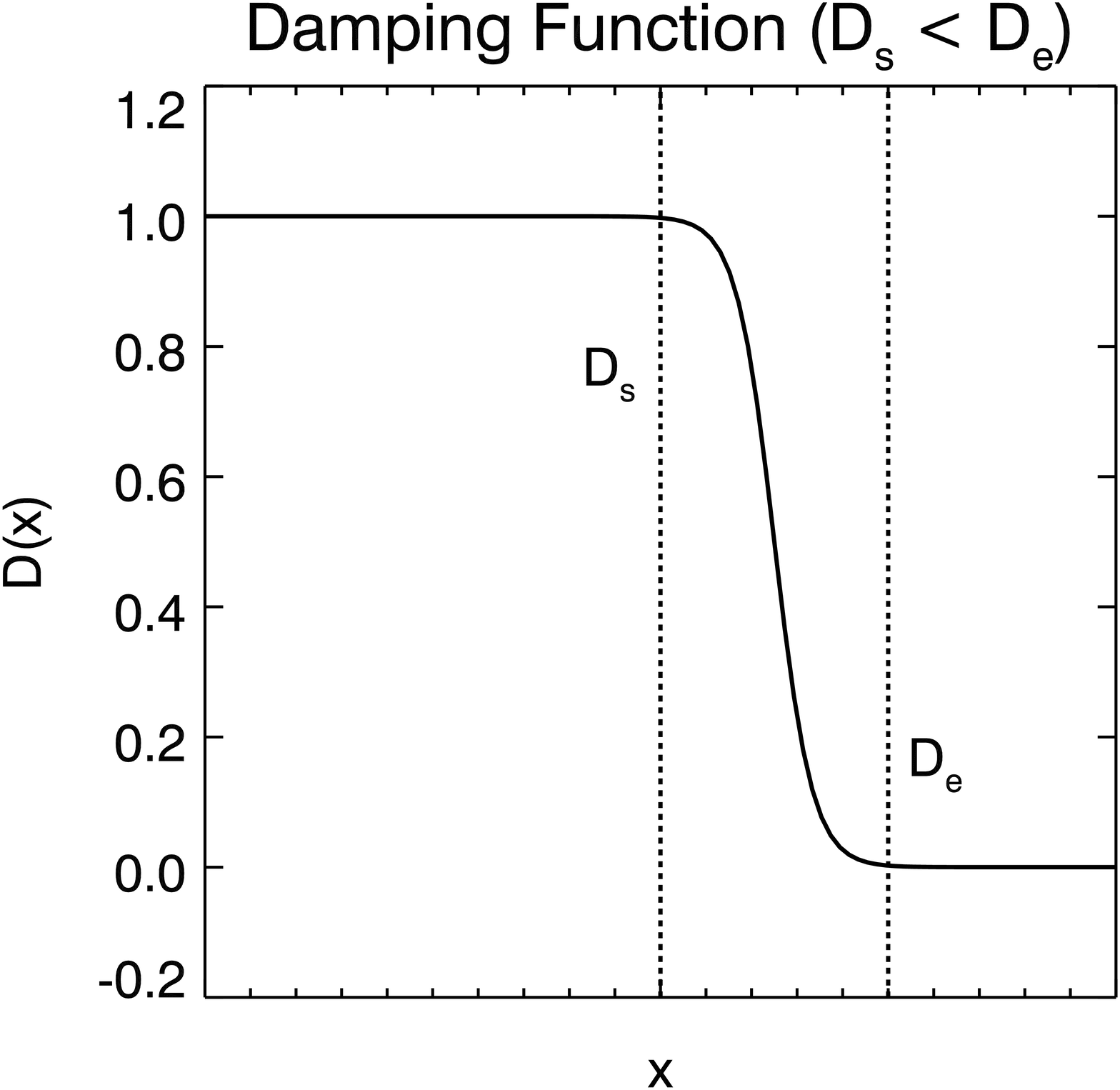}
\caption{Damping function $D(x)$. Left panel shows the situation of
$D_s > D_e$ where the left bounday is damped, while right panel shows
the case with $D_s < D_e$ where the right bounday is damped.}
\label{fig02}
\end{figure}

The damping zone is a range where the MHD waves and matter motions
can be damped to the initial condition. It is a kind of boundary
condition because it is usually adjacent to the physical boundary.
The goal of the zone is to stablize the boundary and to remove the
non-physical inflows and outflows produced by the numerical error.
Given a start position ($D_s$) and an end position ($D_e$) of the
damping zone, the one dimensional damping function $D(x)$ is

\begin{equation}
D(x) = \frac{1}{2}\left(1 - \tanh \left(\frac{6}{D_e - D_s}\left(x -
\frac{D_s + D_e}{2}\right)\right)\right)
\end{equation}

\begin{equation}
u(x) = u_0(x) (1 - D(x)) + u(x) D(x),
\end{equation}

\noindent
where $u_0$ is the initial value of the variable $u$.
Fig.~\ref{fig02} shows the function of $D(x)$ with the situations
$D_s > D_e$ and $D_s < D_e$, which correspond to the different
damping directions, namely, physical waves and motions will vanish
at the lower boundary for $D_s > D_e$ and at the higher boundary
$D_s < D_e$. It is the same way to treat the damping zone in multiple
dimensions. One can do it dimension by dimension or treat $x$, $y$ and $z$
directions together. The damping function only has effect on the density
($\rho$), velocities ($v_x, v_y, v_z$) and pressure ($p$), but not
to the magnetic field. Because the modification of magnetic field will
change the magnetic topological configuration which may destroy the
divergence free condition leading to some non-physical results. It
is noted that the damping zone can also produce some weak reflected
waves. That is inevitable for most of the boundary conditions and
these reflected waves have little effect on the results in our
tests.

\subsection{Multiple spatial dimensions}
All we discussed above is in 1D space. It is easy to
extend the schemes to two or three dimensions. However, the
method in extending the numerical scheme to multiple dimensions depends on the
schemes. We treat the WENO scheme dimension by
dimension in 2D and 3D. For the 2D case, the numerical flux in
$x$-direction is calculated first and then the flux in $y$-direction
by using the same variables at the old time. After that, the variables are
updated to the new time by taking the numerical flux in $x$- and
$y$- directions simultaneously. If the TVD Runge-Kutta method is
available, the variables need to be calculated twice to update all variables.
For TVD-MMC and TVD-LF methods, the dimension by dimension method is not
necessary, since we can do all directions together. The formulae for
2D or 3D are easy to derive following the procedure in Sections~\ref{mmc}
and~\ref{lf}. The time step in multiple dimensional problems are shown
as follows:

\begin{equation}
\Delta t = c_{cfl} \frac{\min(\Delta x, \Delta y, \Delta z)}{n_{dim}\alpha_g} \,\, ,
\end{equation}

\noindent the $n_{dim}$ is the number of the dimensions, i.e.
$n_{dim} = 1$ for 1D, $n_{dim} = 2$ for 2D and $n_{dim} = 3$ for 3D
and $\alpha_g$ is the global maximal wave speed. In Section~\ref{test},
we show several 2D and 3D tests in some special test problems.

\section{AMR parallelization strategy}
\label{AMR} The schematic chart (Fig. \ref{fig01}) shows the main
process in the MAP code. The AMR parallelization is more difficult
than the usual Eulerian meshes. The main difficults
are: (1) how to arrange the structure for AMR hierarchical blocks?
(2) how to guarantee the load balance when new blocks are generated? (3)
how to apply the boundary conditions between different processors
and between different refinement levels? We use the two dimensional
MAP code to explain what we did in the AMR algorithm for
simplicity and intuition. The aim of this section is to give a
detailed explanation of the MAP code. Another purpose is to let the
readers who have no idea about the AMR to know what is AMR and how
to write an AMR code by themselves. MAP code implements 
the MPI parallelization which supports the MPICH2 and
OpenMPI softwave, which are full MPI-2 standards. The other 
parallelization softwares like OpenMP which requires
a shared-memory computer are not supported in our code.

\subsection{Hierarchical structure}
\label{hierarchical_structure} 
\begin{figure}[htbp]
\centering
\includegraphics[scale=0.20]{\figures/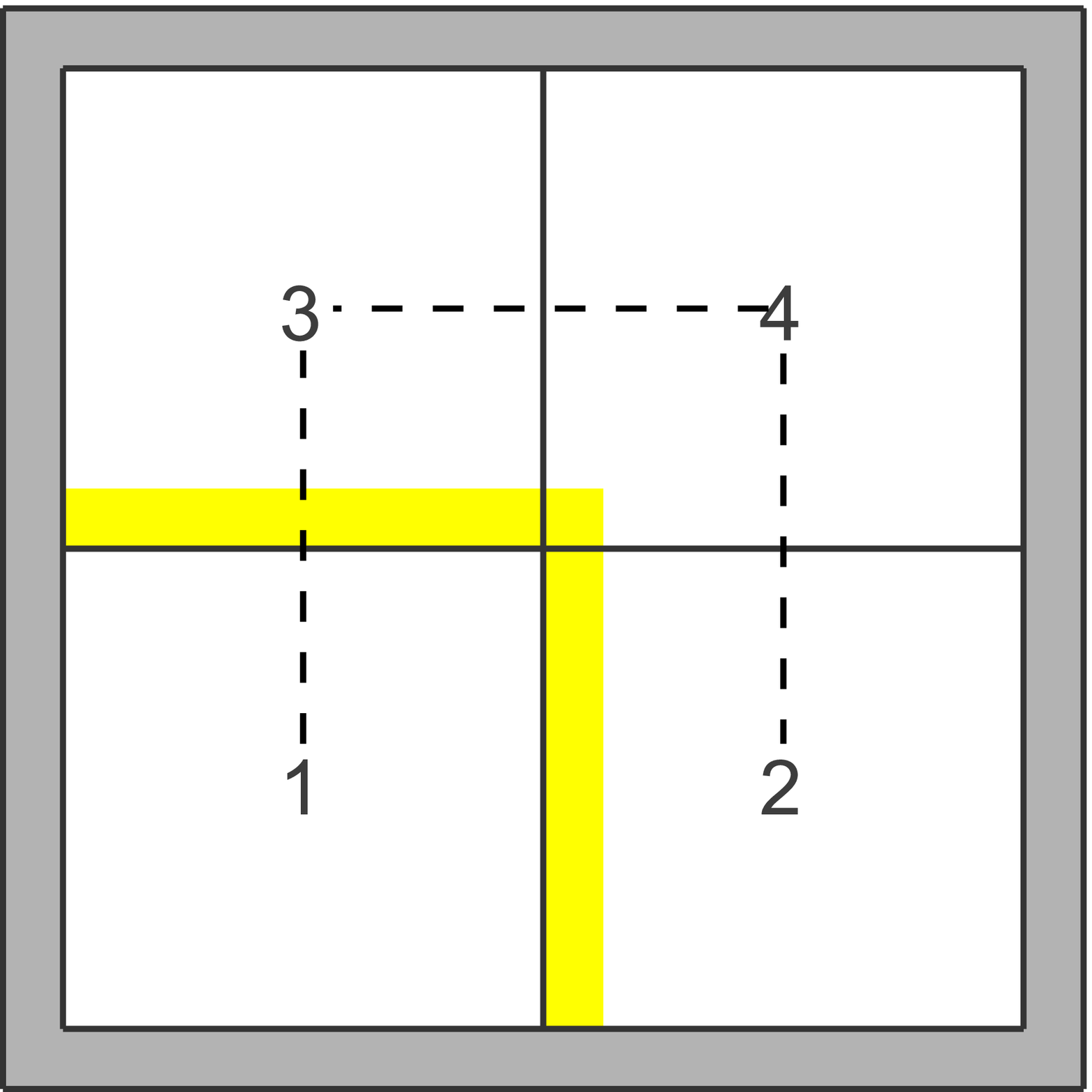}
\includegraphics[scale=0.20]{\figures/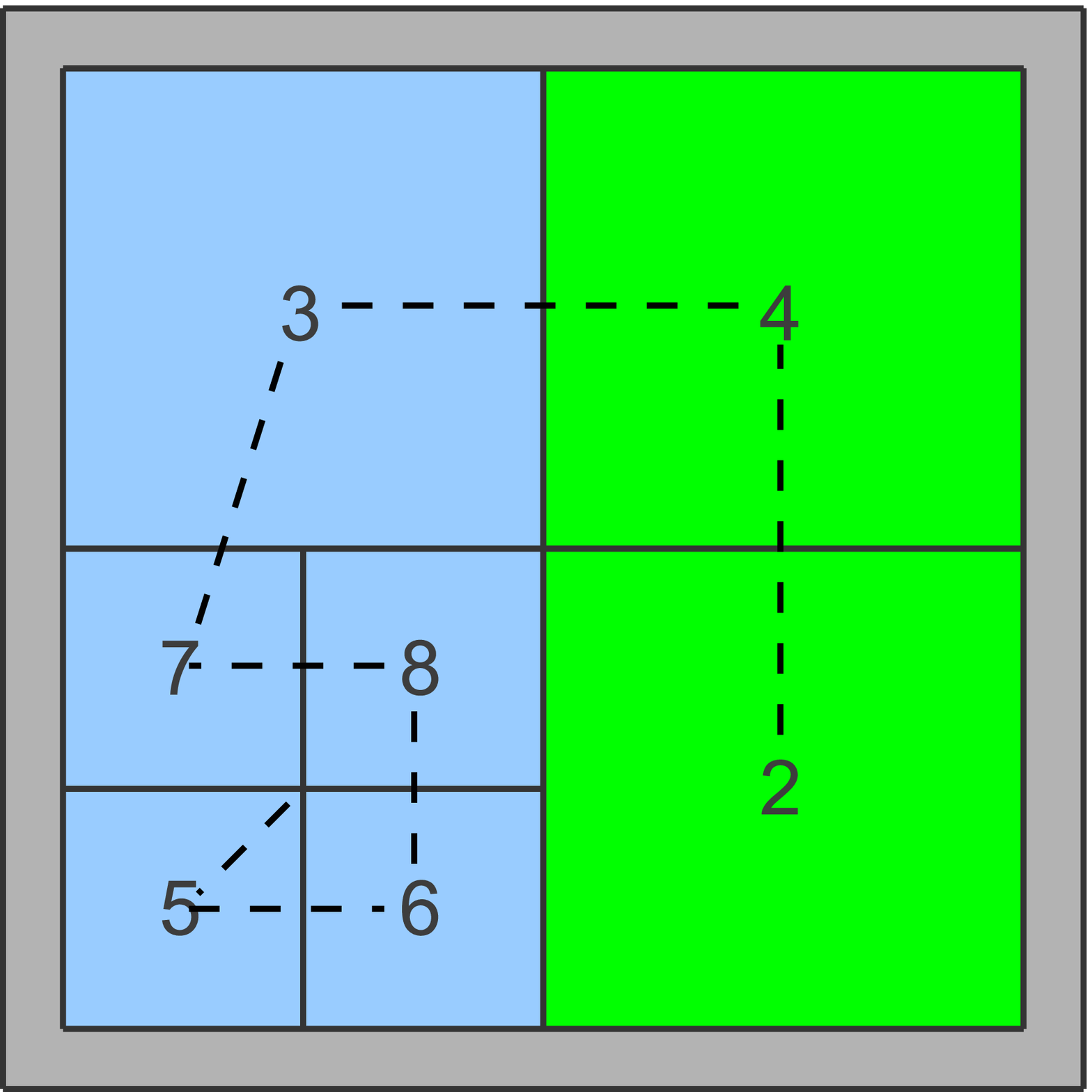}
\includegraphics[scale=0.24]{\figures/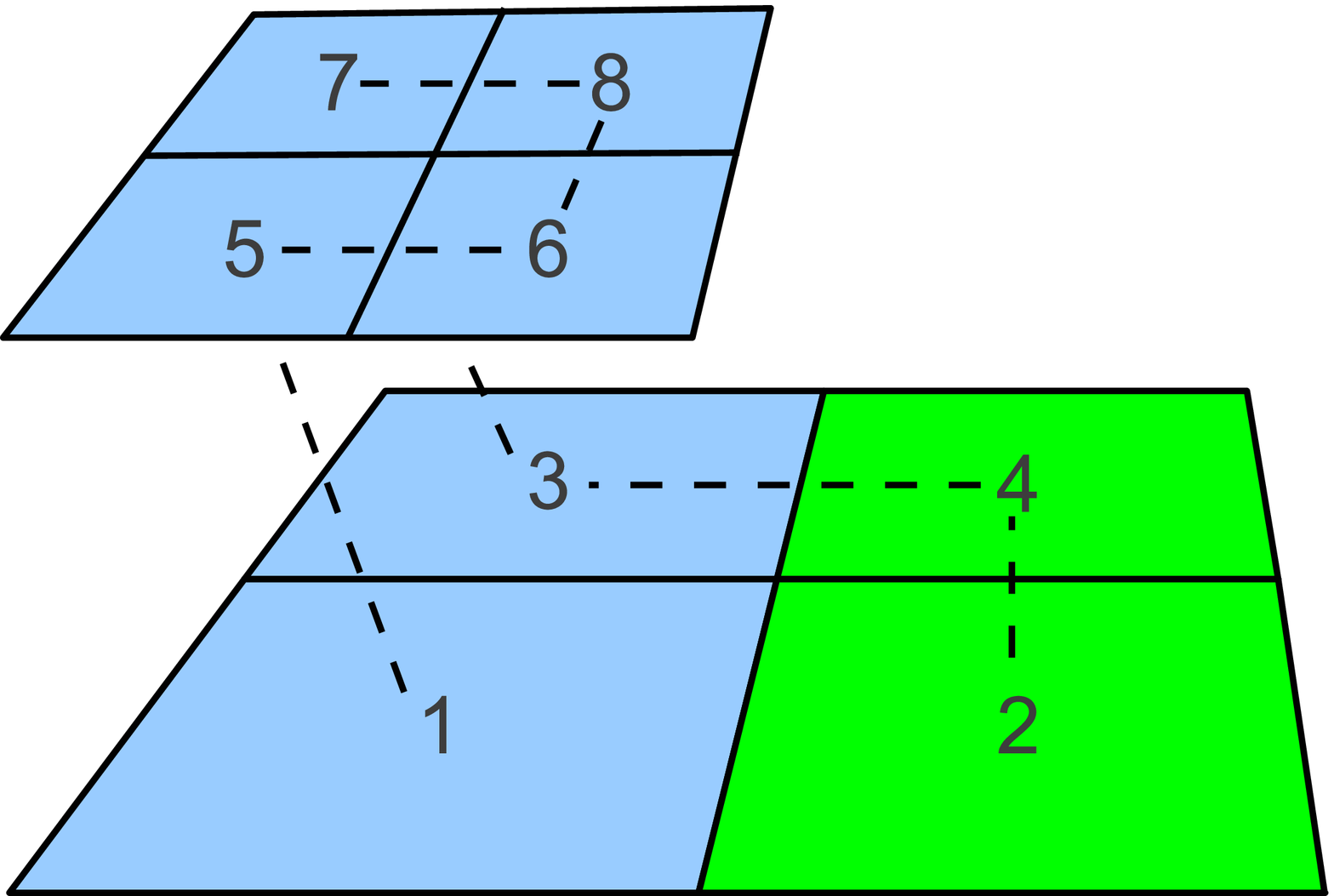}
\includegraphics[scale=0.22]{\figures/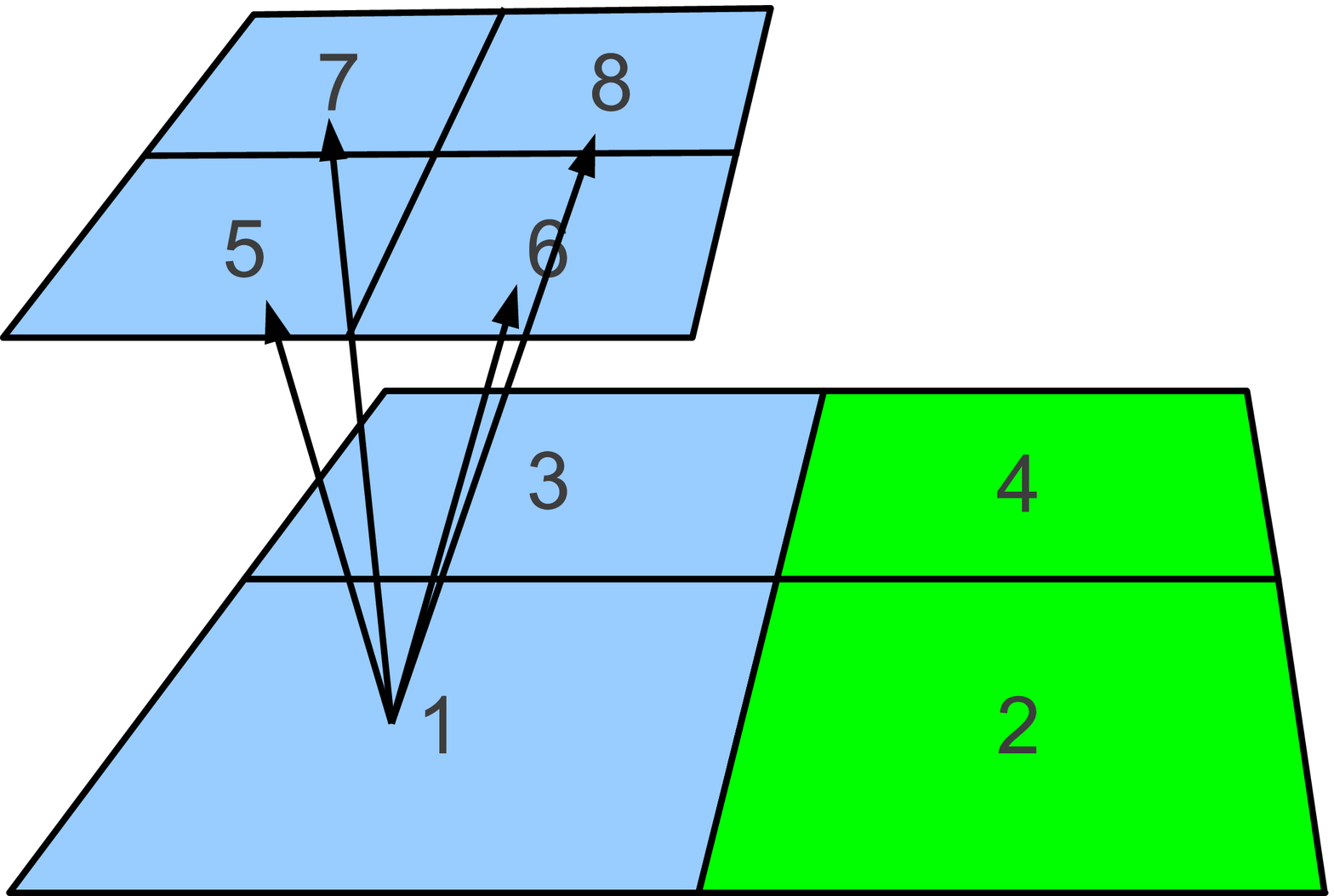}
\includegraphics[scale=0.22]{\figures/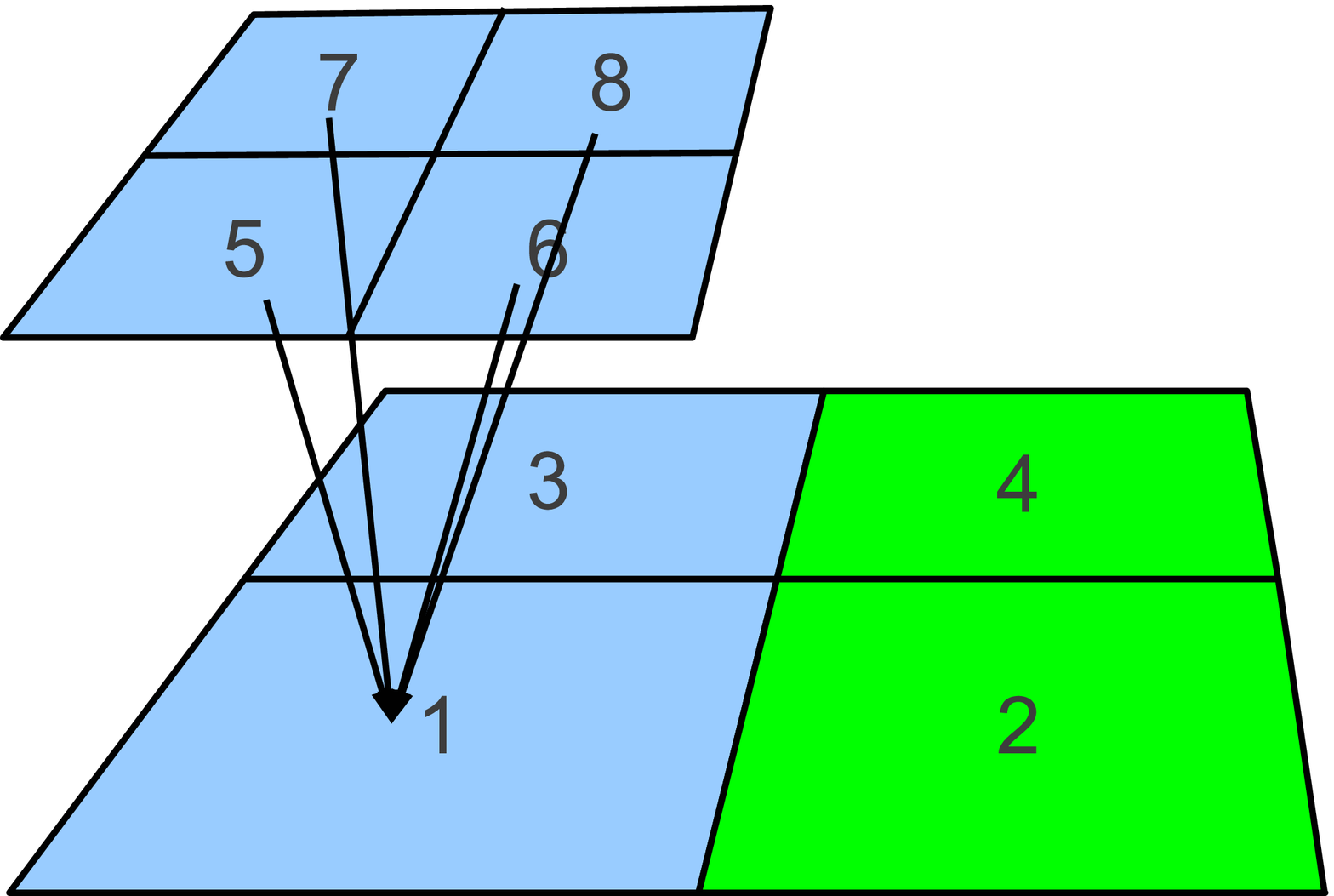}
\includegraphics[scale=0.22]{\figures/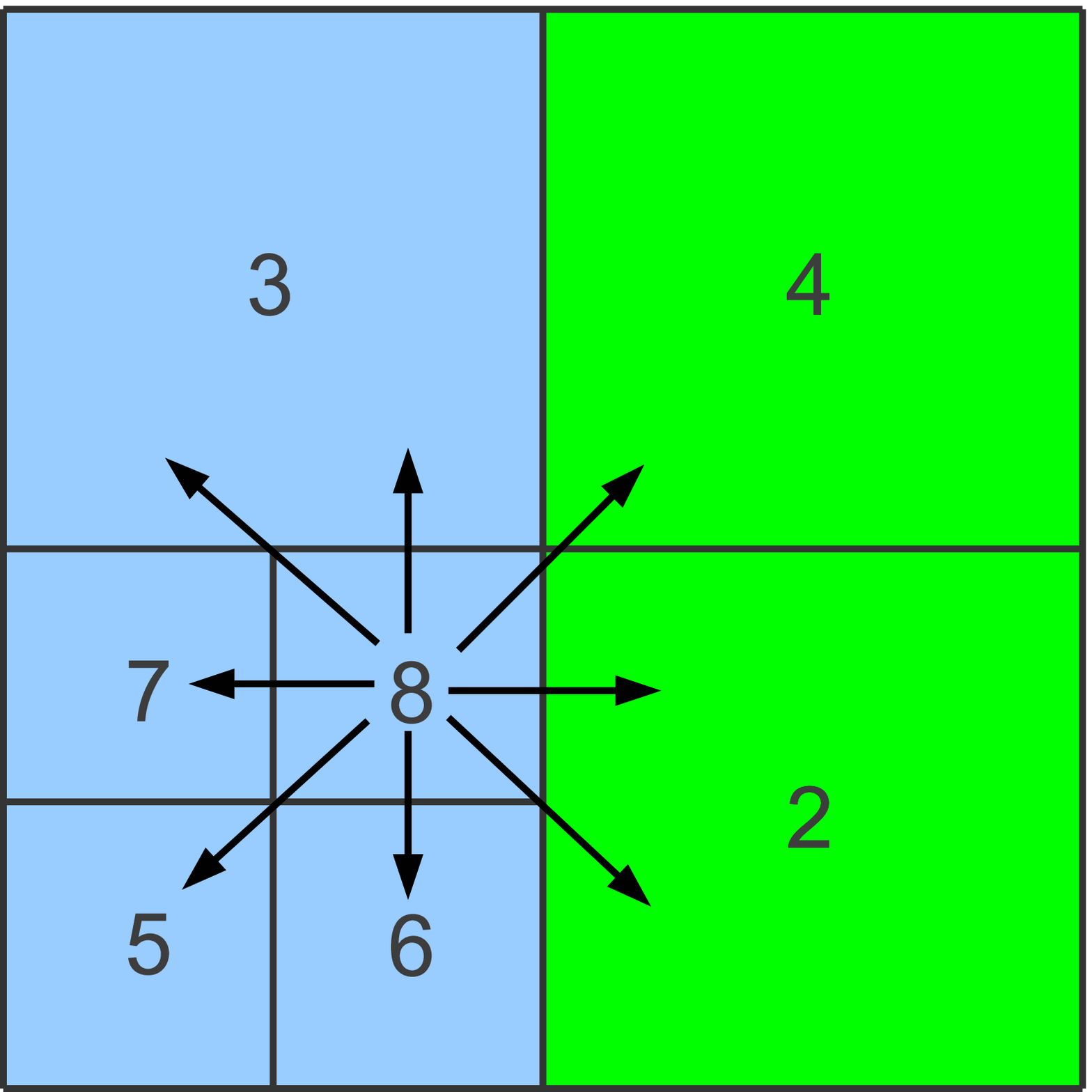}
\caption{The AMR hierarchical structure. Upper left panel: The computational
domain has four blocks with cells $4 \times 4$ and the regions in gray
and yellow are physical and inner boundaries. Only the inner boundaries
of block 1 have been plotted in this panel. Note that the inner boundaries
only exist between blocks. Upper middle panel: Blocks treated in processor 0 are
in blue and those in processor 1 are in green. Block 1 has been refined to
blocks 5, 6, 7 and 8. The inner boundaries have been omitted in this panel.
Upper right panel: Another viewing angle for the middle panel without the physical
and inner boundaries. The dashed lines in three panels are the
H-curves. Lower left panel: the $child$ points of block 1 point to blocks 5, 6, 7 and 8.
Lower middle panel: the $parent$ pointers of blocks 5, 6, 7 and 8 point to block 1.
Lower right panel: The eight $neigh$ pointers of block 8 point to blocks 5, 6, 2, 2, 
4, 3, 3 and 7, respectively.} \label{fig03}
\end{figure}

Supposing that we have a 2D
computational domain with the total mesh cells of $8 \times 8$, the
domain initially is divided into 4 subdomains, labeled by 1, 2, 3,
4. Every block has the same cells $4 \times 4$ and are surrounded by
physical boundaries in gray and inner boundaries in 
yellow as shown in the upper left panel of Fig. \ref{fig03} (note that
the inner boundaries exist only between blocks). The numerical
scheme discussed in Section~\ref{scheme} will solve the four
blocks in turn. Then we can get the solution at the new time by
gathering the data from all blocks.

\begin{figure}[htbp]
\centering
\includegraphics[height=160pt]{\figures/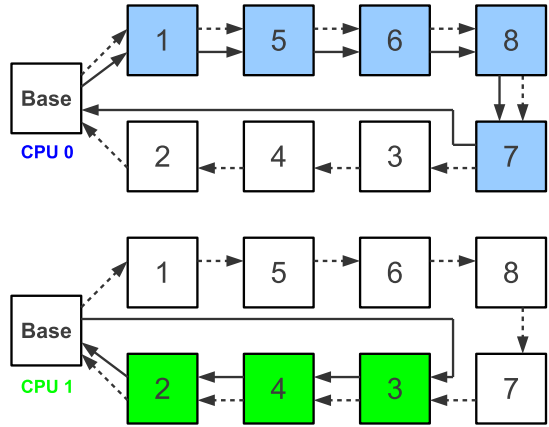}
\caption{The global links (dashed arrows) and local links (solid
arrows) between different blocks and different processors.}
\label{fig04}
\end{figure}

Assuming that blocks 1 and 3 are in processor 0, blocks 2 and 4 in
processor 1 and block 1 has already been refined as shown by
the upper middle and upper right panels of Fig.~\ref{fig03}. The number of
cells in blocks 5, 6, 7, 8 is the same as the block 1, i.e. $4
\times 4$, which means that the amount of calculation of the blocks
5, 6, 7, 8 is the same as the blocks 1, 2, 3, 4. All blocks are
connected by the so-called link list. Every process includes two
kind of link lists, one for collecting information in the local processor
and the other one for global information, i.e. (1) global link list and
(2) local link list. The local one refers to the links in the local
processor while the global one links all blocks which exist in all
processors as shown in Fig.~\ref{fig04}. That is, the local
ones are totally different in individual processors but the global
ones are exactly the same. Actually, the global link is enough to
complete the AMR algorithm, because one can judge which block in the
link is located in the local processor. However, as we know, the
global link sometimes is very long and may include several hundred
thousand blocks, so the frequently used searching operations may take a
long time. The order of these two links are arranged by the Hilbert
space filling curve (H-curve) which is shown by the dashed lines in
the upper panels of Fig.~\ref{fig03}. The H-curve maps multidimensional data to one
dimension while preserving locality of the data points. The order of
the H-curve in the finer level blocks is determined by the
parameters ($s\_pos$, $e\_pos$, $s\_point$ in Table~\ref{list_info})
of the H-curve in their parent blocks.

Every block, which is a STRUCTURE type variable in FORTRAN language,
includes some information and data. The contents of information are
listed in Table~\ref{list_info}. Some examples: the $lv$ of
block 8 is 2 and the $processor\_id$ of it is 0. The $id$ of
block 8 is $\{\{1, 1\}, \{2, 2\}\}$ and block 5 is $\{\{1, 1\},
\{1, 1\}\}$. The $child$ pointers of block 1 point to blocks
5, 6, 7 and 8 (lower left panel of Fig. \ref{fig03}), respectively. 
The $parent$ pointers of blocks 5, 6, 7, 8 point to block 1 
(lower middle panel of Fig. \ref{fig03}). The eight $neigh$ pointers 
of block 8 point to blocks 5, 6, 2, 2, 4,
3, 3, 7 (lower right panel of Fig. \ref{fig03}), respectively. 
The $neigh$ pointers have some identical
values because, for instance, the right and lower right neighbors
are the same.

\begin{table}[htbp]
\caption{The block information.\label{list_info}}
\begin{tabular}{cc}
\hline
\hline
Variable         & Interpretation \\
\hline
$lv$             & Level of current block.                           \\
$processor\_id$  & Processor rank the current block belong to.             \\
$nx, ny, nz$     & Grid points or cells in $x$, $y$, $z$-direction of the current block.\\
$x, y, z$        & $x$, $y$, $z$-coordinates of the current block.              \\
$h\_number$      & Position of the current block in the Hilbert curve.    \\
\hline
$s\_pos, e\_pos, s\_point$  &  \begin{tabular}{c}
                             Used to generate the Hilbert curve in the finer level  \\
                             (array).         \\
                             \end{tabular}   \\
$id$             & \begin{tabular}{c}
                 Position of the current block in the total    \\
                 AMR hierarchical levels (array).   \\
                 \end{tabular}                     \\
\hline
$parent$         & Link to the parent block of the current block (pointer). \\
$child$          & \begin{tabular}{c}
                 Link to the children blocks of the current block,   \\
                 if no point to NULL (pointer array).  \\
                 \end{tabular}   \\
$neigh$          & \begin{tabular}{c}
                 Link to the neighbor blocks of the current block,   \\
                 if no point to NULL (pointer array).  \\
                 \end{tabular}   \\
$next$           & Link to the next block in the local link list (pointer). \\
$framework\_next$ & Link to the next block in the global link list (pointer). \\
\hline
\end{tabular}
\end{table}

\subsection{Load balance}
Before introducing what is the load balance, we first introduce the
definition of $framework$, which is very important to find a simple
way to do the load balance. Just as its name
implies, $framework$ is the framework of the AMR hierarchical
structure. It is the global link list plus the all blocks'
information listed by Table~\ref{list_info}. Like the global link
list, the $framework$ in all processors should be exactly the same.
In this case, block 8 in the processor 0 knows which is its
right neighbor block located in the processor 1 as shown in
Fig. \ref{fig03}. The processors have the same $framework$s, but
they need not store all the variables of the blocks which belong
to other processors. That is to say the blocks which belong to other 
processor have only the information listed by Table 1. 
In Fig. \ref{fig04}, the blocks in blue and
green colors denote that only these blocks have allocated the memory
for variables. It is noted that the blocks without the variable data 
can still be refined or destroyed. The $child$ blocks also have no 
variable data. And the memory required to store the entire global link 
list is very small.

The $framework$ structure has some advantages: (1) load balance can be done all at once 
rather than iteratively. That is to say, all blocks can be 
refined or destroyed to a suitable level 
according to the regrid algorithm. However, there is a special 
situation we have to carry on the regriding and load balance level by level
(iteratively). When only a very small region has to refine to a very high 
level, while this region locates in only one processor, then the code has to 
allocate a huge number of memory in one processor. If the memory in one 
node is not enough, then the code crashes. Therefore, in our MAP code we 
can choose whether carrying on the regrid and load balance processes 
iteratively or not. (2) The data exchanging is simple. When carrying on the 
load balance, for instance, Processor 1 has to send the data of block A to processor 2, 
the processor 1 only needs to send the data of variables to the processor 2. 
And the block A (it already exists in processor 2 but without variable data)
in processor 2 will be filled by this data. It needs not to send the neighbors 
information to processor 2 simultaneously. 
Because of the so-called $framework$, processors know the number of
blocks in different processors and the total number of blocks in all
processors. Thus, the code can calculate the load in different
processors and balance the load by sending the block data from one
processor to other one. The load is measured by the number of blocks
which will be solved using the schemes described in
Section~\ref{scheme}. For instance, the block 3 in the upper middle panel of
Fig.~\ref{fig03} will to sent to processor 1 by MPI calls because
there are five blocks needed to be calculated in the processor 0 (the
block 1 can be updated by the average values of its child blocks)
and two blocks in the processor 1. This is not a rigorous balance because of the
number of blocks which will be solved, the parent blocks (e.g. the
block 1) which will be updated, and the boundary data which will be
exchanged are different. However, this difference is very small
when the total number of blocks is much greater than the number of
processors.

After the load balance is done, the code has to do another
important job: neighbor statistics. This is a preparation for the boundary
conditions between different processors in the next several or
dozens of time steps as shown in the schematic chart
(Fig.~\ref{fig01}). In this procedure, the code will record the
number and the $id$ of blocks which need to reflux, the blocks which need to
send data from fine blocks to coarse blocks, and the blocks which need to send
data to the blocks at the same level. Once the number of the blocks
is known, the code can allocate enough memory to store the data
which will be sent or received in advance. When the block $id$ is
known, the block will be found immediately and store its data to the
allocated array or update its data from the allocated array. This
can reduce the operations of allocating and freeing memory, so that the
performance of the code can be greatly improved.

\subsection{Boundary condition}

The neighbor statistics can provide enough information for processing
the boundary conditions quickly. As we described above, for boundary
conditions we need to do: (1) data updating from the child blocks to
the parent blocks; (2) reflux; (3) data exchange between blocks in
the same level; (4) data exchange between different levels. We have
several steps to do these. Every step has two small sub-steps:
inter-processor first and then intra-processor. The inter-processor
means the boundary conditions between different processors and the
intra-processor stands for the boundary conditions inside one
processor. The procedure is: (1) to update the parent data. In this
step, we should care about shareblocks in the inter-processor case.
The shareblock was firstly introduced by~\cite{Ziegler2008}, which
refers to the child blocks of one block located in different
processors. Suppose that block 8 in Fig.~\ref{fig03} belongs
to the processor 1 while its parent block belongs to the processor
0, so the block 8 is called the shareblock. The shareblock has to
send all of its data (without the ghost region) to the corresponding position 
in the global link list of processor 0
in order to update the data of block 1; (2) to reflux the data
between the fine-coarse boundaries as shown by the upper right panel of
Fig.~\ref{fig03}. Because the data of the coarse block 1 is
updated by the fine blocks 5, 6, 7, 8 and the flux between the
blocks 1 and 2 is changed, the block 2 has to use the new numerical
flux to update its boundary data as suggested by~\cite{Berger1984}
and \cite{Berger1989}; (3) to exchange the $x$-direction data in the
same level; (4) to exchange the $y$-direction data in the same level;
(5) to exchange the $z$-direction data in the same level. The MAP code
does not need extra requirement of MPI calls for the corner data (as
shown by Fig.~\ref{fig05}). The steps (3) - (5) can be
explained by the left panel of Fig.~\ref{fig05}. In this figure,
the blue dashed ghost region will be updated by the data of the blue
dashed box in the block 2; Then, the ghost region of the block 3
will be updated by the data of the red dashed box (already included
the corner data) in the block 1. Of course, the steps (3) - (5) also
treat the physical boundary conditions at the same time; (6) to update
the ghost region of the fine block by interpolating the data of the
coarse block at the fine-coarse boundaries. The only important thing
here is that we should use a conservative interpolation. The formula
is taken from~\cite{Matsumoto2007}:

\begin{figure}[htbp]
\centering
\includegraphics[height=160pt]{\figures/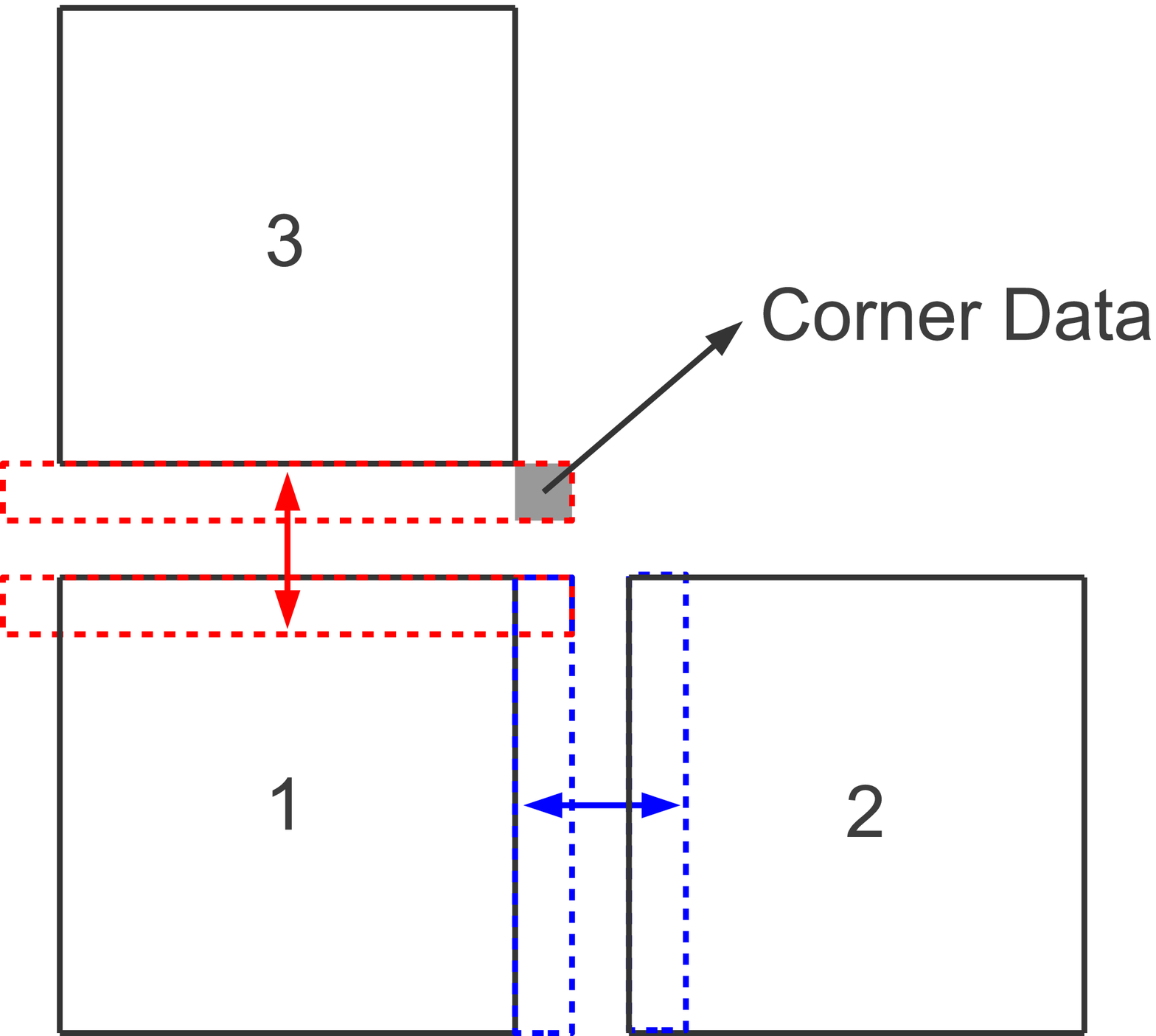}
\includegraphics[height=160pt]{\figures/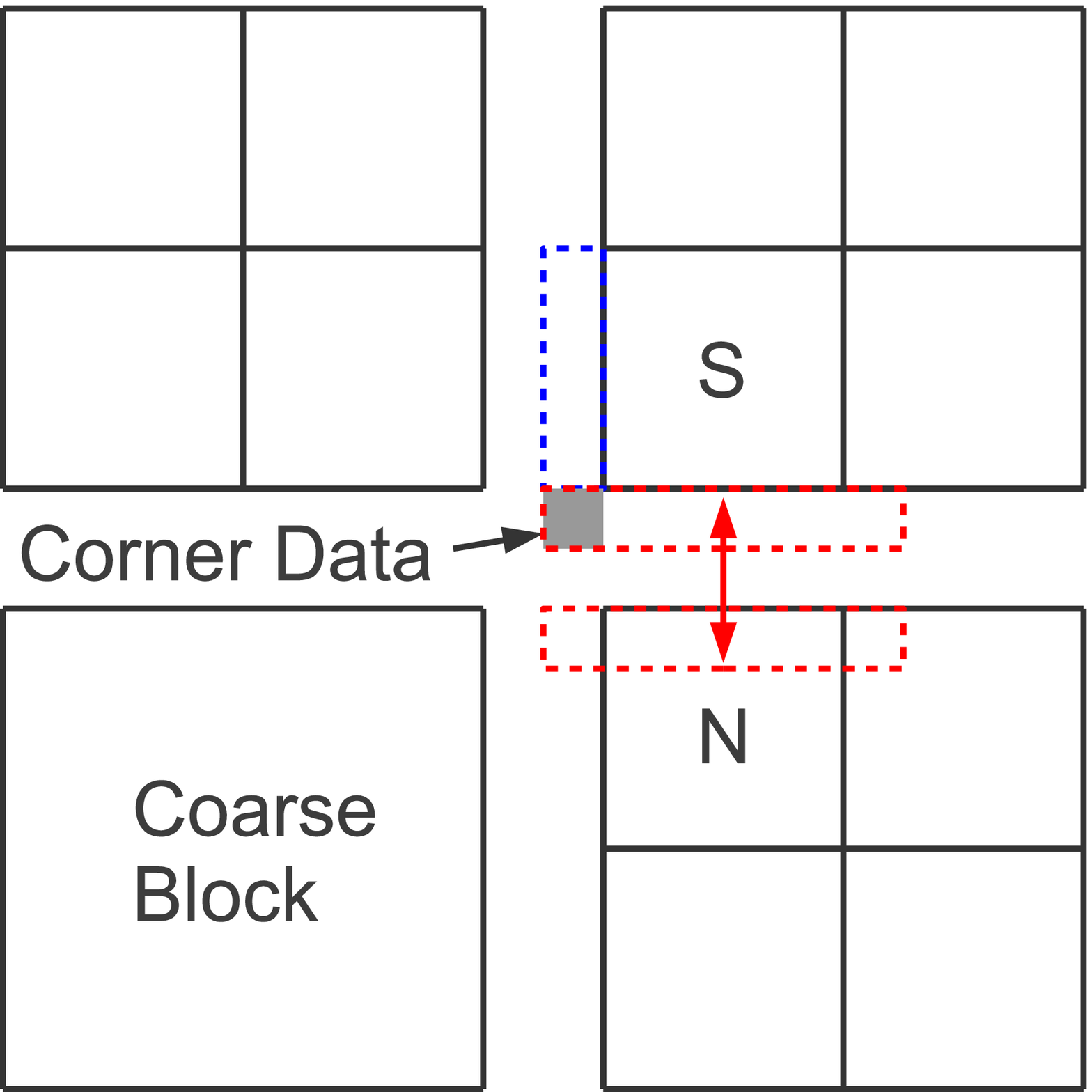}
\caption{Left panel shows the data exchange at the same level. In
the panel, we exchange the data in the $x$-direction first and then the
$y$-direction. The corner data are also updated. The right panel
shows a special situation in which the corner data can not be
updated correctly.} \label{fig05}
\end{figure}

\begin{eqnarray}
u^f_{i, j, k} = u^c_{i^c, j^c, k^c} + \mathrm{minmod}\left
      (u^c_{i^c + 1, j^c, k^c} - u^c_{i^c, j^c, k^c},
       u^c_{i^c, j^c, k^c} - u^c_{i^c - 1, j^c, k^c}\right)\frac{x^f(i) - x^c(i^c)}{\Delta x^c} \,\,\, \notag \\
                                 + \mathrm{minmod}\left
      (u^c_{i^c, j^c + 1, k^c} - u^c_{i^c, j^c, k^c},
       u^c_{i^c, j^c, k^c} - u^c_{i^c, j^c - 1, k^c}\right)\frac{y^f(j) - z^c(j^c)}{\Delta y^c} \,\,\, \\
                                 + \mathrm{minmod}\left
      (u^c_{i^c, j^c, k^c + 1} - u^c_{i^c, j^c, k^c},
       u^c_{i^c, j^c, k^c} - u^c_{i^c, j^c, k^c - 1}\right)\frac{z^f(k) - z^c(k^c)}{\Delta z^c} \,, \notag
\end{eqnarray}

\noindent where the fine block ($f$) is a child of the coarse block
($c$), the grids index ($i^c, j^c, k^c$) is for the coarse block and
the coordinate ($x, y, z$) is defined at the center of the cell. The
function of $\mathrm{minmod}$ is given by Eq.  (\ref{minmod});
(7) to fix the corner data. The steps (1) - (6) are successful to exchange
almost all boundary data except one special situation shown by the
right panel of Fig.~\ref{fig05}. Only the lower left neighbor of the
special block ($S$) is a coarse block. Before the fine-coarse step
(6), the non-updated corner data of the normal block ($N$) are sent
to the block ($S$) since the steps (3) - (5) is the advance step
(6). Thus the corner data of the special block ($S$) need a fix
step. In the MAP code, we just look for these so-called $S$ blocks and
do the steps (4) - (5) again.

\subsection{Framework synchronization}

As we mentioned above, every processor knows the total information
of the AMR hierarchical structure, it is possible to accomplish the
load balance and the boundary conditions quickly and effectively.
However, the next question is how to keep the $framework$ systems
synchronous for every processor.

The framework synchronization is something like the cloning
technique. The processors can generate the same $framework$ by the
same $gene$. The local sequence of $gene$ is built by recording the
$id$ of every block which will be regrided (including refining and
destroying procedures) in a regridding operation in the current processor.
The global sequence of $gene$ is obtained by merging all the local
sequences. The MPI communication of $mpi\_allgatherv$ is necessary
for such a merging. The criterion for refining or destroying will
be discussed in Section~\ref{regird}. With the assumption that we have
the same $framework$ in the current state, we can get the same $gene$
after regriding and then the code should generate the exactly same
$framework$ at the next time step.

\subsection{Regrid}
\label{regird} The error estimation formula is taken
from~\citet{Matsumoto2007, Ziegler2008}, which is determined by the first and second
derivatives in the current level:

\begin{equation}
\left(\frac{B ||\Delta u||_2}{|u| + \epsilon} +
\frac{(1-B)||\Delta^2 u||_2}{||\Delta u||_2 + F \cdot \left(|u| +
\epsilon\right)}\right) R ^ {\xi (L - 1)} \,\, ,
\end{equation}

\begin{equation}
||\Delta u||_2 = \frac{\sqrt3}{2}\left(\left(u_{i + 1, j, k} - u_{i
- 1, j, k}\right)^2 + \left(u_{i, j + 1, k} - u_{i, j - 1,
k}\right)^2 + \left(u_{i, j, k + 1} - u_{i, j, k -
1}\right)^2\right)^{1/2} \,\, ,
\end{equation}

\begin{eqnarray}
||\Delta^2 u||_2 = (\left(u_{i + 1, j, k} - 2u_{i, j, k} + u_{i - 1, j, k}\right)^2 + \,\,\, \notag \\
\left(u_{i, j + 1, k} - 2u_{i, j, k} + u_{i, j - 1, k}\right)^2 + \,\,\, \\
\left(u_{i, j, k + 1} - 2u_{i, j, k} + u_{i, j, k - 1}\right)^2)^{1/2} \, \, , \notag
\end{eqnarray}

\noindent where the value of the filter $F$ is 0.05, $R$ is the
refinement ratio and $L$ the level of the current block, and
$\epsilon$ is taken as $10^{-12}$ for $\rho, p, T$ and 
$0.1$ for $\mathbf{v},\mathbf{B}$. The parameter $B = 0.6$ specifies a bias toward the
first derivatives for the criterion. The value $\xi$ makes the
threshold for higher levels higher ($\xi < 0$) or lower ($\xi > 0$).
When the error exceeds the given threshold values, the current block
will be refined. If the errors estimated in every cell in the
current block are larger than the threshold values, the child block
of the current block can be destroyed. We also try to use the
Richardson error estimation, however, this method usually takes a long
time and is not so effective. For this reason, the Richardson error
estimation is not included in the MAP code.

\begin{figure}[htbp]
\centering
\includegraphics[height=160pt]{\figures/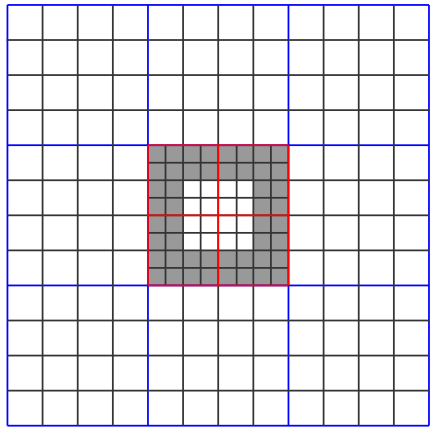}
\caption{The blocks with blue borders show the coarse blocks and the
red ones show the fine blocks. Assuming that the ghost cell is $1$,
the buffer zone for the next regridding is the shadowy cells in the fine
blocks since the MAP code checks every cell in the current block. }
\label{fig06}
\end{figure}

After several time steps the regridding operation will be carried
on. The estimation interval for the next regridding is

\begin{equation}
\Delta t_r = c_r \frac{2 n_g \min(\Delta x, \Delta y, \Delta
z)}{\alpha_g}= c_r\frac{2 n_g n_{dim} \Delta t}{c_{cfl}},
\label{regrid_time_1}
\end{equation}

\noindent
where $\Delta t_r$ is the time interval for the next regridding and
$\Delta t$ is the time step of the simulation, which is taken from the
finest level, and $\alpha_g$ is the global maximum wave speed. The
different time steps for different levels will be achieved in the
future work. Since the code checks the errors for every cell
covering the ghost region so the length of buffer region in which
the waves or discontinuities do not propagate to the coarse block
before the next regridding time is $2 n_g$ (Fig.~\ref{fig06}),
where $n_g$ is the number of ghost cells.

\section{Numerical tests} \label{test}

Our MAP code is examined with a variety of test and application problems
in this section. The 1D problems test the difference between
different schemes and the accuracy of them. The 2D and 3D problems
check the AMR parallel performance of our code. In all our test cases, 
we use the primitive variables to describe the initial conditions.

\subsection{1D accuracy test}
We test the accuracy of the three schemes, namely, MMC, LF and 
WENO by simplest advection problem: 
$(\rho, v_x, v_y, v_x, B_x, B_y, B_z, p) = ((\sin(2 \pi x)
+ 2)/3, 1, 0, 0, 0, 0, 0, 1)$ in the domain $[-0.5, 0.5]$ with the
periodic boundary. Lax-Friedrichs splitting for MMC and 
Lax-Friedrichs flux for LF and WENO are used in this test.
The adiabatic index is $\gamma = 1.4$. The $L^1$, $L^2$ and
$L^\infty$ errors are listed in Table~\ref{list_error_1d}, which are
defined as:

\begin{equation}
L^1(u) = \frac{1}{N} \sum_{i=1}^{N}|u_i^n-U_i| \, \, ,
\end{equation}
\begin{equation}
L^2(u) = \left(\frac{1}{N} \sum_{i=1}^{N}|u_i^n-U_i|^2\right)^{1/2} \, \, ,
\end{equation}
\begin{equation}
L^\infty(u) = \max_{1 \le i \le N}(u_i^n-U_i) \, \, ,
\end{equation}

\noindent where $U$ is the exact solution and $N$ the grid
number. From Table~\ref{list_error_1d}, it can be seen that these schemes are second order
accuracy and WENO scheme is better than the other two. This is why
we mainly use WENO for most of our 2D and 3D tests. 

\begin{table}[htbp]
\caption{$L^1$, $L^2$ and $L^\infty$ error and order in 1D accuracy test.\label{list_error_1d}}
\begin{tabular}{ccccccccc}
\hline
\hline
Scheme  & \vline &  $N$  & $L^1$ error       & $L^1$ order  & $L^2$ error & $L^2$ order      & $L^\infty$ error & $L^\infty$ order \\
\hline
        & \vline & $128$ & $5.47 \times 10^{-3}$ & $  -   $ & $7.27 \times 10^{-3}$ & $ -  $   & $1.74 \times 10^{-2}$ & $ -  $  \\
        & \vline & $256$ & $1.55 \times 10^{-3}$ & $ 1.82 $ & $2.34 \times 10^{-3}$ & $1.64$   & $7.18 \times 10^{-3}$ & $1.27$  \\
MMC     & \vline & $384$ & $7.37 \times 10^{-4}$ & $ 1.84 $ & $1.20 \times 10^{-3}$ & $1.65$   & $4.25 \times 10^{-3}$ & $1.30$  \\
        & \vline & $512$ & $4.32 \times 10^{-4}$ & $ 1.85 $ & $7.42 \times 10^{-4}$ & $1.66$   & $2.91 \times 10^{-3}$ & $1.31$  \\
        & \vline & $640$ & $2.84 \times 10^{-4}$ & $ 1.88 $ & $5.12 \times 10^{-4}$ & $1.66$   & $2.17 \times 10^{-3}$ & $1.31$  \\
\hline
        & \vline & $128$ & $4.64 \times 10^{-3}$ & $  -   $ & $6.32 \times 10^{-3}$ & $ -  $   & $1.72 \times 10^{-2}$ & $ -  $  \\
        & \vline & $256$ & $1.31 \times 10^{-3}$ & $ 1.82 $ & $2.03 \times 10^{-3}$ & $1.64$   & $7.12 \times 10^{-3}$ & $1.27$  \\
LF      & \vline & $384$ & $6.20 \times 10^{-4}$ & $ 1.85 $ & $1.04 \times 10^{-3}$ & $1.65$   & $4.22 \times 10^{-3}$ & $1.29$  \\
        & \vline & $512$ & $3.62 \times 10^{-4}$ & $ 1.87 $ & $6.44 \times 10^{-4}$ & $1.66$   & $2.90 \times 10^{-3}$ & $1.30$  \\
        & \vline & $640$ & $2.38 \times 10^{-4}$ & $ 1.89 $ & $4.45 \times 10^{-4}$ & $1.66$   & $2.17 \times 10^{-3}$ & $1.31$  \\
\hline
        & \vline & $128$ & $2.99 \times 10^{-3}$ & $-     $ & $4.62 \times 10^{-3}$ & $ -  $   & $1.24 \times 10^{-2}$ & $ -  $  \\
        & \vline & $256$ & $7.04 \times 10^{-4}$ & $ 2.09 $ & $1.35 \times 10^{-3}$ & $1.77$   & $4.68 \times 10^{-3}$ & $1.41$  \\
WENO    & \vline & $384$ & $2.96 \times 10^{-4}$ & $ 2.14 $ & $6.54 \times 10^{-4}$ & $1.79$   & $2.61 \times 10^{-3}$ & $1.44$  \\
        & \vline & $512$ & $1.59 \times 10^{-4}$ & $ 2.16 $ & $3.89 \times 10^{-4}$ & $1.81$   & $1.73 \times 10^{-3}$ & $1.44$  \\
        & \vline & $640$ & $9.79 \times 10^{-5}$ & $ 2.17 $ & $2.60 \times 10^{-4}$ & $1.81$   & $1.25 \times 10^{-3}$ & $1.46$  \\
\hline
\end{tabular}
\end{table}

\subsection{1D Brio-Wu shocktube test}
\begin{figure}[htbp]
\centering
\includegraphics[height=160pt]{\figures/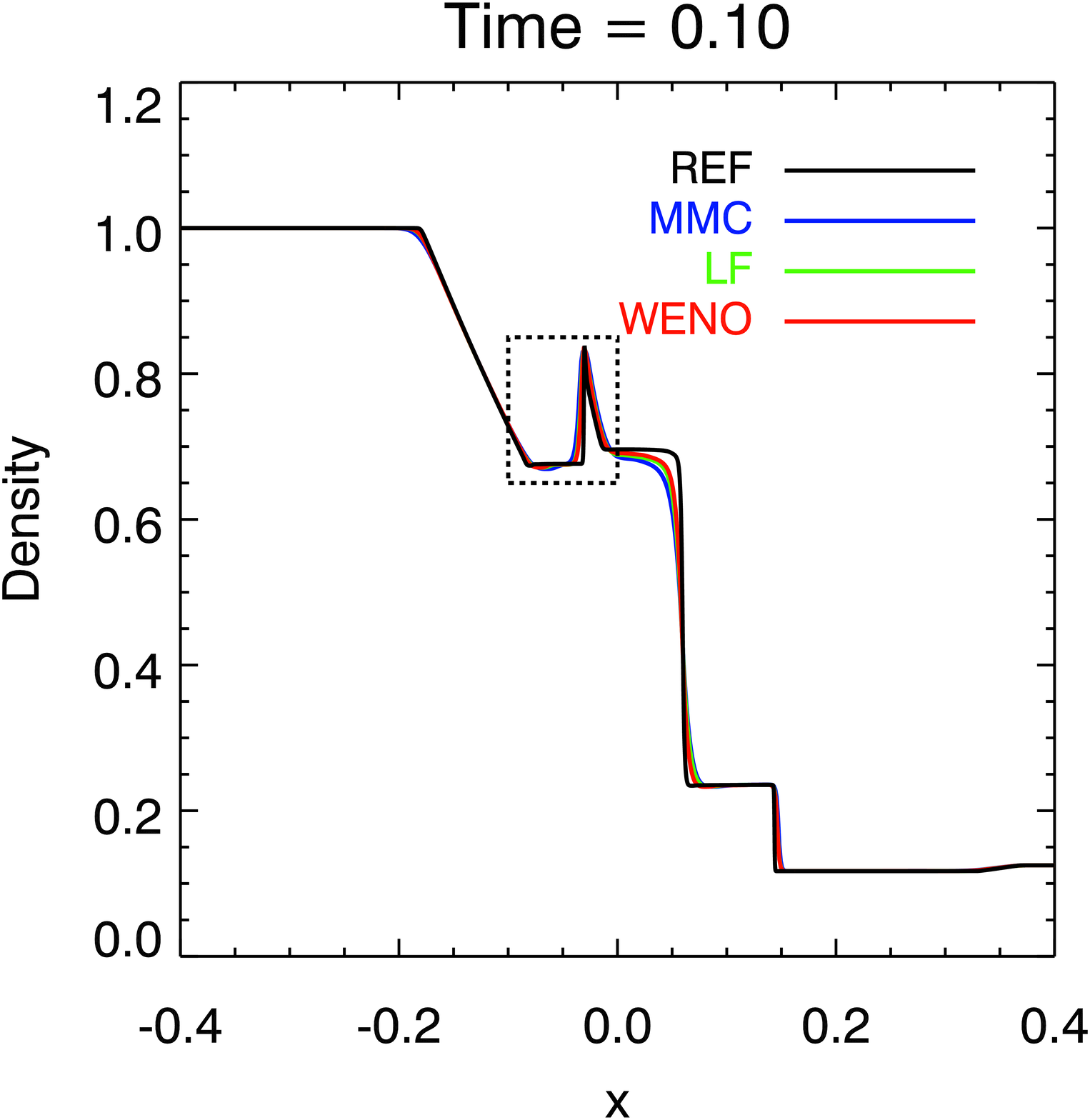}
\includegraphics[height=160pt]{\figures/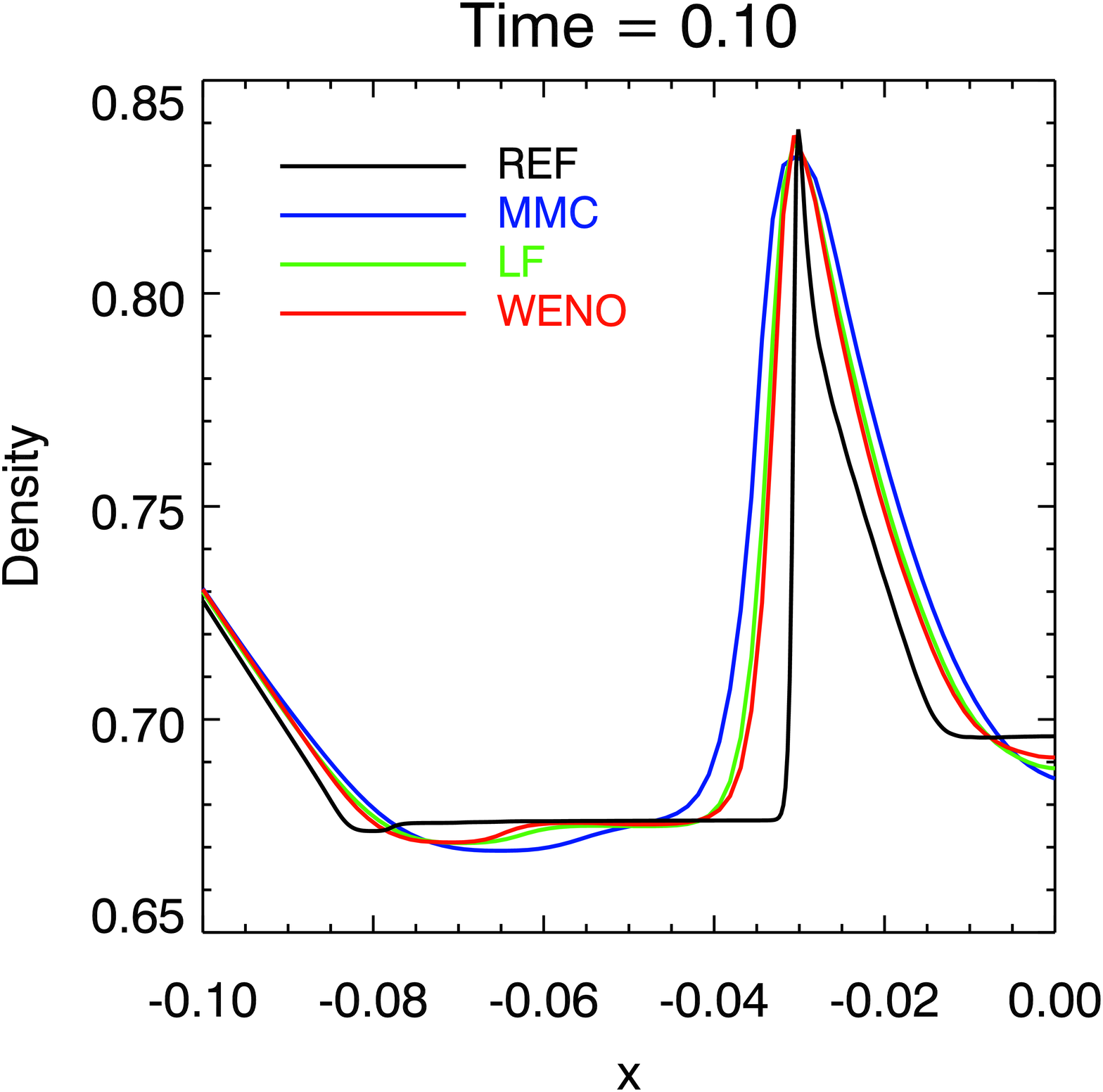}
\includegraphics[height=160pt]{\figures/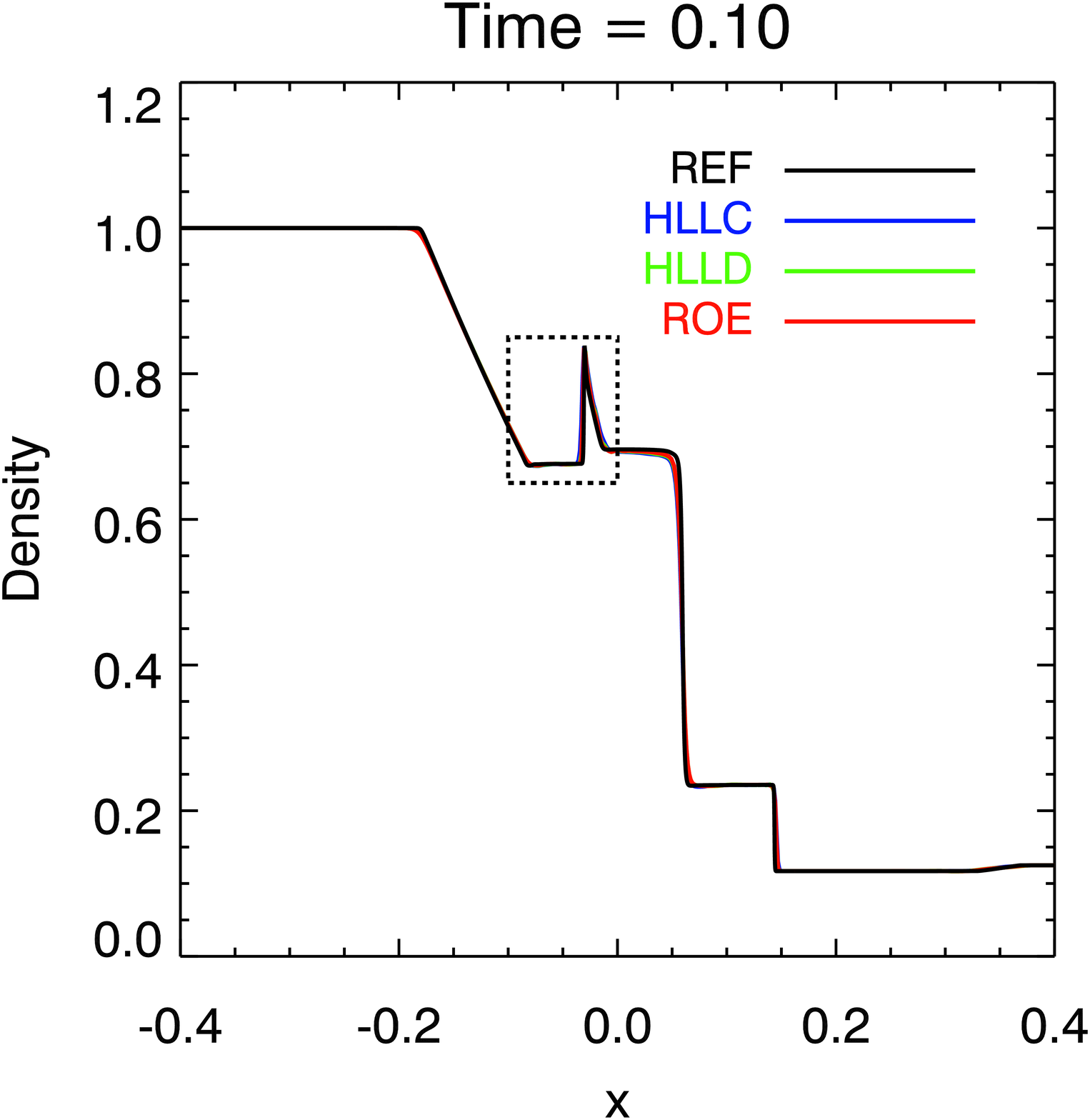}
\includegraphics[height=160pt]{\figures/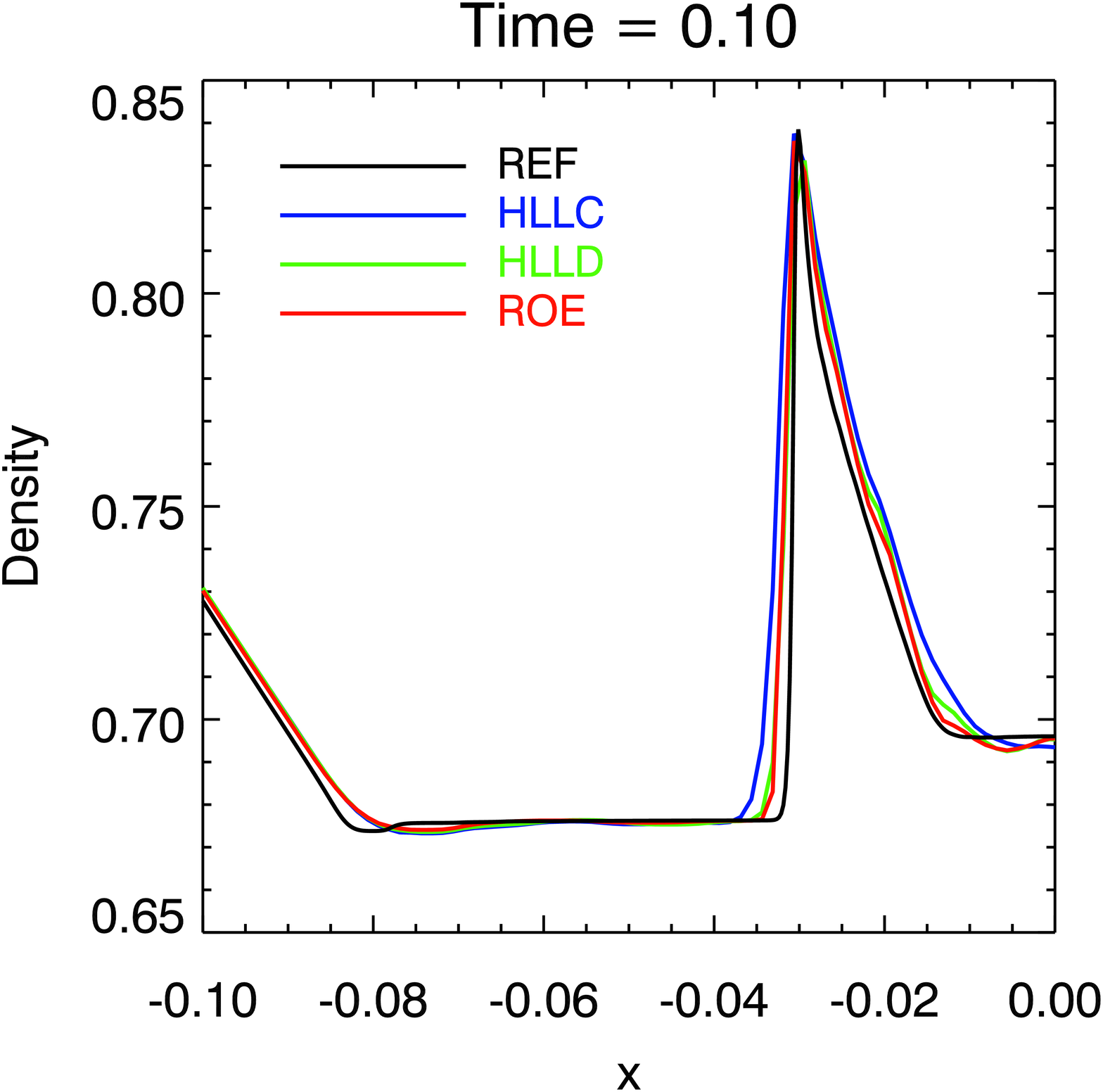}
\caption{Comparsion between the modified Mac Cormack Scheme (MMC),
Lax-Fridrichs scheme (LF), and weighted essentially non-oscillatory
(WENO) scheme in the 1.5D MHD shocktube problem. The figure shows the
range from $-0.4$ to $0.4$ at time $0.10$. The solid line is a
reference solution with $4000$ cells, while the MMC, LF, WENO
schemes are computed with $800$ cells. The lower panels are
calculated based on the WENO scheme with different approximate Riemann
solvers, i.e. HLLC, HLLD and Roe solver. The right panel is a zoom-in
view of the dashed box in the left panel. } \label{fig07}
\end{figure}

This test is taken from~\cite{Brio1988}, which is a standard 1.5D
(one dimensional multi components) MHD shocktube problem. The
initial conditions of the left ($x < 0$) and right ($x > 0$) states are
$(\rho, v_x, v_y, v_x, B_x, B_y, B_z, p)_L = (1, 0, 0, 0, 0.75, 1, 0, 1)$ 
and $(\rho, v_x, v_y, v_x, B_x, B_y, B_z, p)_R = (0.125, 0,
0, 0, 0.75, -1, 0, 0.1)$ with the adiabatic index $\gamma = 2$. The
length of computational domain is 1 and $x \in [-0.5, 0.5]$. The
upper panels of Fig.~\ref{fig07} show the comparison between the
three kinds of schemes without the approximate Riemann solvers 
(Lax-Friedrichs splitting for MMC and 
Lax-Friedrichs flux for LF and WENO) while the
lower panels show the same scheme (WENO) but with different Riemann
solvers. From this figure, we found that the WENO and LF schemes are
much better than the MMC scheme (upper panels). Although the MMC scheme has the
same accuracy as the LF scheme, the former is not suitable to treat the MHD
problem. HLLD and Roe Riemann solvers are more accurate than HLLC (lower panels).
However, in our multi-dimensional tests, the HLLD and Roe solvers
fail sometimes, thus we mainly use HLLC for our 2D and 3D test
problems.

\subsection{2D accuracy test}
As the 1D advection test, we also test the convergence of the 
three schemes in 2D advection problem: 
$(\rho, v_x, v_y, v_x, B_x, B_y, B_z, p) = ((\sin(2 \pi (x + y / \sqrt{3}) + pi / 2)
+ 2)/3, 1, \sqrt{3}, 0, 0, 0, 0, 1)$ in the domain 
$[-0.5, 0.5] \times [-\sqrt{3} / 2, \sqrt{3} / 2]$ with the
periodic boundary. The velocity has angle of $\pi / 6$ with the $y$-axis. 
Lax-Friedrichs splitting for MMC and 
Lax-Friedrichs flux for LF and WENO are used in this test. 
The adiabatic index $\gamma = 1.4$. The $L^1$, $L^2$ and
$L^\infty$ errors are listed in Table~\ref{list_error_2d}. 
As shown by this table, we got almost the same results with the 1D test. 
Moreover, we found that the extended 3D advection test also given the similar 
result, which is no longer necessary to discuss in this paper.

\begin{table}[htbp]
\caption{$L^1$, $L^2$ and $L^\infty$ error and order in 2D accuracy test.\label{list_error_2d}}
\begin{tabular}{ccccccccc}
\hline
\hline
Scheme  & \vline &  $N$  & $L^1$ error       & $L^1$ order  & $L^2$ error & $L^2$ order      & $L^\infty$ error & $L^\infty$ order \\
\hline
        & \vline & $64$  & $1.89 \times 10^{-2}$ & $  -   $ & $2.30 \times 10^{-2}$ & $ -  $   & $4.82 \times 10^{-2}$ & $ -  $  \\
        & \vline & $128$ & $6.35 \times 10^{-3}$ & $ 1.59 $ & $8.34 \times 10^{-3}$ & $1.48$   & $2.09 \times 10^{-2}$ & $1.22$  \\
MMC     & \vline & $192$ & $2.95 \times 10^{-3}$ & $ 1.90 $ & $4.32 \times 10^{-3}$ & $1.63$   & $1.26 \times 10^{-2}$ & $1.25$  \\
        & \vline & $256$ & $1.76 \times 10^{-3}$ & $ 1.80 $ & $2.70 \times 10^{-3}$ & $1.65$   & $8.75 \times 10^{-3}$ & $1.27$  \\
        & \vline & $320$ & $1.16 \times 10^{-3}$ & $ 1.90 $ & $1.86 \times 10^{-3}$ & $1.67$   & $6.58 \times 10^{-3}$ & $1.29$  \\
\hline
        & \vline & $64$  & $2.07 \times 10^{-2}$ & $  -   $ & $2.49 \times 10^{-2}$ & $ -  $   & $5.16 \times 10^{-2}$ & $ -  $  \\
        & \vline & $128$ & $6.99 \times 10^{-3}$ & $ 1.58 $ & $9.09 \times 10^{-3}$ & $1.47$   & $2.25 \times 10^{-2}$ & $1.21$  \\
LF      & \vline & $192$ & $3.35 \times 10^{-3}$ & $ 1.83 $ & $4.67 \times 10^{-3}$ & $1.65$   & $1.35 \times 10^{-2}$ & $1.25$  \\
        & \vline & $256$ & $1.96 \times 10^{-3}$ & $ 1.87 $ & $2.91 \times 10^{-3}$ & $1.64$   & $9.38 \times 10^{-3}$ & $1.28$  \\
        & \vline & $320$ & $1.29 \times 10^{-3}$ & $ 1.87 $ & $2.01 \times 10^{-3}$ & $1.66$   & $7.04 \times 10^{-3}$ & $1.29$  \\
\hline
        & \vline & $64$  & $1.78 \times 10^{-2}$ & $-     $ & $2.09 \times 10^{-2}$ & $ -  $   & $4.14 \times 10^{-2}$ & $ -  $  \\
        & \vline & $128$ & $4.39 \times 10^{-3}$ & $ 2.05 $ & $6.40 \times 10^{-3}$ & $1.72$   & $1.63 \times 10^{-2}$ & $1.36$  \\
WENO    & \vline & $192$ & $1.90 \times 10^{-3}$ & $ 2.07 $ & $3.15 \times 10^{-3}$ & $1.76$   & $9.26 \times 10^{-3}$ & $1.40$  \\
        & \vline & $256$ & $1.05 \times 10^{-3}$ & $ 2.08 $ & $1.89 \times 10^{-3}$ & $1.78$   & $6.17 \times 10^{-3}$ & $1.42$  \\
        & \vline & $320$ & $6.64 \times 10^{-4}$ & $ 2.06 $ & $1.27 \times 10^{-3}$ & $1.79$   & $4.49 \times 10^{-3}$ & $1.43$  \\
\hline
\end{tabular}
\end{table}

\subsection{2D parallelization efficiency test}
\label{efficiency_test}
\begin{figure}[htbp]
\centering
\includegraphics[height=160pt]{\figures/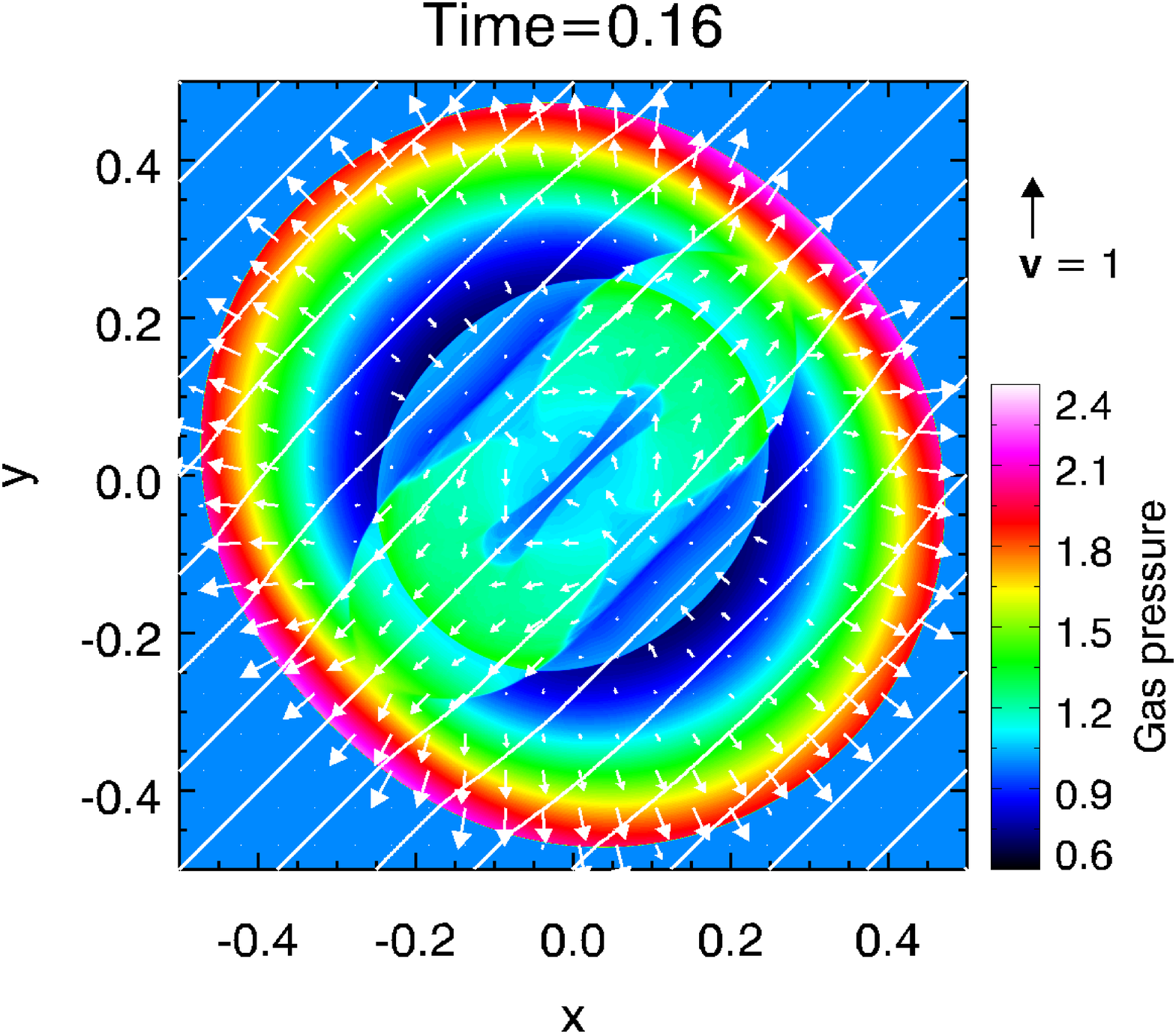}
\includegraphics[height=160pt]{\figures/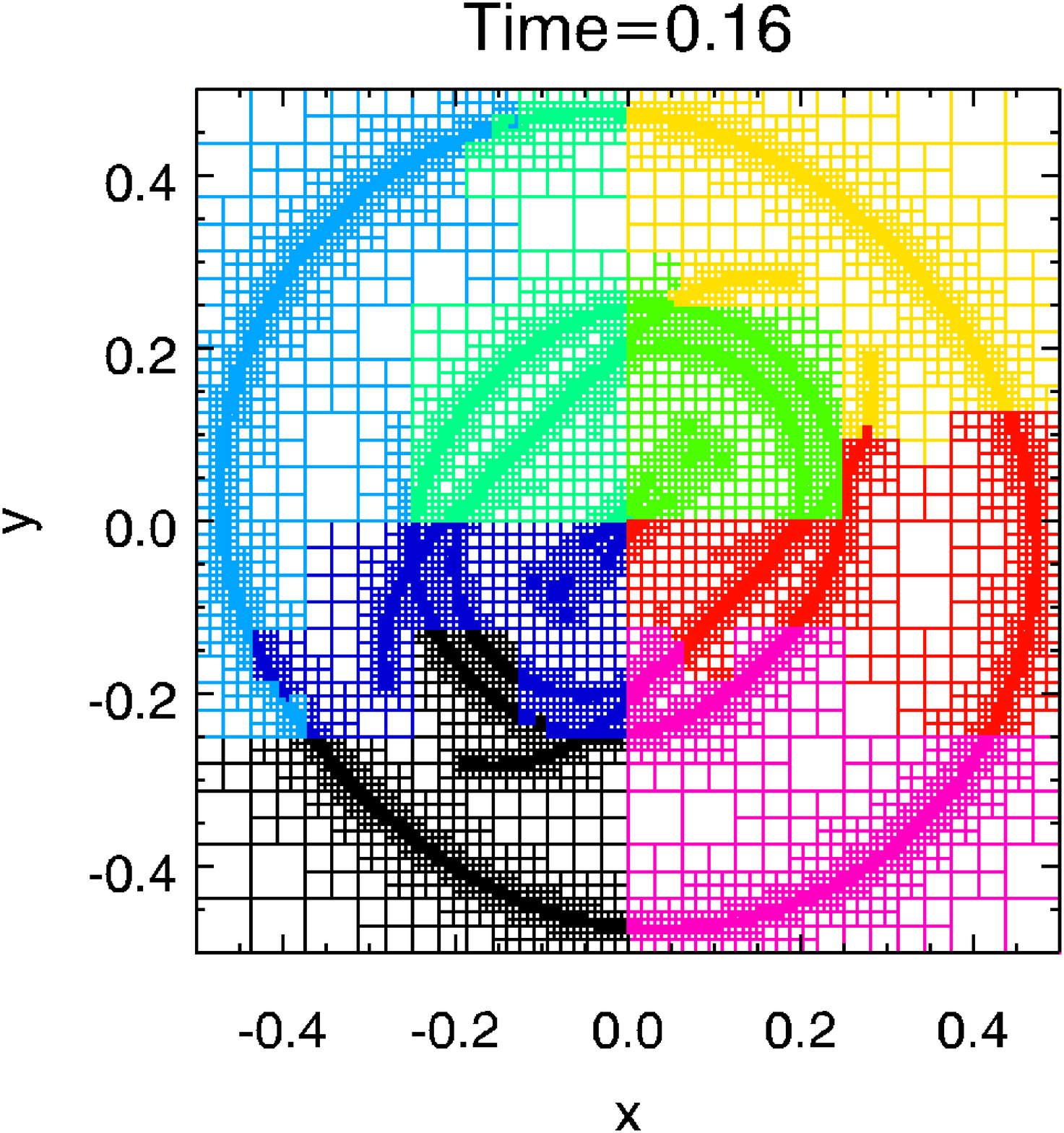}
\caption{Left panel is the gas pressure distribution with the
velocity arrows and magnetic field lines at time 0.16. The right
panel shows the AMR mesh with the refinement level 5. The eight
colors in the right panel stand for the blocks occupied by the eight
processors.} \label{fig08}
\end{figure}

A simple MHD 2D test is a blast wave. The problem can generate
several shocks from the central high gas pressure zone, it is very
suitable to test the AMR algorithm and the parallel efficiency. The
density ($\rho$) and magnetic field ($B_x, B_y, B_z$) are uniform in
the domain $[-0.5, 0.5] \times [-0.5, 0.5]$ with the value $1$ and
$(1 / \sqrt 2, 1 / \sqrt 2, 0)$, respectively. A small hot area is
located at the center of the domain, i.e. $p = 10$ within the circle
$x^2 + y^2 \le 0.1^2$, whereas $p = 1$ out of this circle. Adiabatic
index is $\gamma = 5 / 3$. The left panel of Fig.~\ref{fig08}
shows the pressure distribution at time $0.16$ and the right one
shows the AMR mesh with the refinement level $L=5$. Since the base
resolution is $256 \times 256$, the effective resolution is $4096
\times 4096$. Fig.~\ref{fig08} corresponds to a 8-processor run, the
parallelization efficiency for other number of processors are listed in
Table~\ref{list_efficiency}, where $N_p$ is the number of processors,
$Blocks$ indicates the total number of blocks in all processors at
time 0.16. The efficiencies higher than 1 may due to the speed
fluctuation of the supercomputer.

\begin{table}[htbp]
\caption{Parallelization efficiency for Blast wave test (AMR level
$L=5$).\label{list_efficiency}}
\begin{tabular}{cccc}
\hline
\hline
$N_p$   & Blocks & Time     & Efficiency \\
\hline
$1$     & $15512$  & $88504$ s & $1   $       \\
$2$     & $15512$  & $43595$ s & $1.02$       \\
$4$     & $15512$  & $21941$ s & $1.01$       \\
$8$     & $15512$  & $11327$ s & $0.98$       \\
$16$    & $15512$  & $ 5731$ s & $0.97$       \\
$32$    & $15512$  & $ 2908$ s & $0.95$       \\
$64$    & $15504$  & $ 1521$ s & $0.90$       \\
$128$   & $15504$  & $  831$ s & $0.83$       \\
\hline
\end{tabular}
\end{table}

\subsection{2D divergence free test}
\begin{figure}[htbp]
\centering
\includegraphics[height=150pt]{\figures/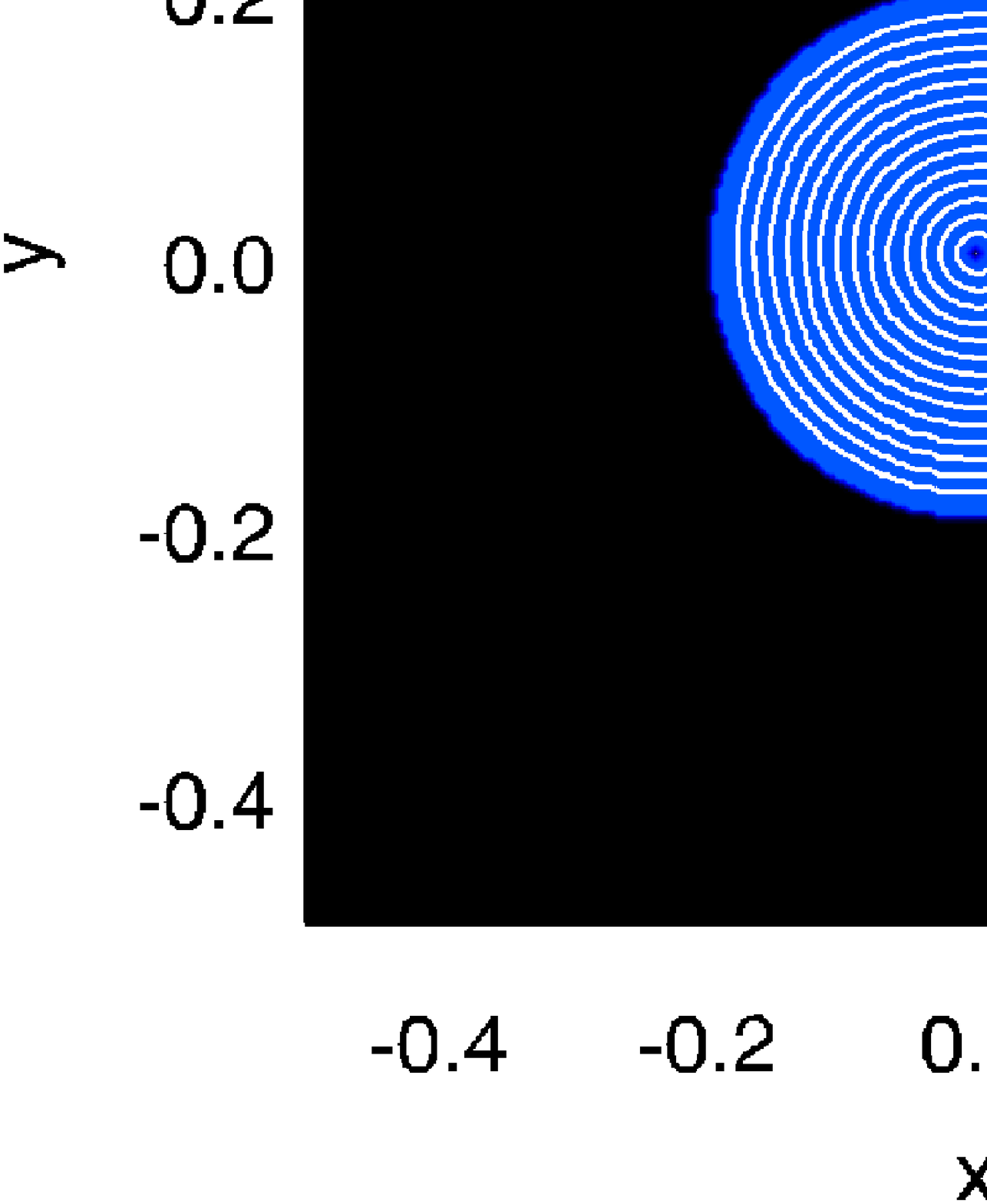}
\includegraphics[height=150pt]{\figures/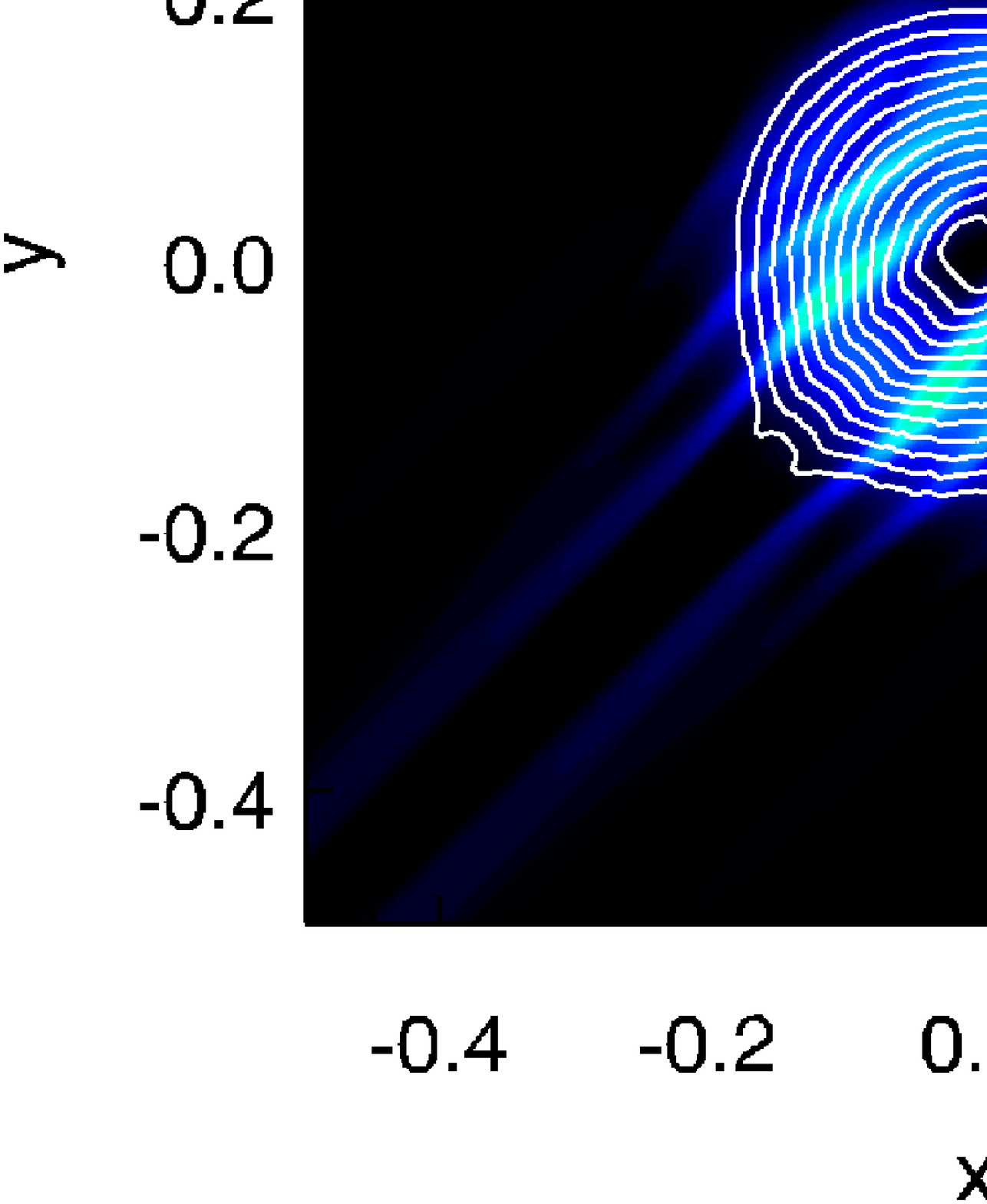}
\includegraphics[height=150pt]{\figures/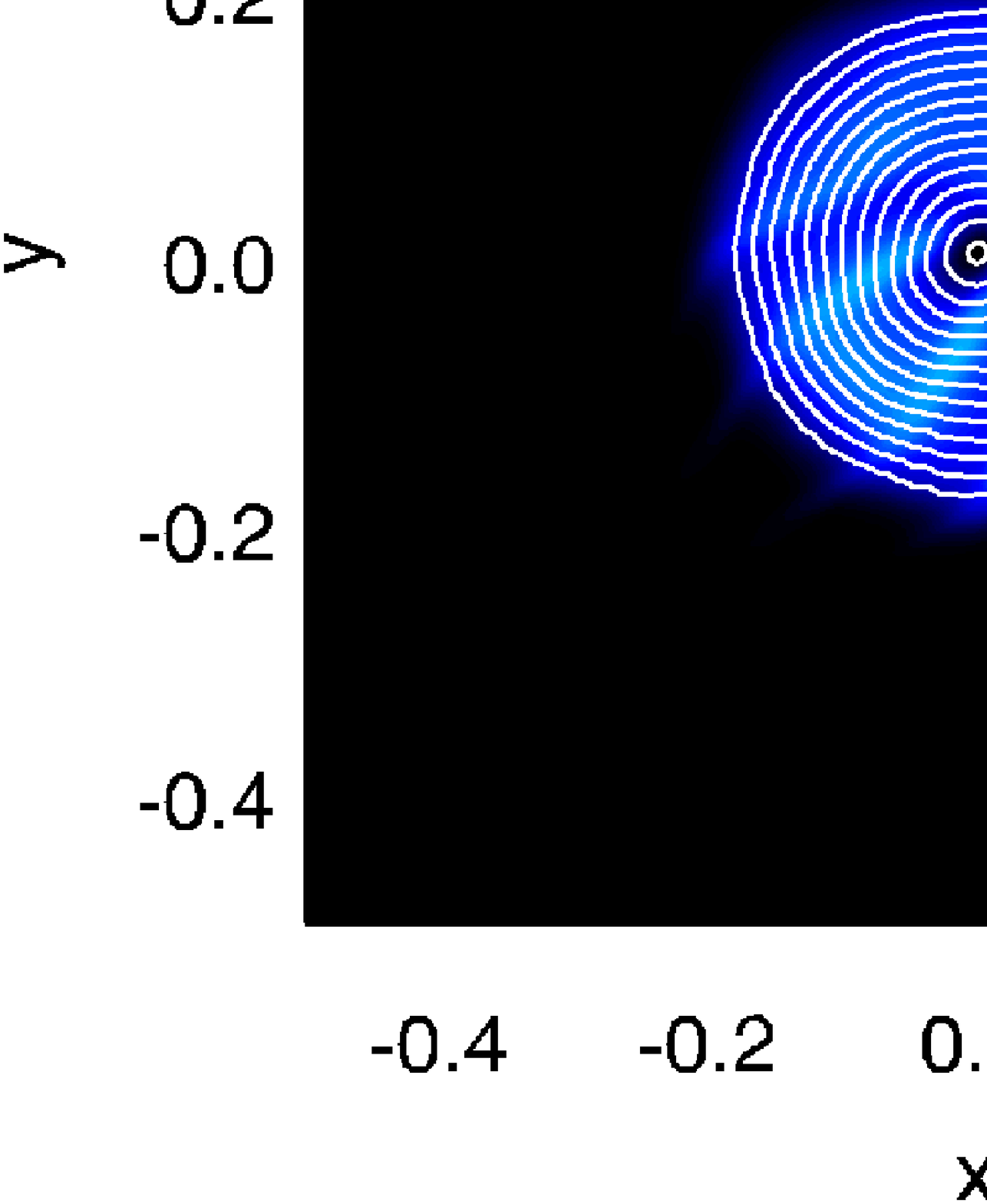}
\caption{Total magnetic pressure distributions with the magnetic
field lines of divergence free test. Left panel: The initial
condition; Middle panel: advection result at time $2$ without
divergence cleanance method; Right panel: advection result at time
$2$ using EGLM-MHD with a coefficient $c_d = 0.18$ as described in
Section~\ref{equations}. This simulation is carried on by using the
WENO scheme with the Lax-Friedrichs flux. The resolution
in this test is $256 \times 256$, no AMR involved.} \label{fig09}
\end{figure}

\begin{figure}[htbp]
\centering
\includegraphics[height=140pt]{\figures/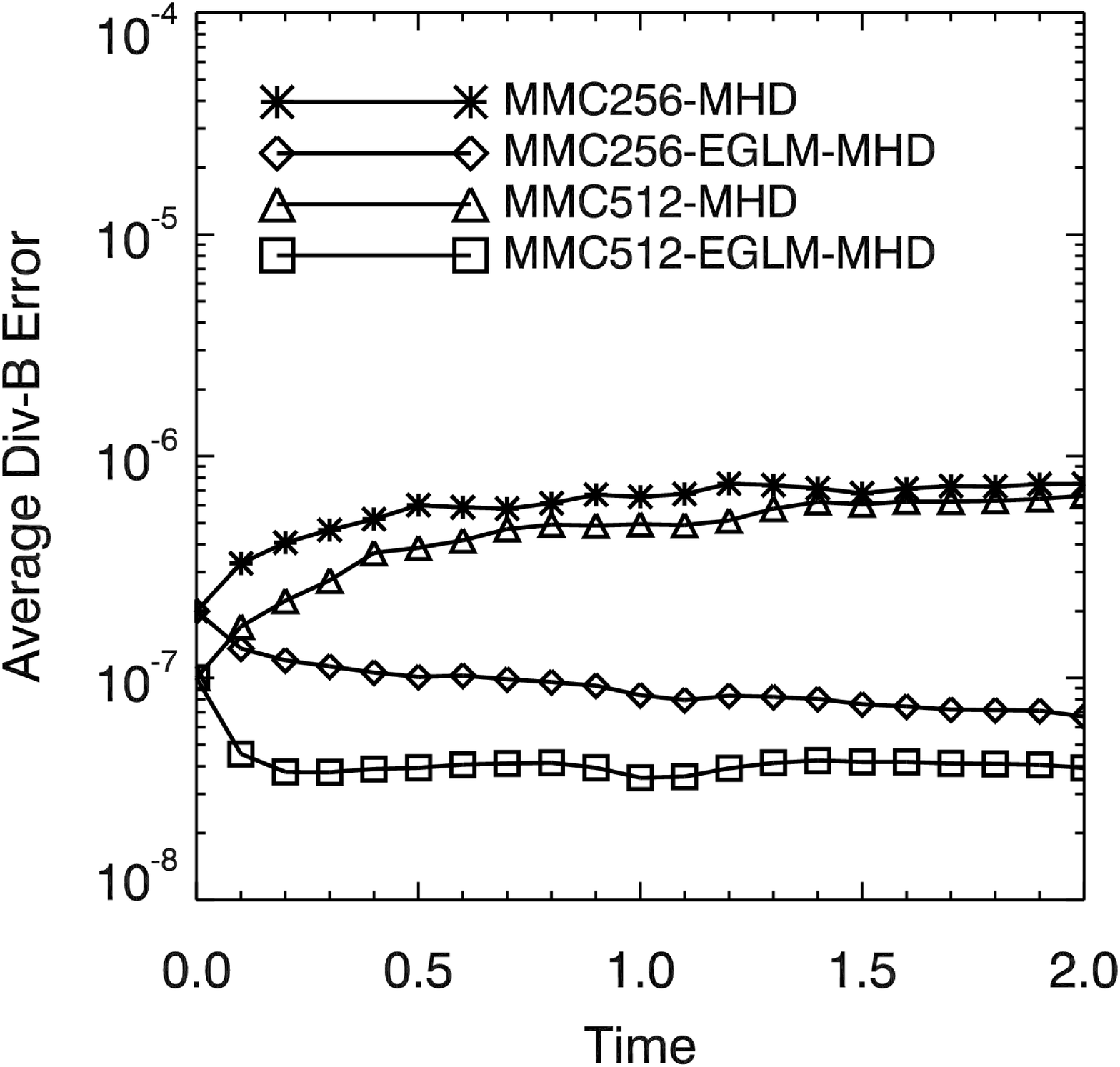}
\includegraphics[height=140pt]{\figures/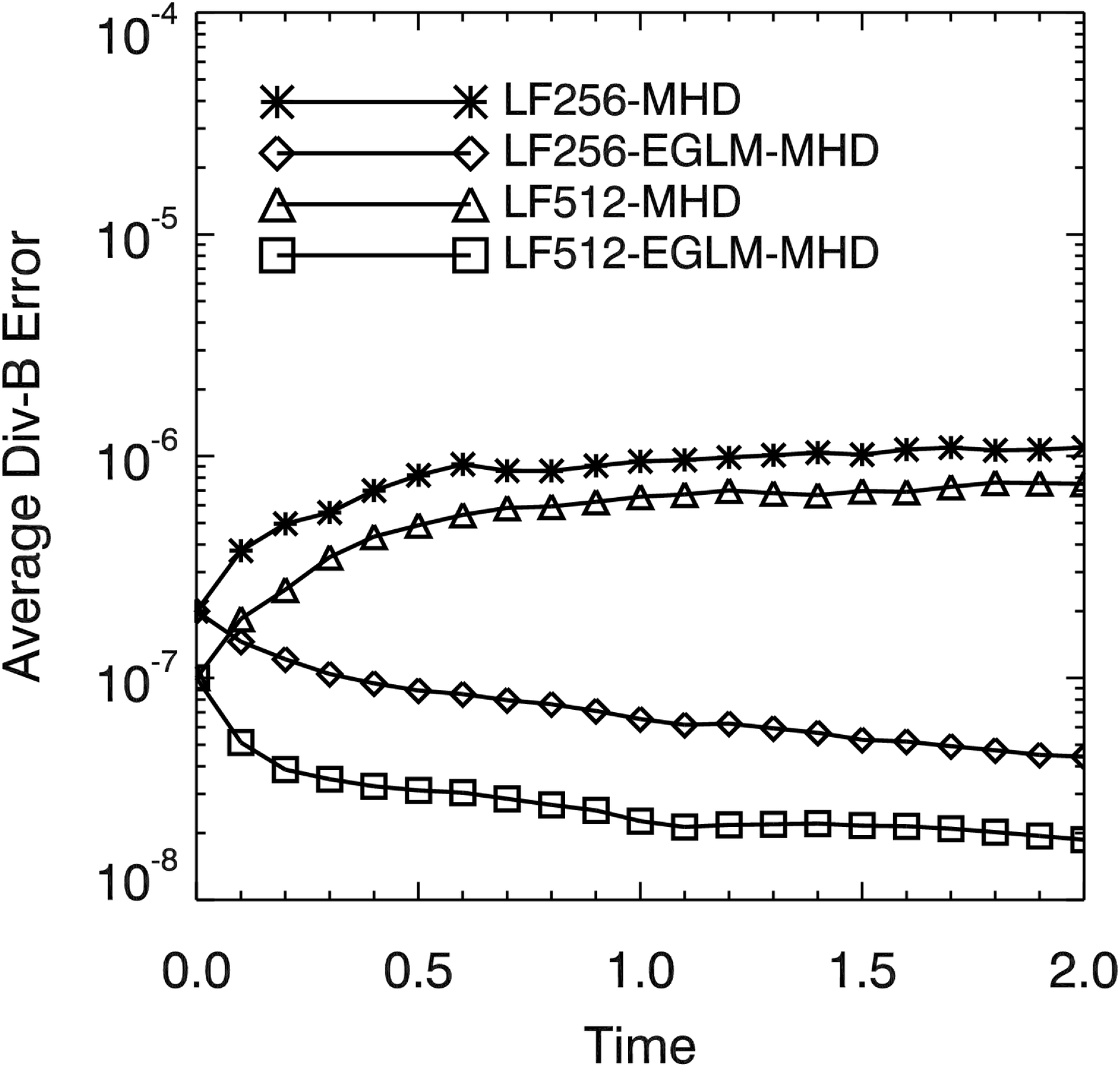}
\includegraphics[height=140pt]{\figures/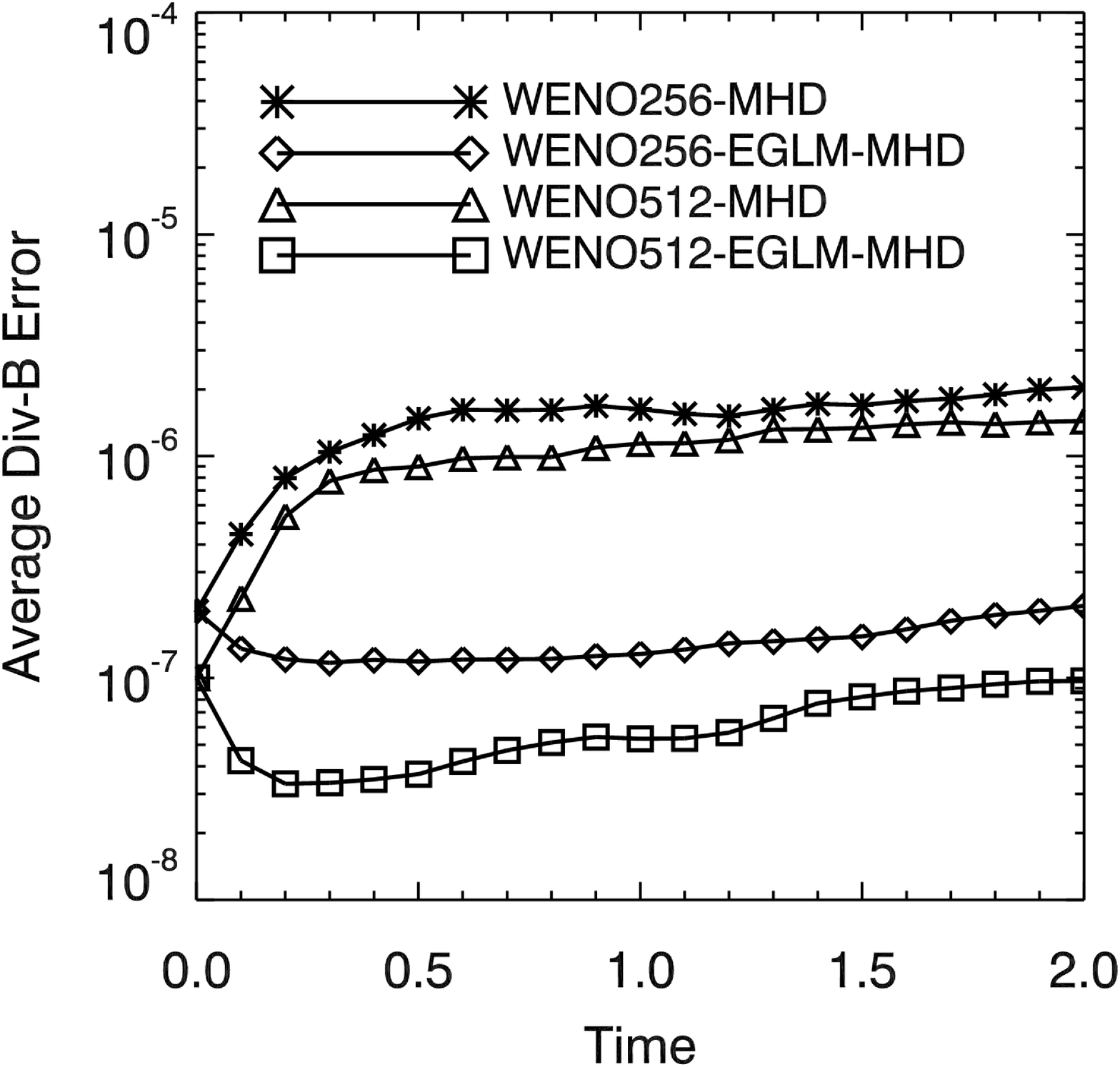}
\caption{The $\nabla \cdot \mathbf{B}$ average errors with a 
function of time for different methods and resolutions (the numbers 
256 and 512 mean the resolutions of $256 \times 256$ and $512 \times 512$). 
Left panel: the MMC scheme; Middle panel: LF scheme; Right panel: WENO scheme.
Lax-Friedrichs splitting for MMC and 
Lax-Friedrichs flux for LF and WENO are used in this test.} \label{fig10}
\end{figure}

The test was introduced by~\cite{Gardiner2005} to check the
conservation of the divergence-free condition. If we had not taken
this condition into account, the code may give us non-physical results in
some MHD applications. The problem domain is $[-0.5,0.5] \times
[-0.5,0.5]$, with the uniform density ($\rho = 1$) and velocity ($v_x
= 1$, $v_y = 1$, $v_z = 0$). The magnetic configuration is given by
$B_x = - B_0 y / r$, $B_y = B_0 x / r$ and $B_z = 0$ in the region
$r = \sqrt{x^2 + y^2} < 0.2$. In this region, we set $p = 1 - B_0 ^
2 / 2$ for magnetostatic equilibrium and $p = 1$ for outside of this
region. Adiabatic index $\gamma = 5 / 3$ and $B_0 = 1 \times
10^{-3}$. The results are displayed in Fig.~\ref{fig09}. The
magnetic loops advect across the boundary twice when the
dimensionless time is $2$. As shown in Fig.~\ref{fig09}, the
magnetic loops without divergence cleanance are distorted ({\it middle
panel}). However, when the divergence cleanance is conducted, the loops
keep their original shapes ({\it right panel}). The $\nabla \cdot \mathbf{B}$ errors 
are plotted with the function of time in the Fig. \ref{fig10}. In this figure, 
we show the performances of the three schemes. As expected, the methods using EGLM-MHD 
equations can maintain the error one order of magnitude smaller than the methods using pure 
MHD. The method with higher resolution also given less $\nabla \cdot \mathbf{B}$ error.
Although it is not so perfect as the CT schemes, EGLM-MHD has its
advantages: (1) it is easy to be equipped in the existed MHD codes; (2) it can
damp the non-divergence error produced by the rapidly-changing
boundaries.

\subsection{2D magnetic rotor}
\begin{figure}[htbp]
\centering
\includegraphics[height=180pt]{\figures/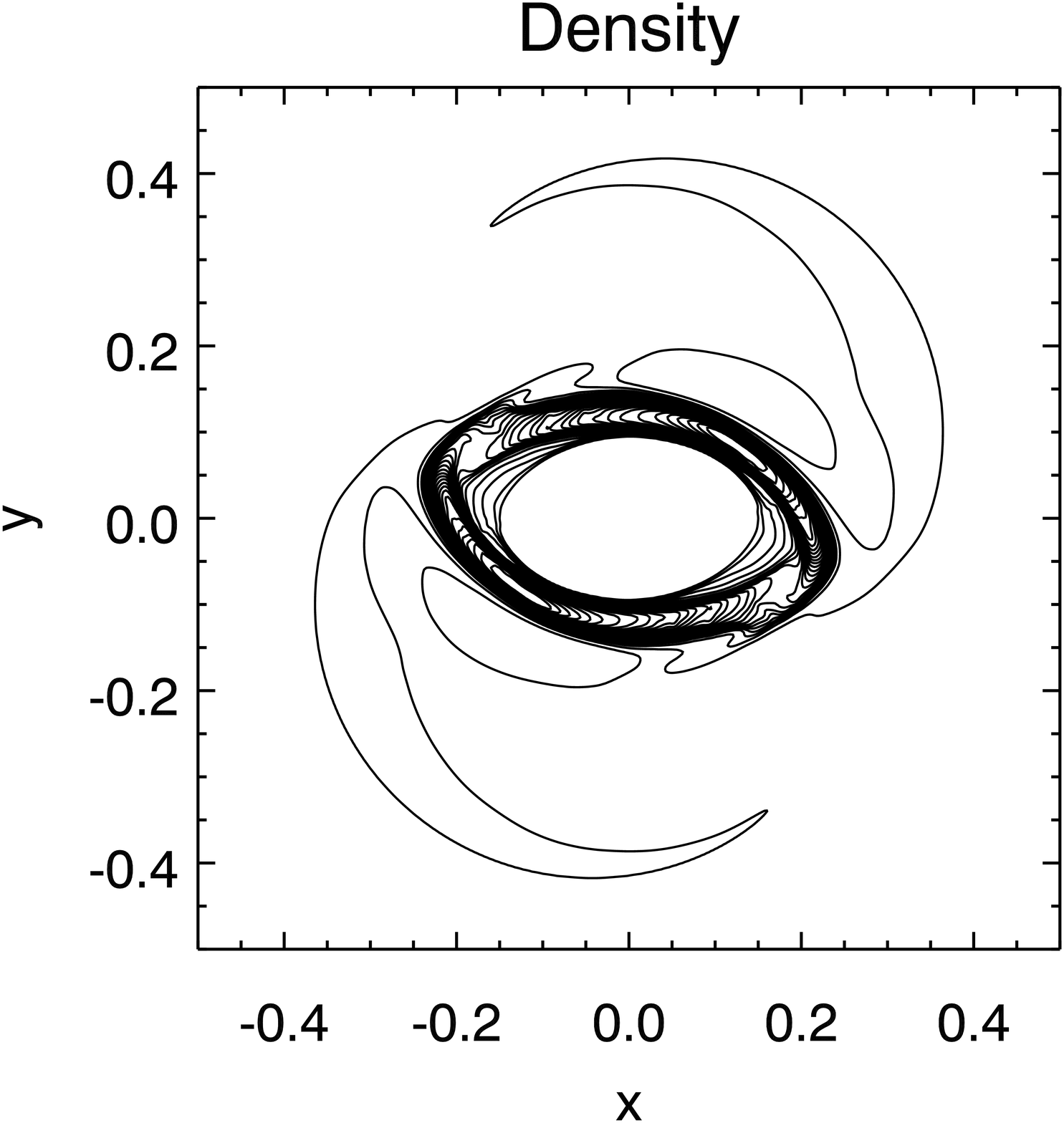}
\includegraphics[height=180pt]{\figures/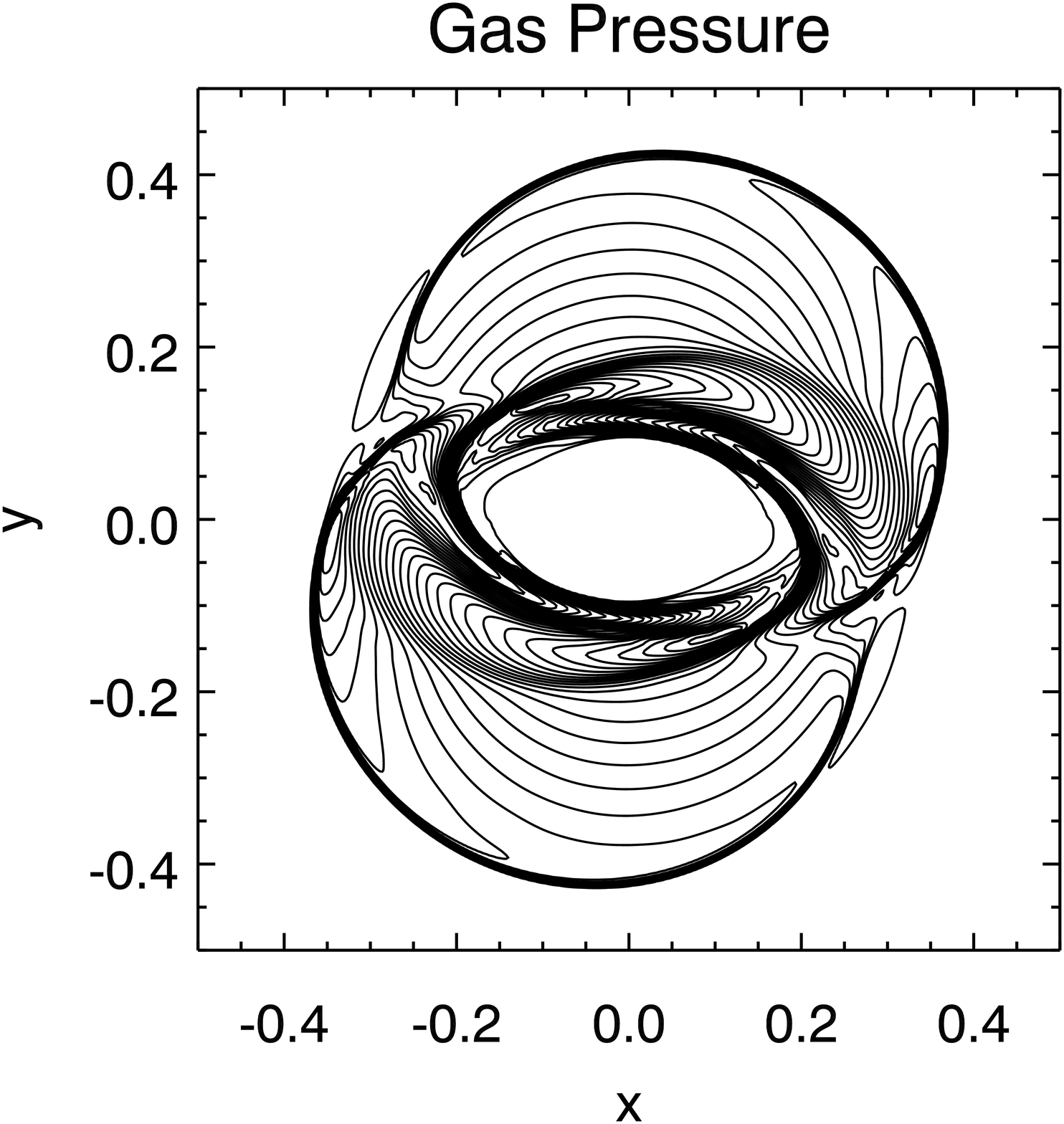}
\includegraphics[height=180pt]{\figures/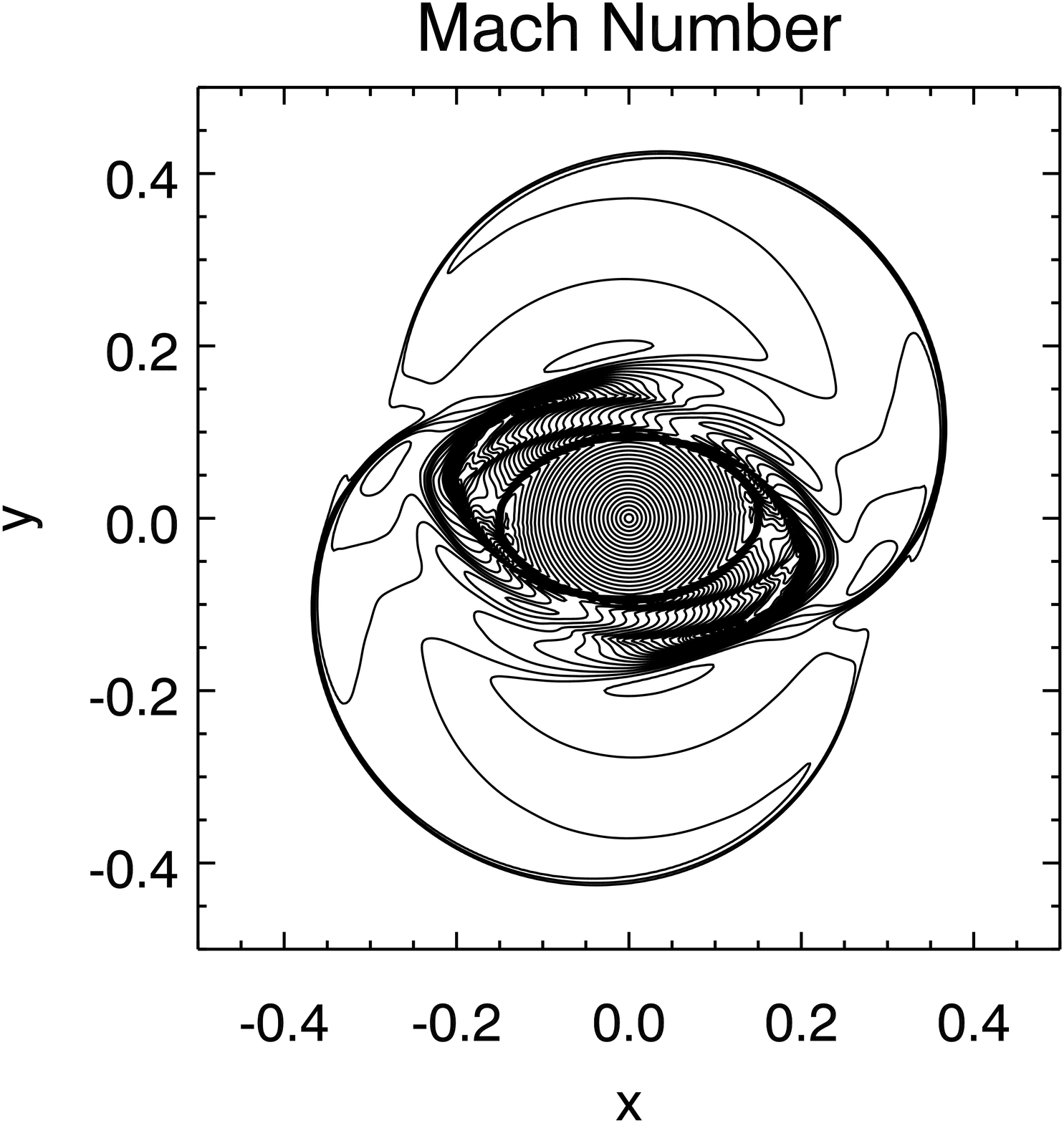}
\includegraphics[height=180pt]{\figures/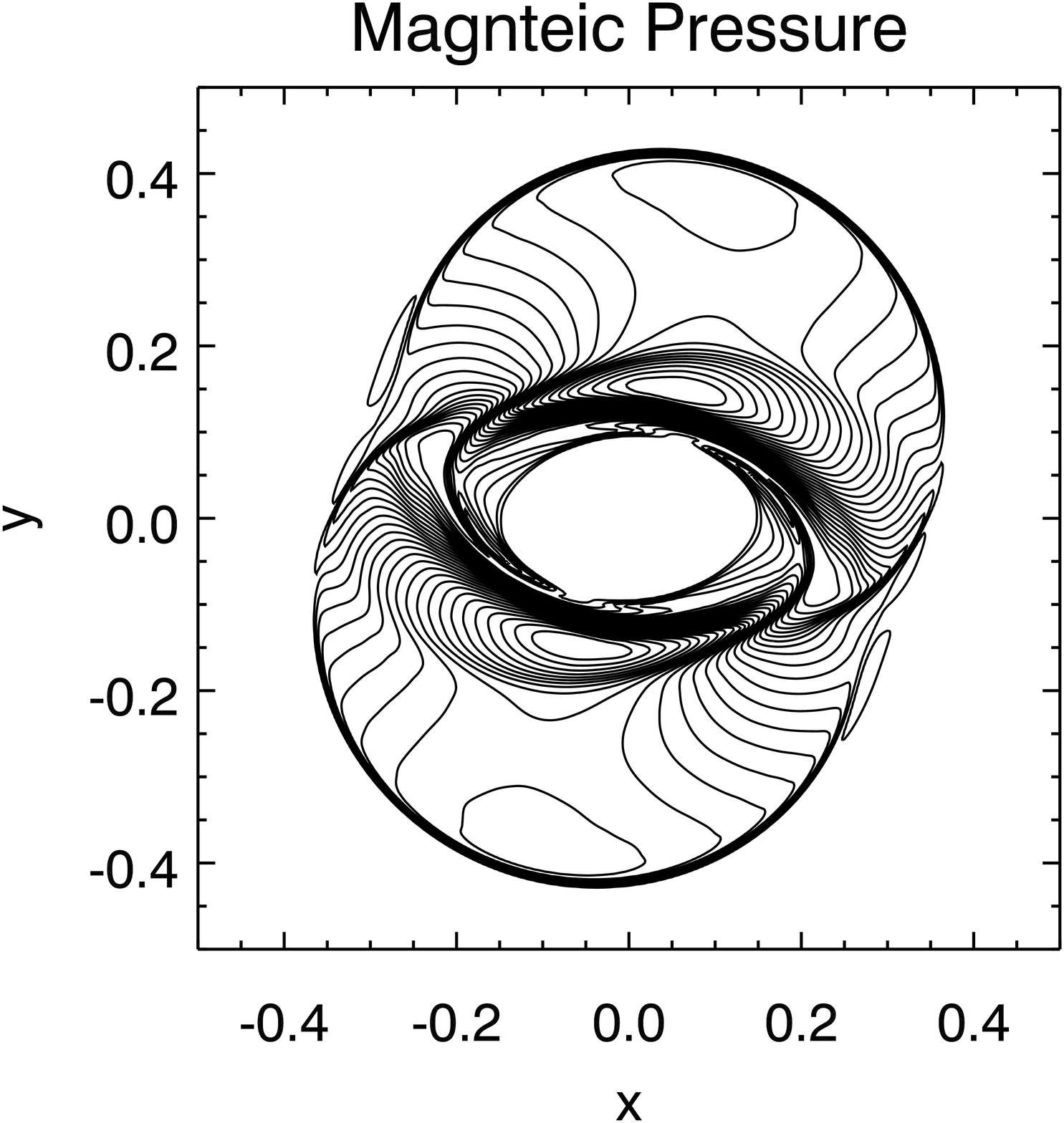}
\caption{The density, gas pressure, Mach number and 
magnetic pressure distributions at Time = 0.15. There are 30 equally spaced contours between 
the maximum and minimum for each plot. This test is based on the WENO scheme with the resolution 
of $400 \times 400$ and the Lax-Friedrichs flux.} \label{fig11}
\end{figure}

This test suite is introduced by \citet{Balsara1999}, which is used to 
test the propagation of strong torsional Alfv\'en waves. The 
rotating disk in the computational center 
generates shocks and waves from the strong shearing surface between the rotating disk 
and the static ambient fluid. We use the 
initial conditions as described by \citet{Toth2000} which involves a
higher initial velocities. It is better to test the robustness of our code. 
The computational domain is $[-0.5,0.5] \times [-0.5,0.5]$, the initial distributions 
are:

\begin{equation}
\rho=\left\{
\begin{array}{ll}
10                             & \ \ \textrm{for} \ \   r  < r_0  \\
1 + 9 f                        & \ \ \textrm{for} \ \   r_0 \le r \le r_1   \\
 1                             & \ \ \textrm{for} \ \   r  >  r_1  \\
\end{array}
\,\, ,
\right.
\end{equation}

\begin{equation}
v_x=\left\{
\begin{array}{ll}
-2 y / r_0                      & \ \ \textrm{for} \ \   r  < r_0  \\
-2 f y / r                      & \ \ \textrm{for} \ \   r_0 \le r \le r_1   \\
 0                              & \ \ \textrm{for} \ \   r  >  r_1  \\
\end{array}
\,\, ,
\right.
\end{equation}

\begin{equation}
v_y=\left\{
\begin{array}{ll}
2 x / r_0                      & \ \ \textrm{for} \ \   r  < r_0  \\
2 f x / r                      & \ \ \textrm{for} \ \   r_0 \le r \le r_1   \\
 0                             & \ \ \textrm{for} \ \   r  >  r_1  \\
\end{array}
\,\, ,
\right.
\end{equation}

\noindent
where, $r = \sqrt{x^2 + y^2}$ and $f = (r_1 - r) / (r_1 - r_0)$ with $r_0 = 0.1$ 
and $r_1 = 0.115$. $f$ function helps to reduce initial transients. The gas pressuse 
$p = 1$ and magnetic field $B_x = 5 / \sqrt{4\pi}, B_y=0$ are uniform. The third components 
of velocity and magnetic field are set to zero.
Adiabatic index $\gamma = 1.4$. The density, 
gas pressure, Mach number and magnetic pressure distributions at Time = 0.15 are shown in 
Fig. \ref{fig11}. The test is based on the WENO scheme with the resolution of $400 \times 400$ and 
no approximate Riemann solvers included. From this figure we can observe that the torsional 
Alfv\'en waves are generated. The EGLM-MHD equations prevent the magnetic monopoles to form 
and the contours of Mach number keep the 
shape of concentric circles. Other methods (LF and MMC) show almost the same results.

\subsection{2D magnetic reconnection application}
\begin{figure}[htbp]
\centering
\includegraphics[height=180pt]{\figures/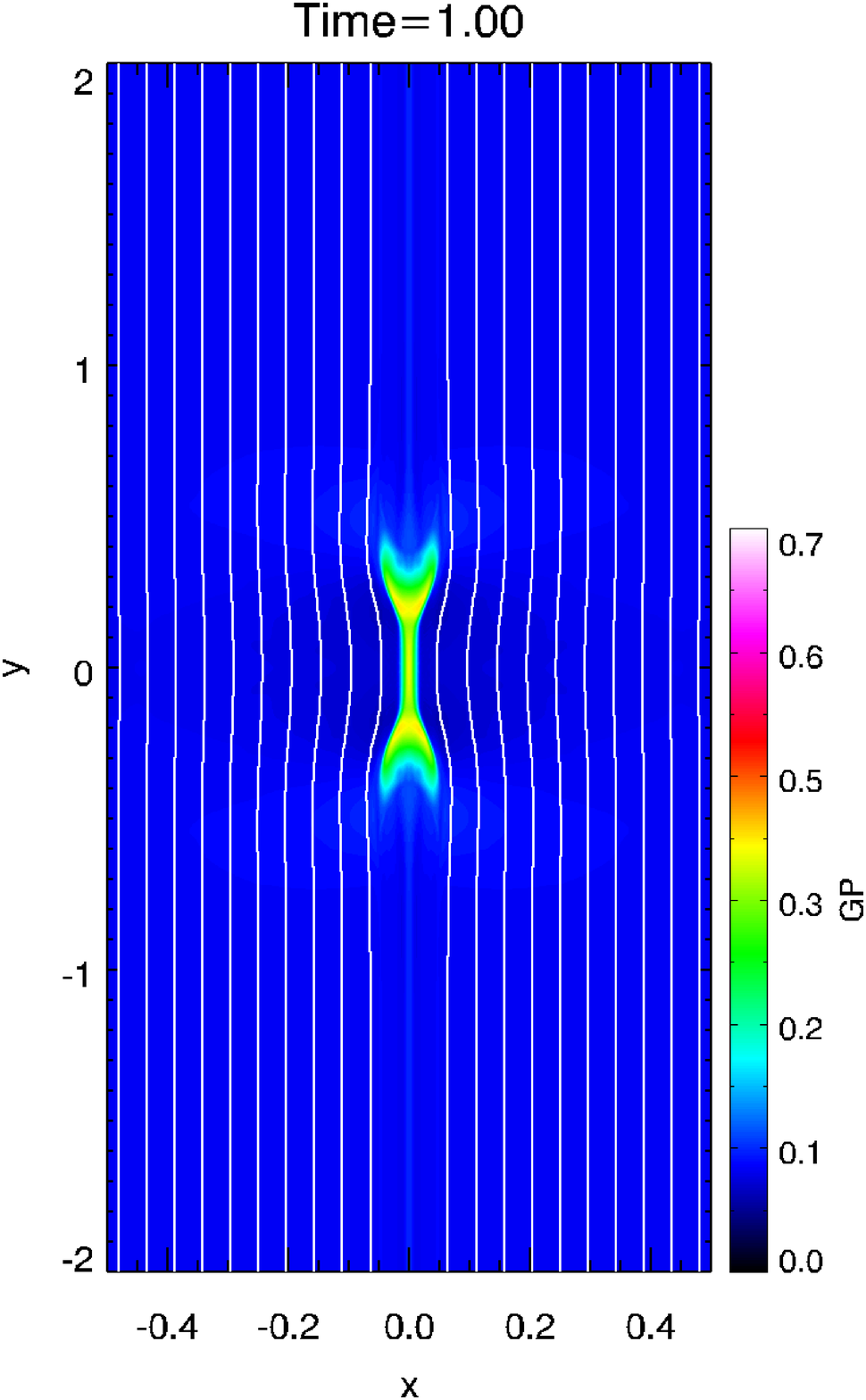}
\includegraphics[height=180pt]{\figures/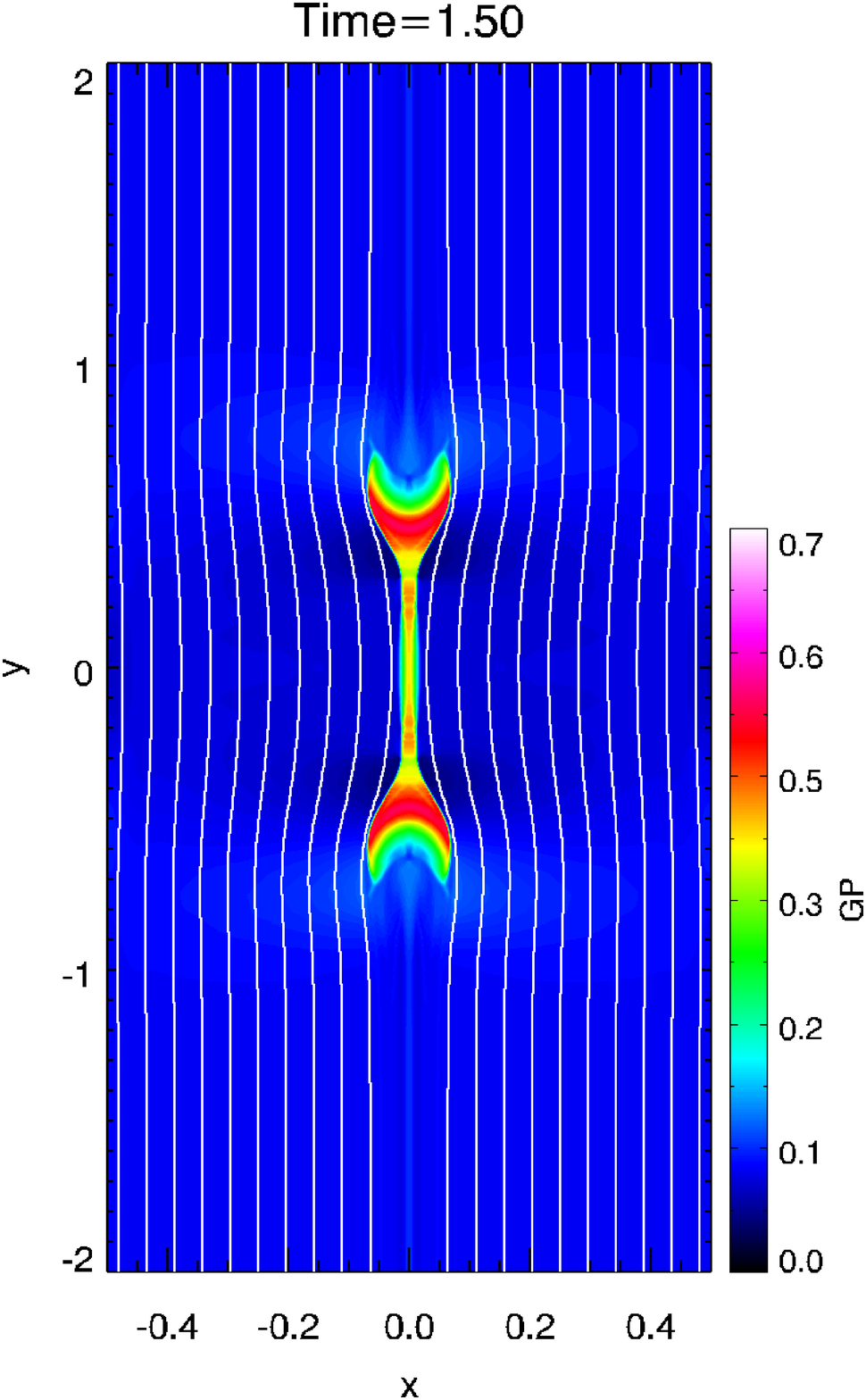}
\includegraphics[height=180pt]{\figures/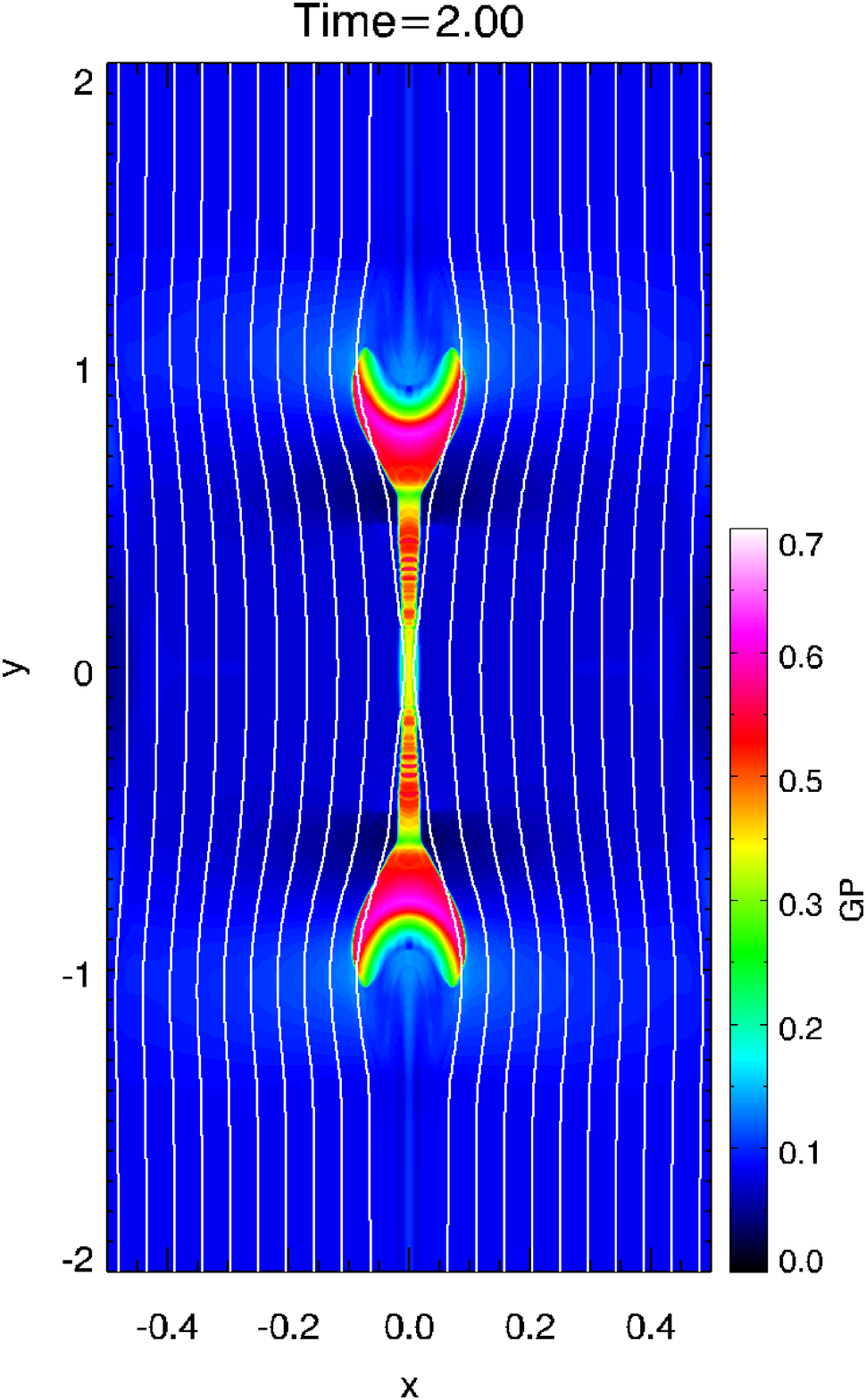}
\includegraphics[height=180pt]{\figures/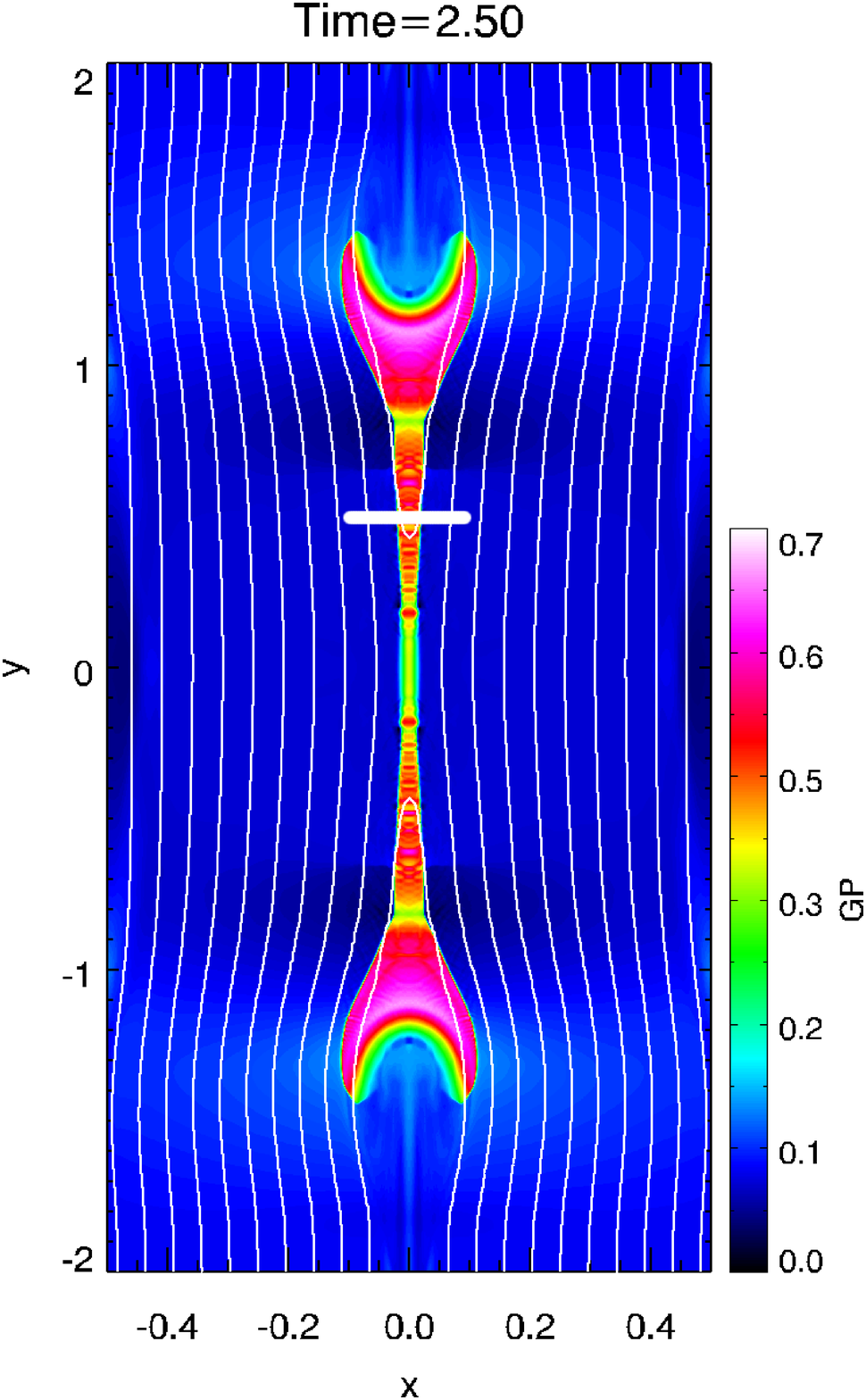}
\caption{Gas pressure distributions with velocity arrows and
magnetic field lines at time 1.0, 1.5, 2.0 and 2.5. This test 
is based on the WENO scheme with the Lax-Friedrichs flux. Base 
resolution is $128 \times 256$ and the maximum refinement level is 5.}
\label{fig12}
\end{figure}

In this subsection we mainly test the effect of resistivity in the 
magnetic reconnection process. This kind 
of resistivity may due to some microscopic instabilities,
although how these instabilities drive the macroscopic reconnection is still not clear. 
The velocities ($v_x$,$v_y$,$v_z$) are zero in the initial state. We set a uniform
density ($\rho=1$) and pressure ($p=0.1$) distributions with an
anti-parallel magnetic configuration (see below) in the
computational domain ($[-0.5, 0.5] \times [-2, 2]$) as follows:

\begin{equation}
B_x=0 \,\, ,
\end{equation}

\begin{equation}
B_y=\left\{
\begin{array}{ll}
-1                             & \ \ \textrm{for} \ \   x  < -L_r  \\
 \sin(\pi x / 2L_r)            & \ \ \textrm{for} \ \  |x| \le  L_r  \\
 1                             & \ \ \textrm{for} \ \   x  >  L_r  \\
\end{array}
\,\, ,
\right.
\end{equation}

\begin{equation}
B_z=\left\{
\begin{array}{ll}
0                              & \ \ \textrm{for} \ \   x  < -L_r  \\
\cos(\pi x / 2L_r)             & \ \ \textrm{for} \ \  |x| \le  L_r \\
0                              & \ \ \textrm{for} \ \   x  >  L_r  \\
\end{array}
\,\, ,
\right.
\end{equation}

\noindent where $L_r = 0.05$ is the half width of the resistivity
region in the $x$-direction. The resistivity has the form $\eta = \eta_0
\left(\cos(\pi x / 0.1) + 1\right)\left(\cos(\pi y / 0.4) + 1\right)
/ 4$ in the small region $[-0.05,0.05] \times [-0.2,0.2]$. 
The reconnection becomes fast when localized resistivity is included. 
The simulation results are shown in Fig.~\ref{fig12}. We get a fast magnetic reconnection
results by a localized resistivity region~\citep{Jiang2010,
Jiang2011} at the center of the domain. The magnetic reconnection
releases the magnetic energy to heat the matter located at the
center of the computation box. The upward and downward outflows
are accelerated to the Alfv\'en speed by the magnetic tension force. This
test is based on the WENO scheme with the Lax-Friedrichs flux. The effective 
resolution is $2048\times4096$ and we can observe some plasmoids formed at the positions 
$x = 0$ and $y=\pm 0.2$.

\begin{figure}[htbp]
\centering
\includegraphics[height=180pt, trim = 35mm 35mm 35mm 35mm, clip]{\figures/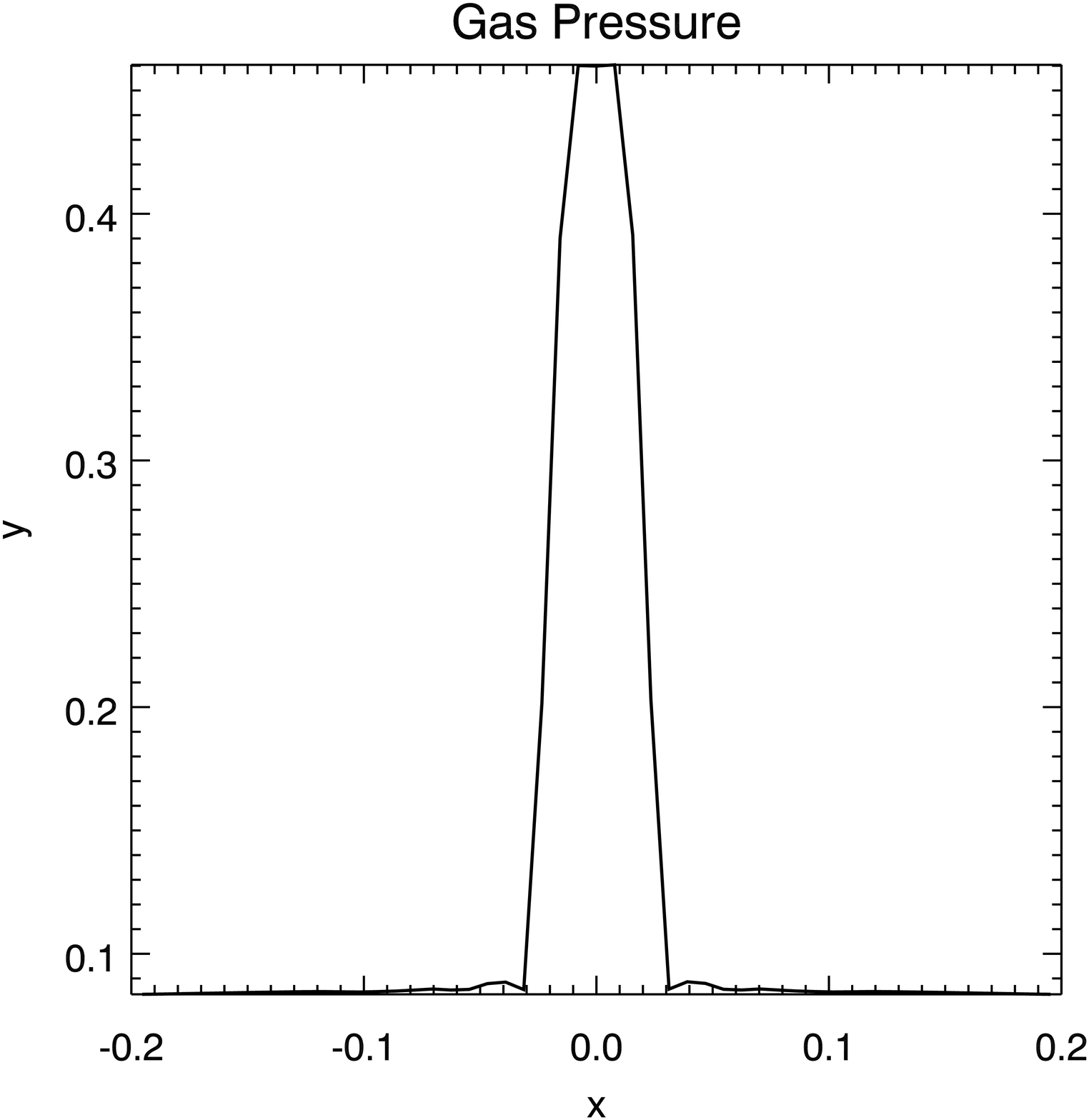}
\includegraphics[height=180pt, trim = 35mm 35mm 35mm 35mm, clip]{\figures/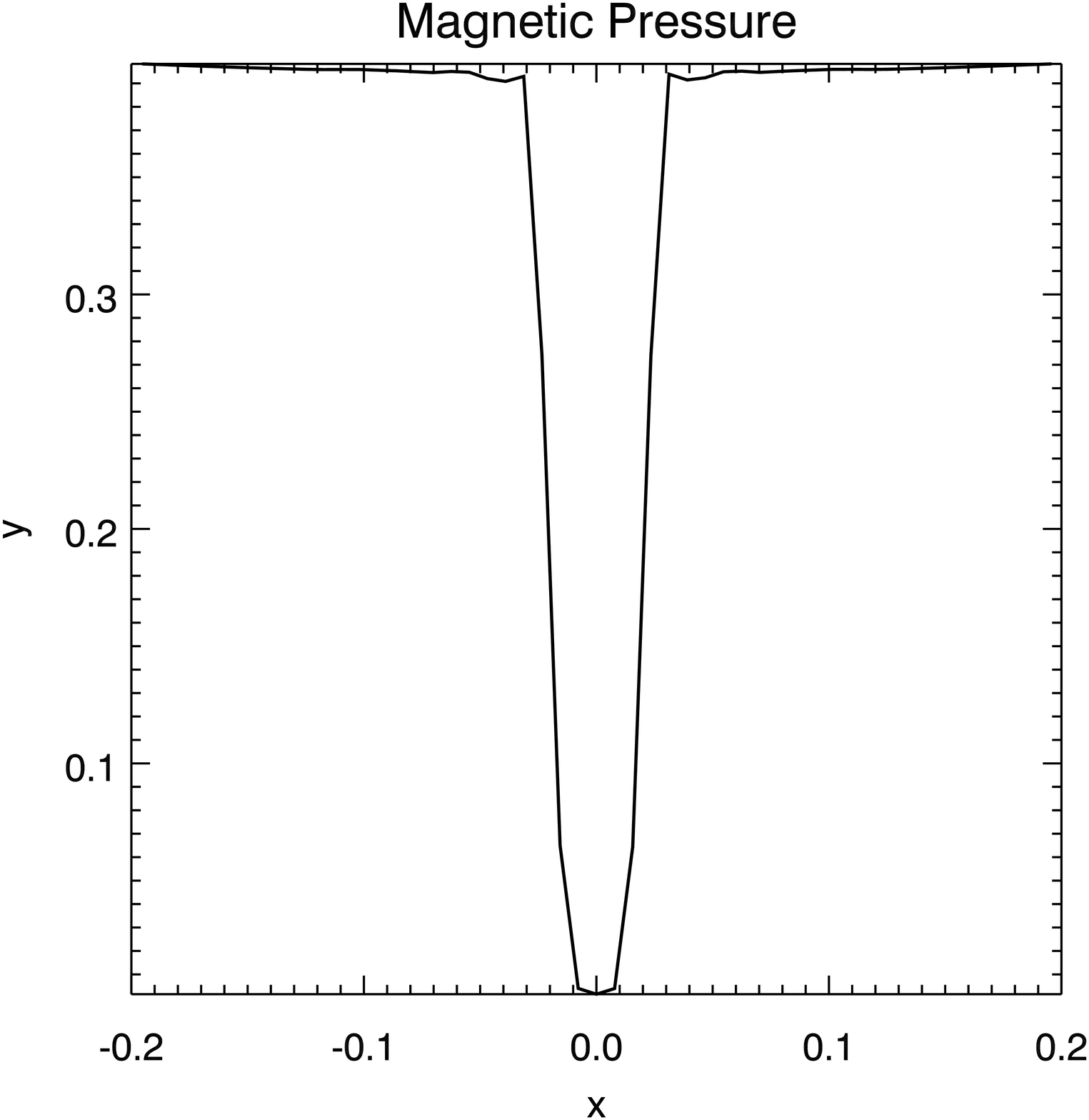}
\includegraphics[height=180pt, trim = 35mm 35mm 35mm 35mm, clip]{\figures/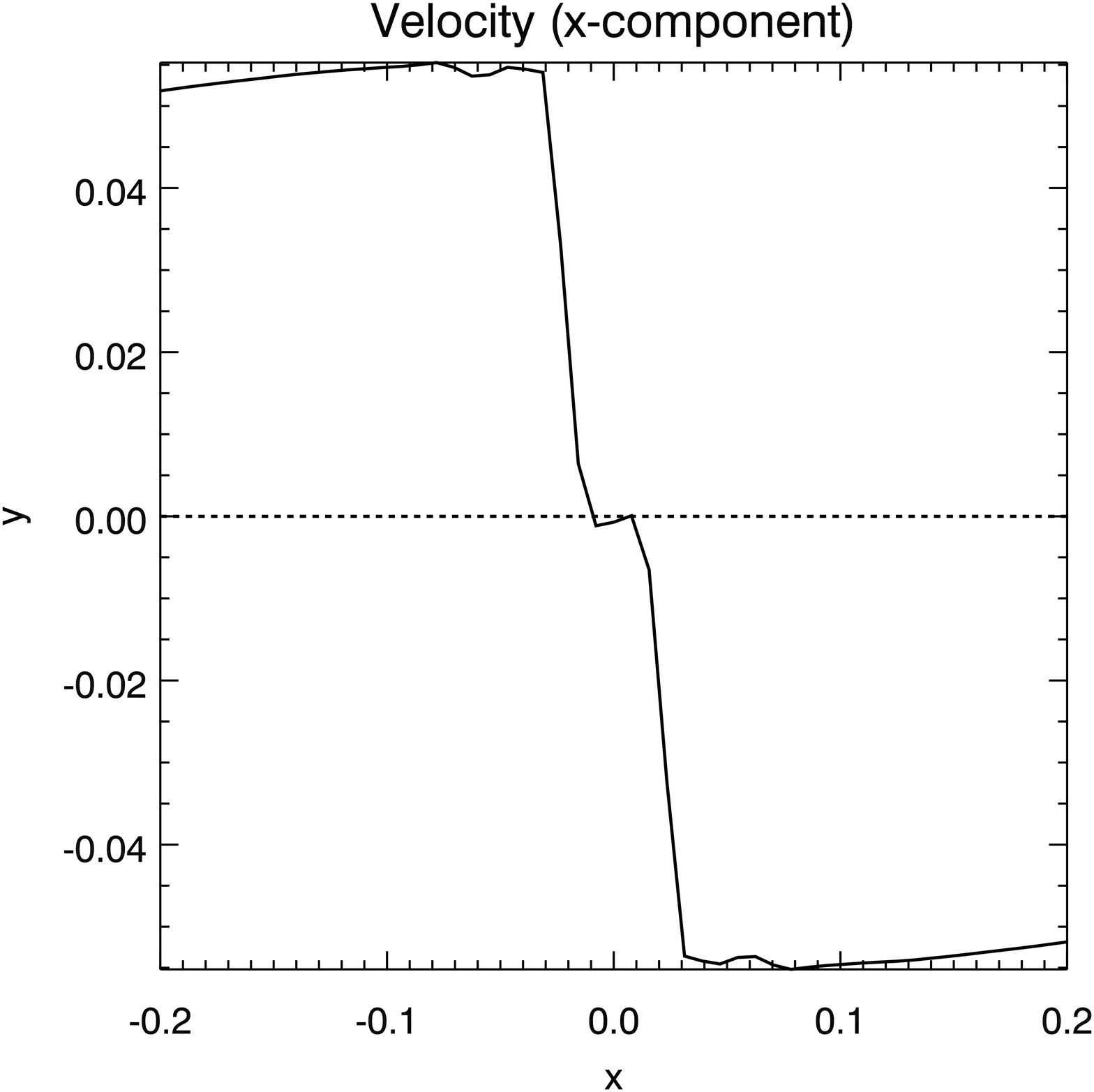}
\includegraphics[height=180pt, trim = 35mm 35mm 35mm 35mm, clip]{\figures/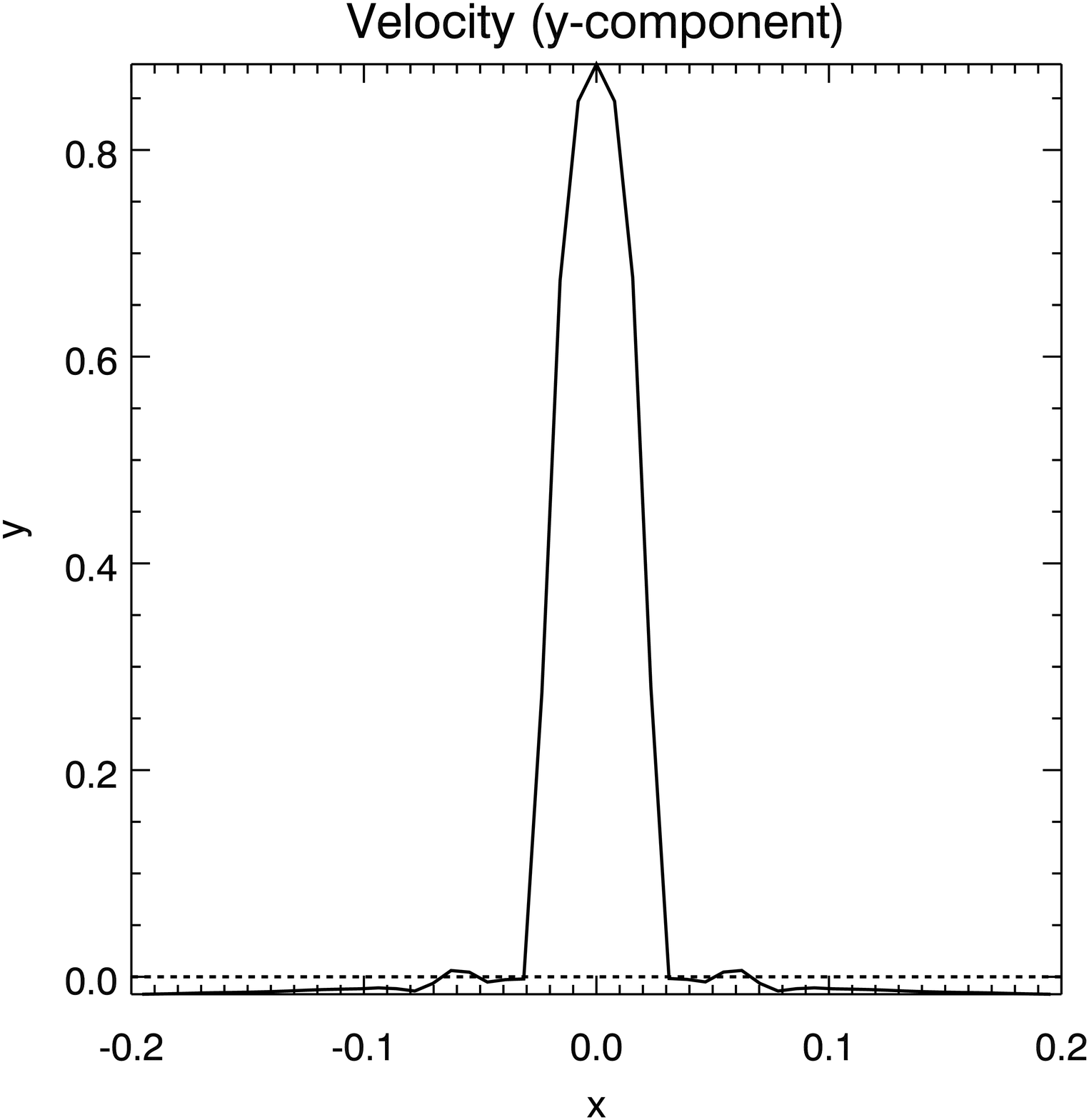}
\includegraphics[height=180pt, trim = 35mm 35mm 35mm 35mm, clip]{\figures/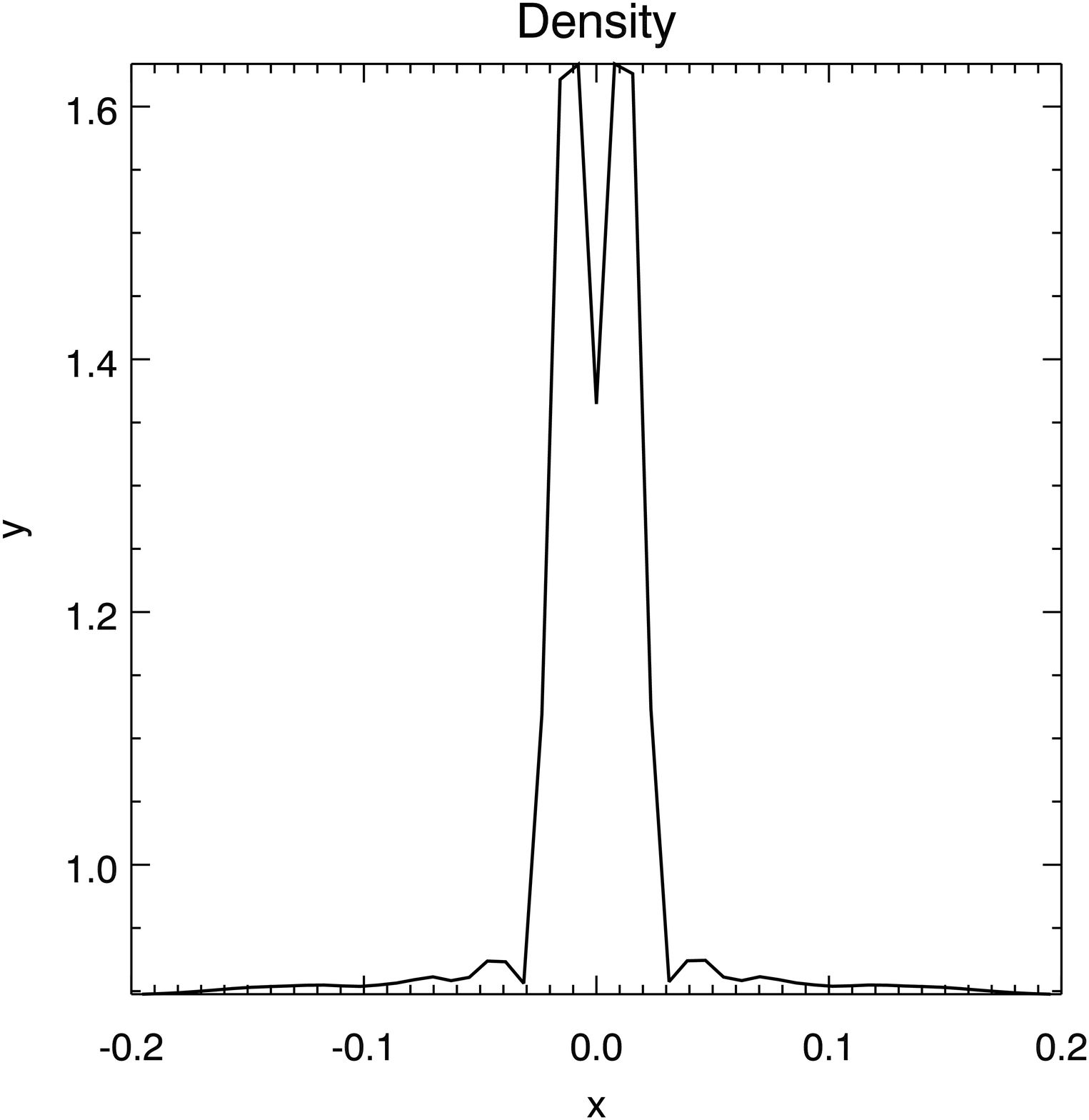}
\includegraphics[height=180pt, trim = 35mm 35mm 35mm 35mm, clip]{\figures/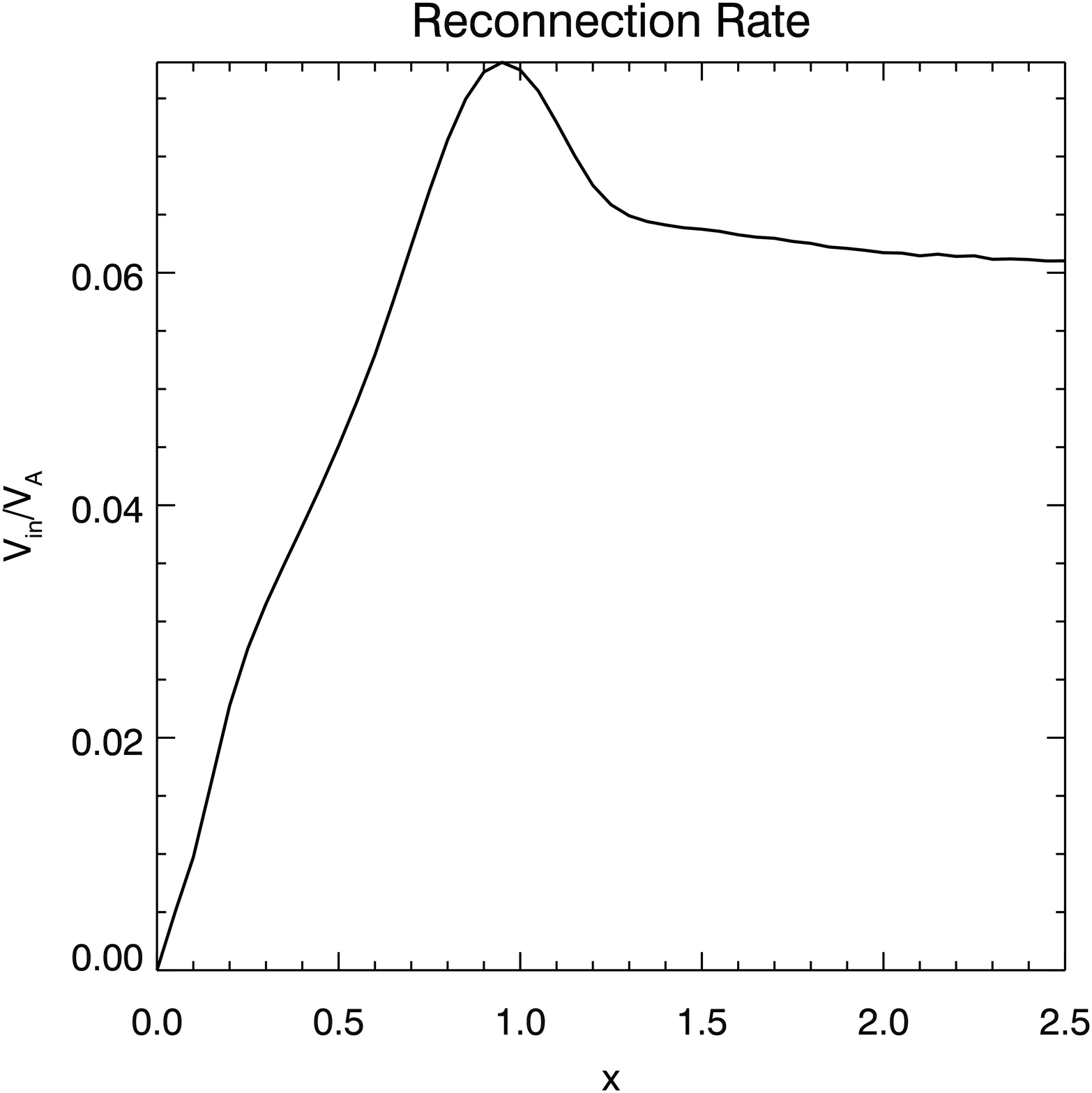}
\caption{The distributions of gas pressure, magnetic pressure, x-component velocity, 
y-component velocity, density along the white line in Fig. \ref{fig12} at time 2.5.
The lower right panel is the magnetic reconnection rate as a function of the time. 
The dashed lines show the velocity value of 0.}
\label{fig13}
\end{figure}

In Fig. \ref{fig13}, we can see five variable distributions 
along the white horizontal line in the right panel of Fig. \ref{fig12} and the reconnection 
rate as a function of time. The variable distributions clearly show the slow 
mode shock in the ranges ($ -0.03 < x < -0.01$ and $ 0.01 < x < 0.03$).
The variables sharply changed between the upstream and the downstream of this shock.
The reconnection rate is calculated by $V_{in} / V_A$, where $V_{in}$ is the 
inflow speed and $V_A$ the Alfv\'en speed. As shown by the lower right of Fig. \ref{fig13}, 
the reconnection rate reaches to the maximum around the time 1.0 and the value of this rate
is about 0.76 which is in the range of $0.01-0.1$ as expected by \citet{Petschek1964}.

\subsection{2D thermal conduction test}
\begin{figure}[htbp]
\centering
\includegraphics[height=120pt]{\figures/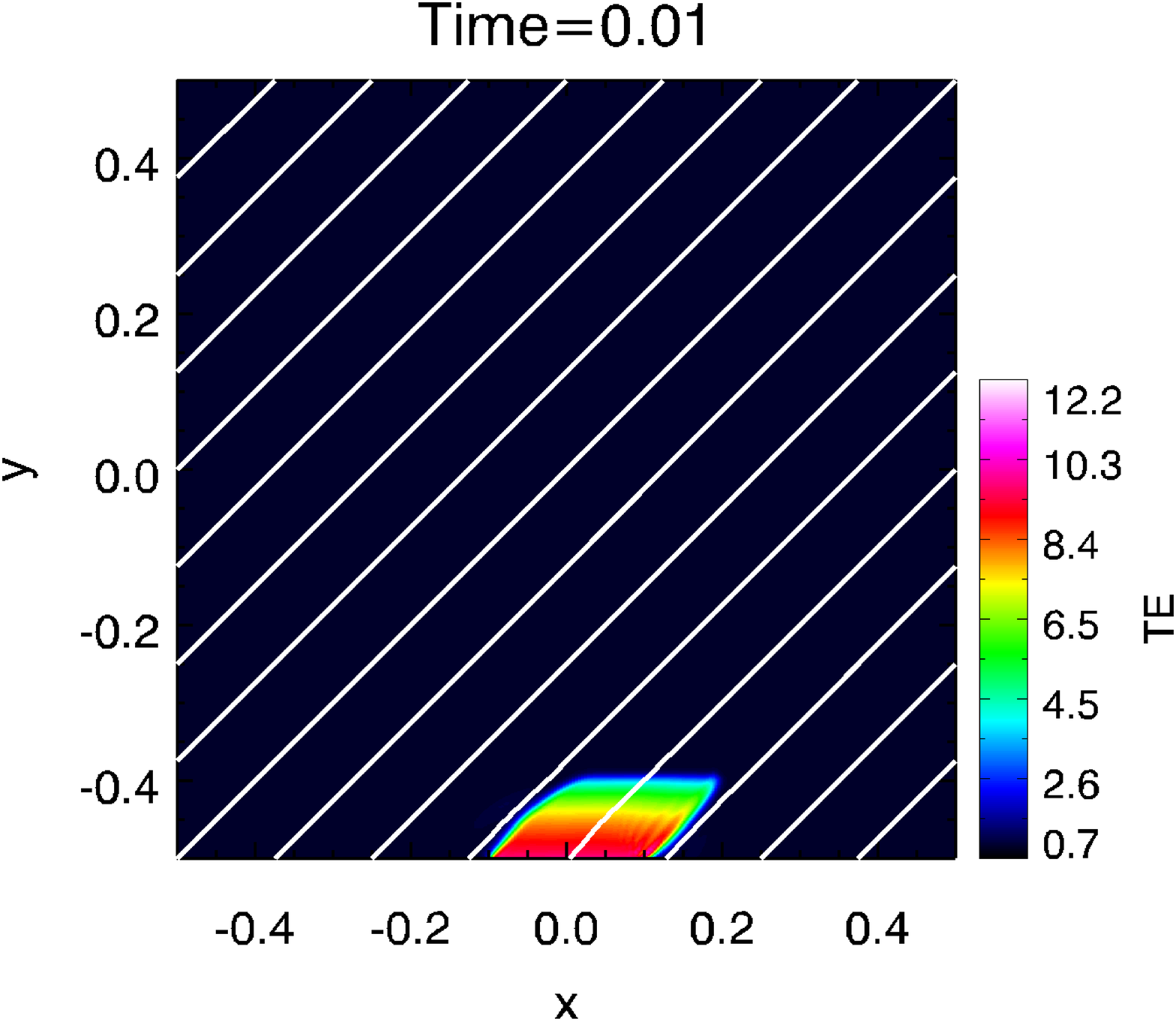}
\includegraphics[height=120pt]{\figures/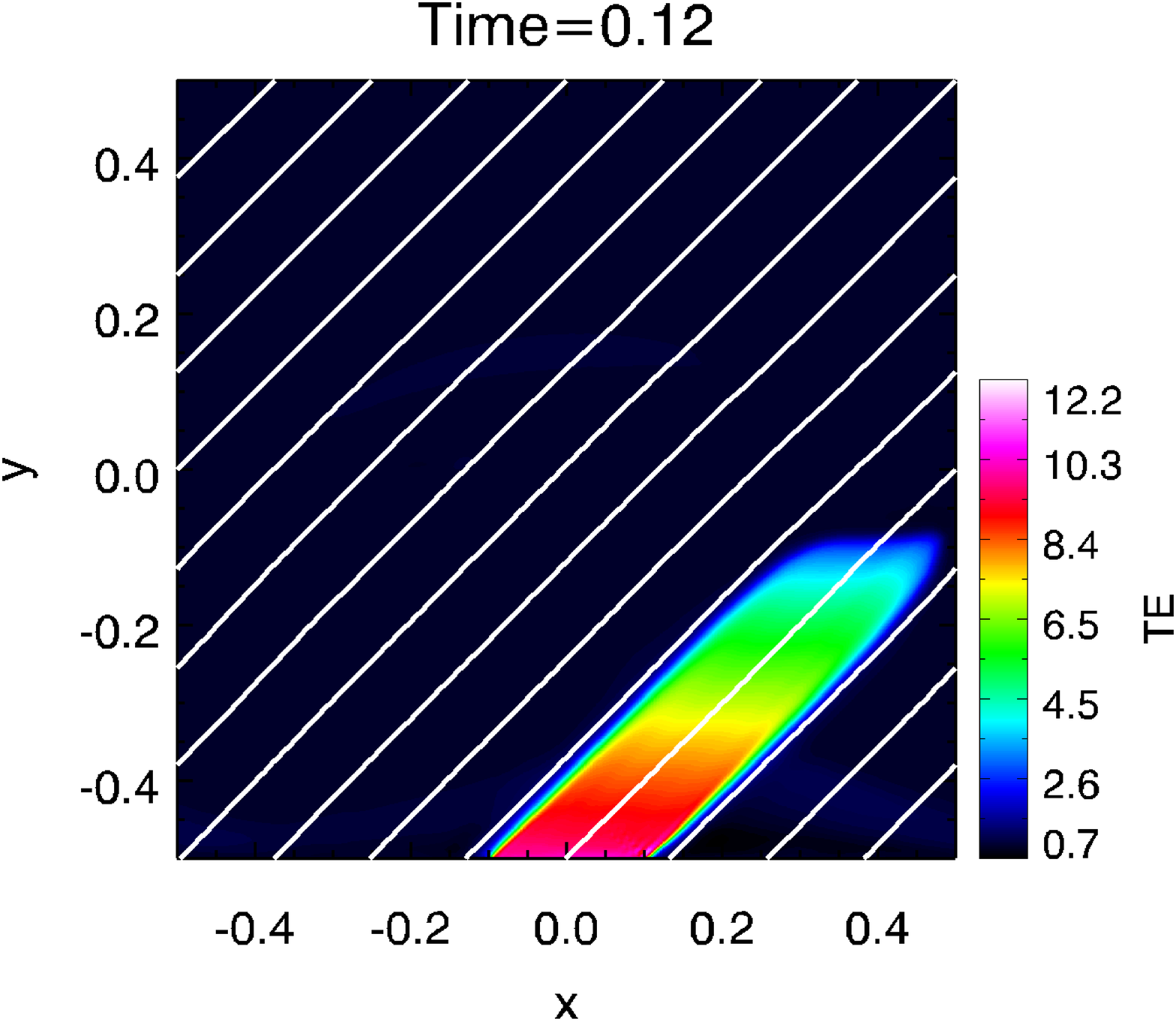}
\includegraphics[height=120pt]{\figures/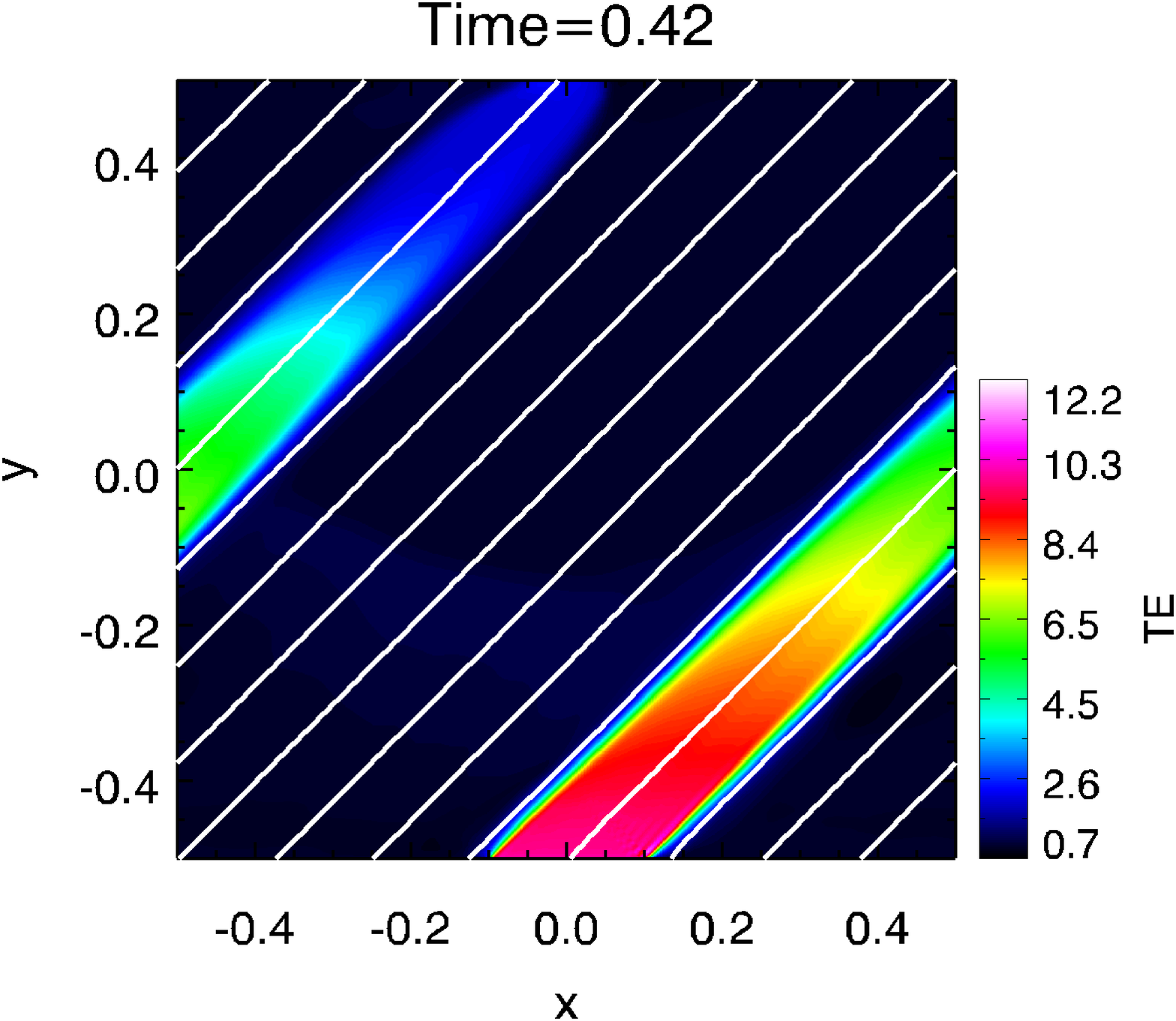}
\includegraphics[height=120pt]{\figures/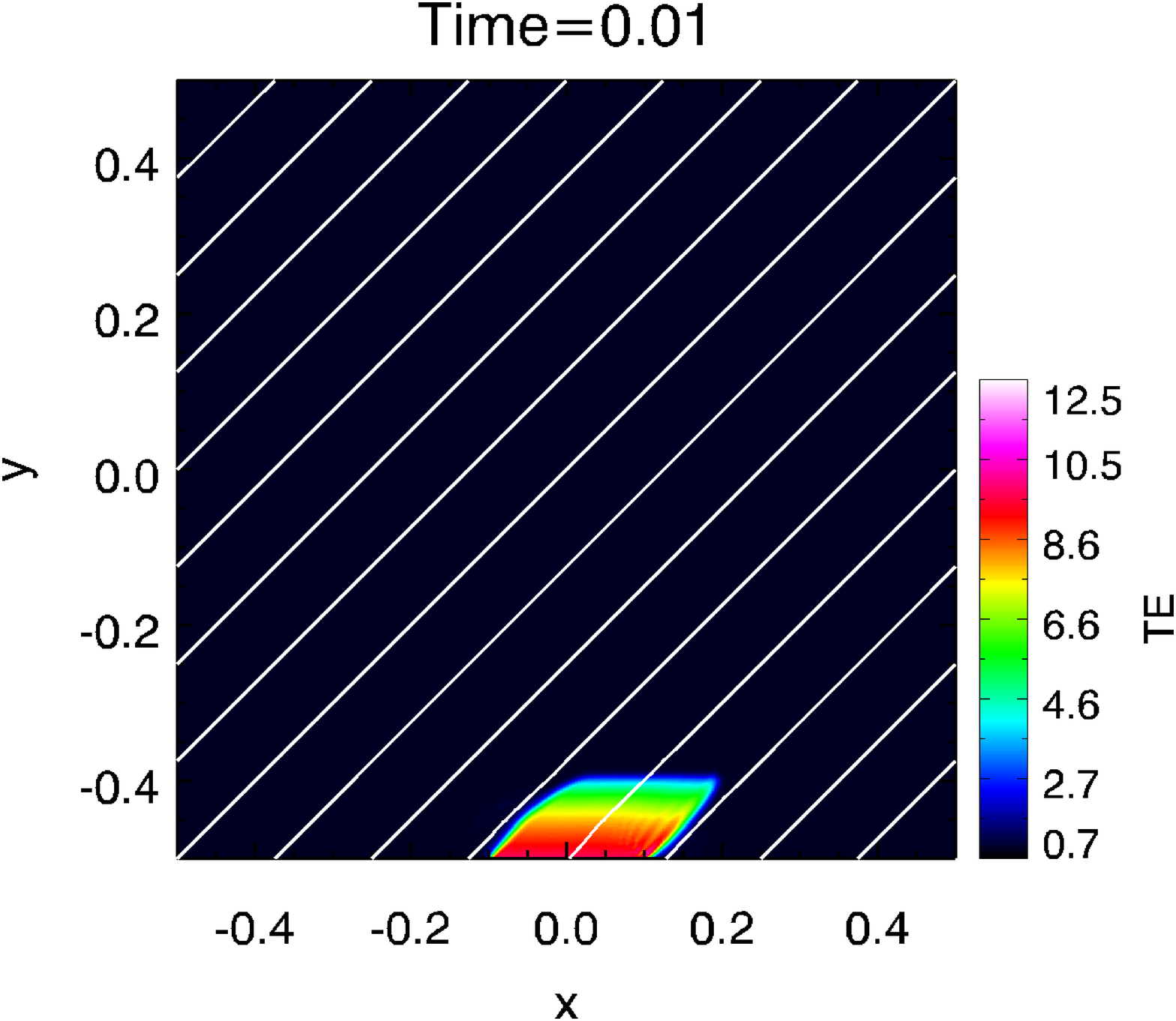}
\includegraphics[height=120pt]{\figures/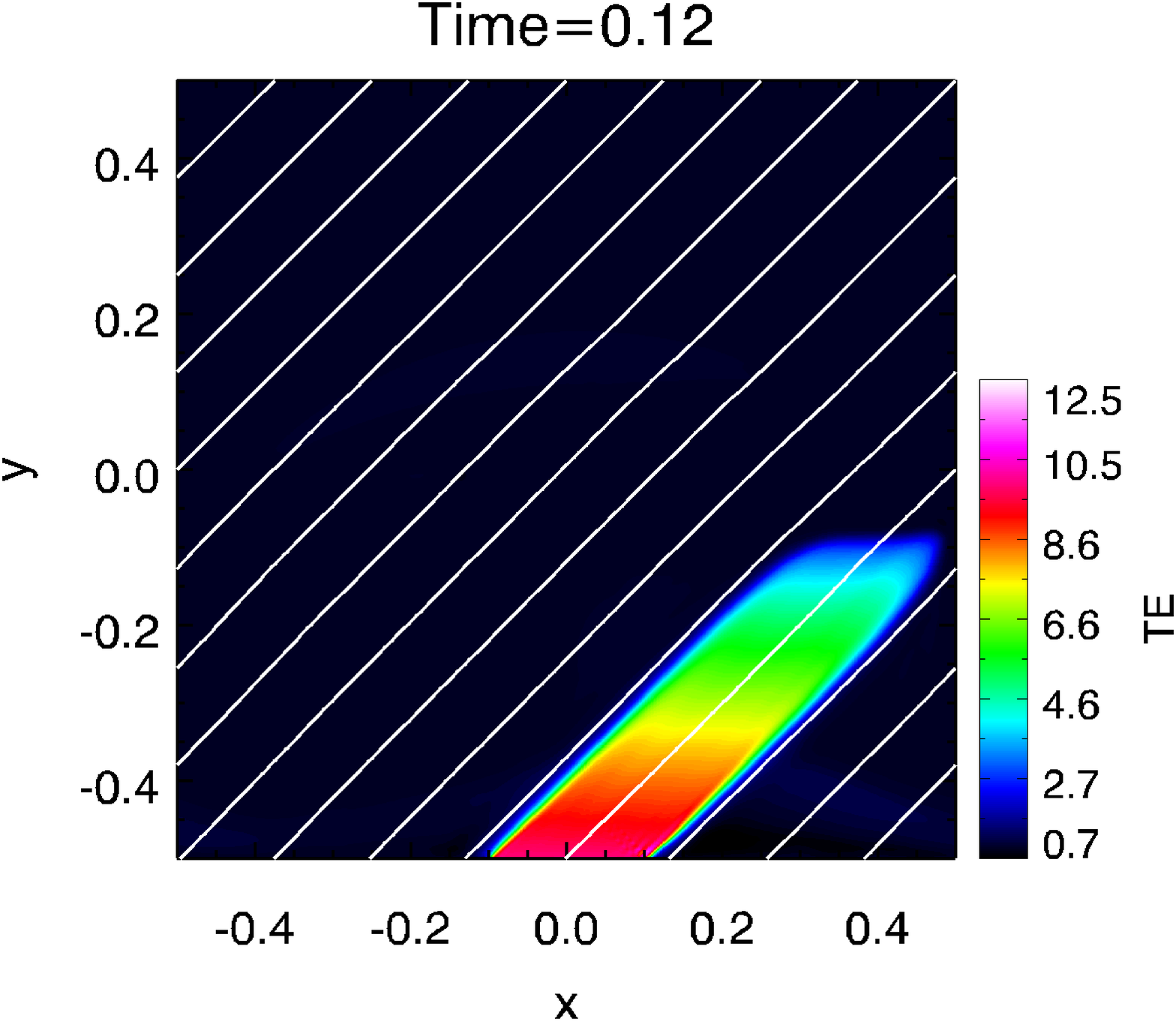}
\includegraphics[height=120pt]{\figures/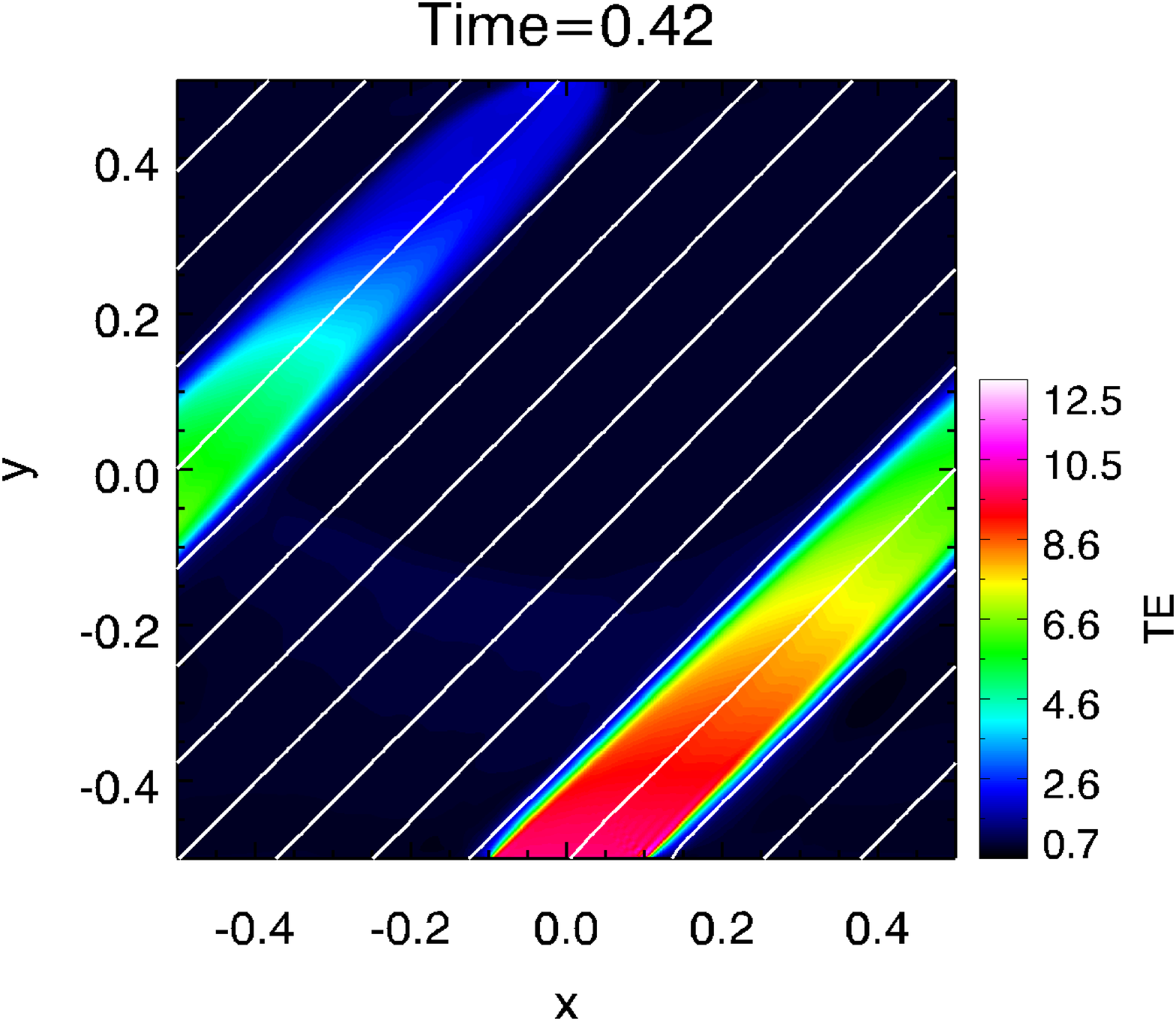}
\caption{Temperature distributions with the
magnetic field lines (white lines) at time 0.01, 0.12 and 0.42. The upper and lower panels show the 
results with subcycle and without subcycle, respectively. This test is based on the WENO
scheme with the Lax-Friedrichs flux.}
\label{fig14}
\end{figure}

In this test, we test the thermal condction effect with the method of subcycle and 
without the subcycle. Considering
fully ionization plasma ($\gamma = 5/3$), the thermal conduction can
only transfer the energy along the magnetic field lines. The
conduction term $\nabla \cdot \left (\kappa \nabla T \right)$ in the MHD
energy equation (Eq. (\ref{MHD-4})) is changed to the form $\nabla \cdot
\left(\kappa_0 T^{5 / 2}(\mathbf{B} \cdot \nabla T) \mathbf{B} / B ^
2 \right)$, where $\kappa_0$ is a coefficient for thermal conduction
(we take $\kappa_0 = 0.001$ in this test). The initial condition is very simple. 
In the computational domain ($[-0.5, 0.5] \times [-0.5, 0.5]$), we have uniform distributions:
$(\rho, v_x, v_y, v_x, B_x, B_y, B_z, p) = (0.1, 0, 0, 0, 1, 1, 0, 0.1)$. Besides, 
we set a small hot range ($\rho(x, 0) = 0.01$ in the range $|x|<0.1$) 
at the bottom of the boundary and the heat will be transferred 
along the oblique magnetic field lines as shown by Fig. \ref{fig14}. The bottom 
boundary is maintained as its initial value in the total evolution, while the left
and right boundaries are periodic. Finally, the top boundary is free.

\begin{figure}[htbp]
\centering
\includegraphics[height=240pt]{\figures/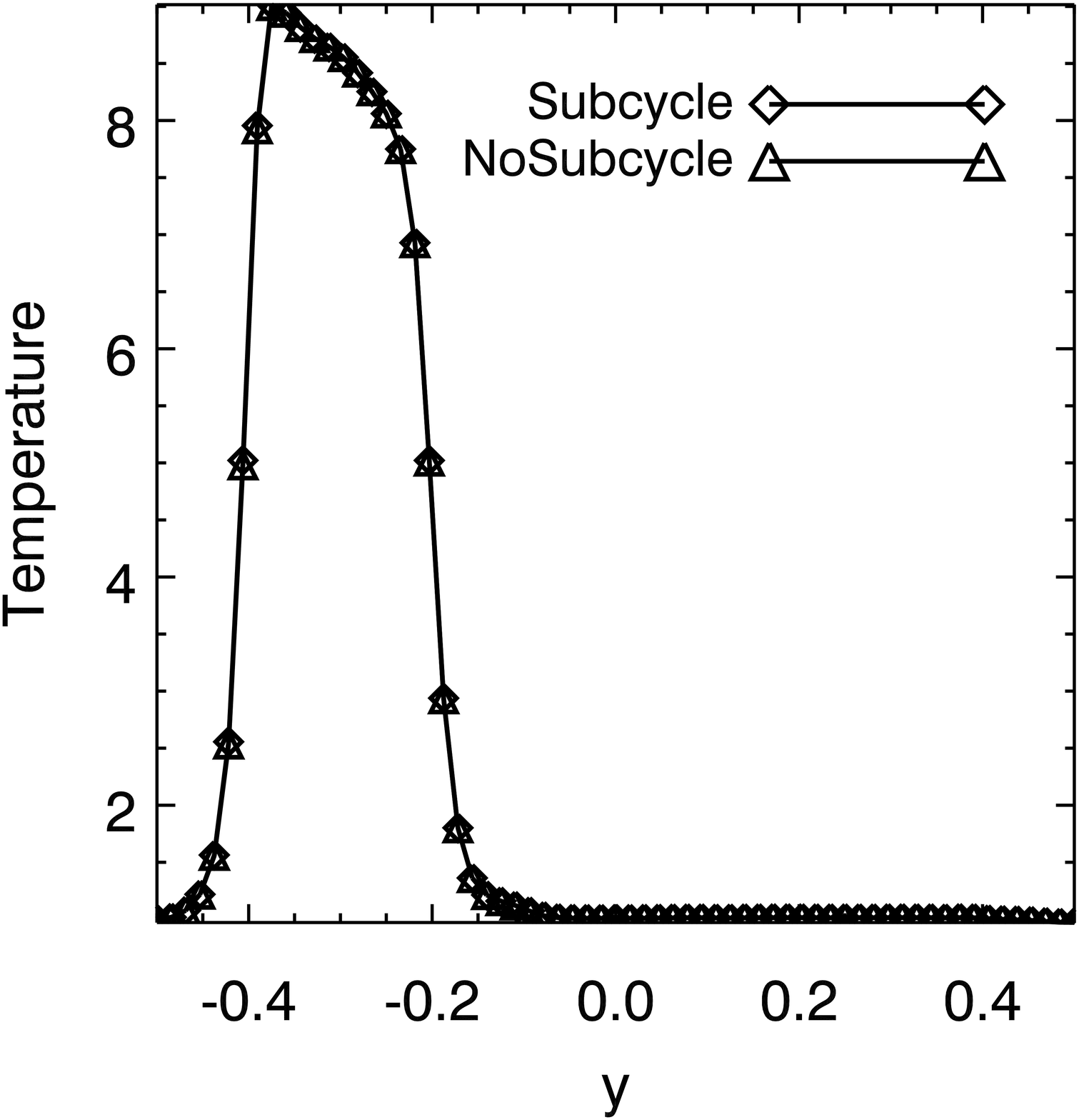}
\caption{Comparison of temperature distributions for the test with and without subcycle. 
The distributions are along the line $x=0.4$ at the Time = 0.42 for both cases shown in Fig. \ref{fig14}.}
\label{fig15}
\end{figure}

Fig. \ref{fig14} shows that the results with and without subcycle almost have the same 
results. Additional comparison is given in Fig. \ref{fig15} which taken the temperature
distributions along a vertical line ($x=0.4$) for both the upper right and the lower right panels. 
We can see that the shapes of two results are identical. We checked the values for both cases and found 
the difference between two case is less than $0.5\%$. However, the time spent in both cases are not in 
the same order of magnitude. For instance, in this test, the total elapsed time is 2225.27 seconds for 
the subcycle case while 28770.04 seconds for the case without subcycle. That is to say, we can get a 
almost identical result in a relatively short time by using the subcycle method.

\subsection{3D Rayleigh Taylor instability application}
\begin{figure}[htbp]
\centering
\includegraphics[height=150pt]{\figures/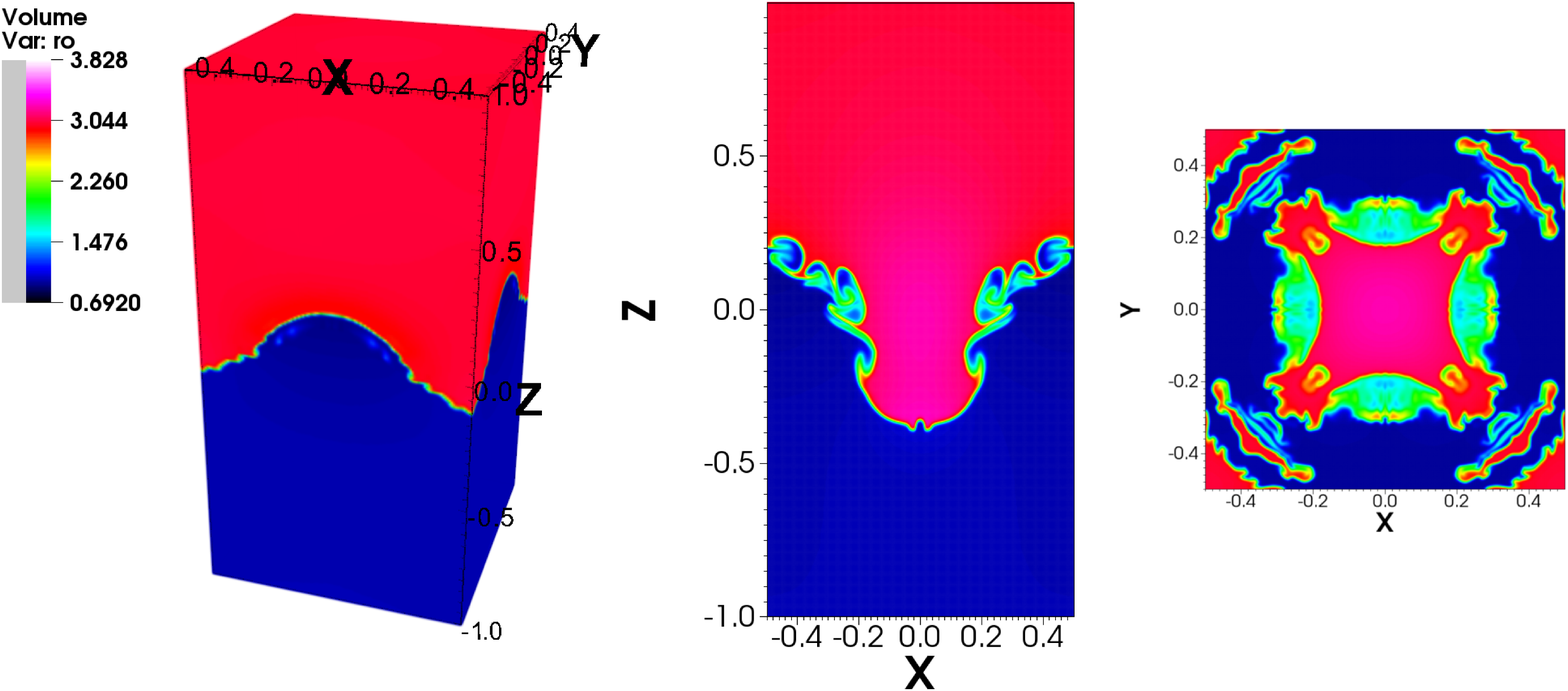}
\includegraphics[height=150pt]{\figures/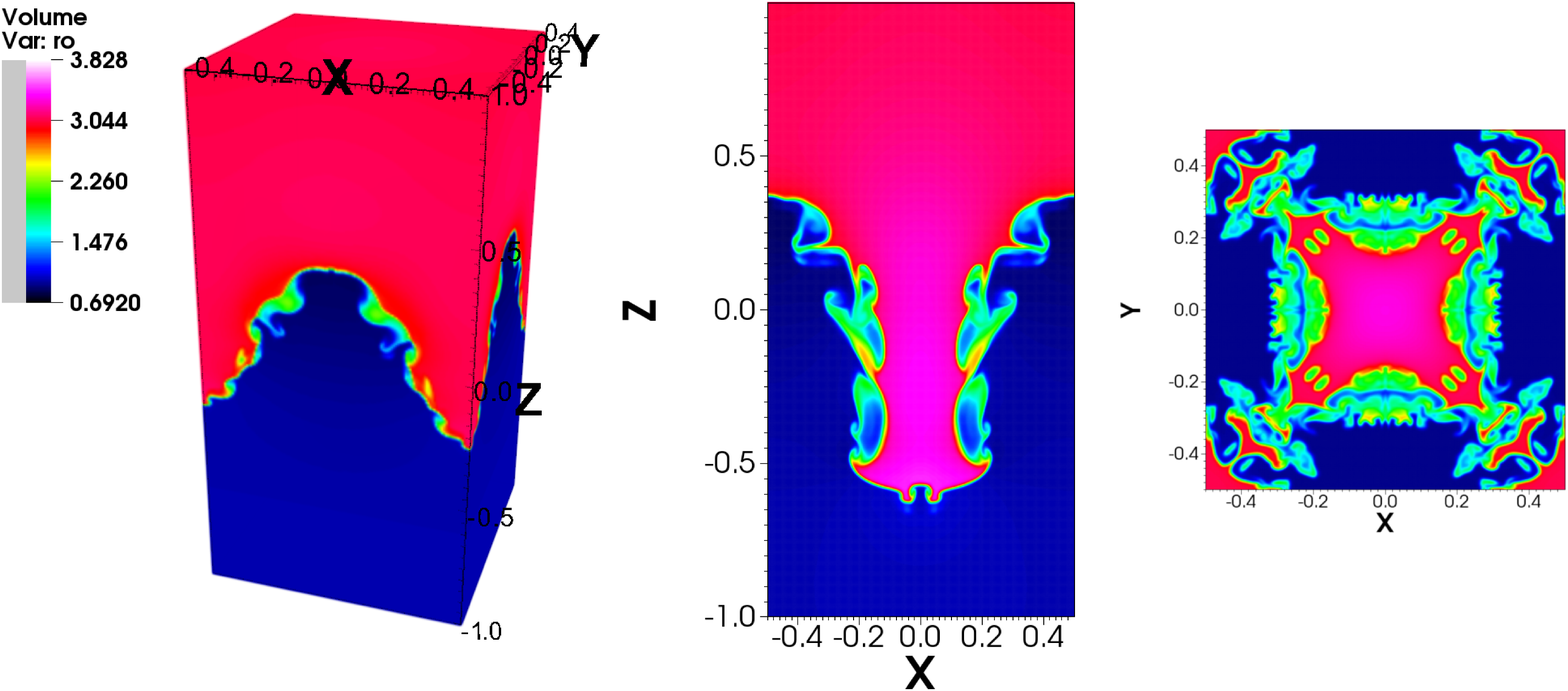}
\includegraphics[height=150pt]{\figures/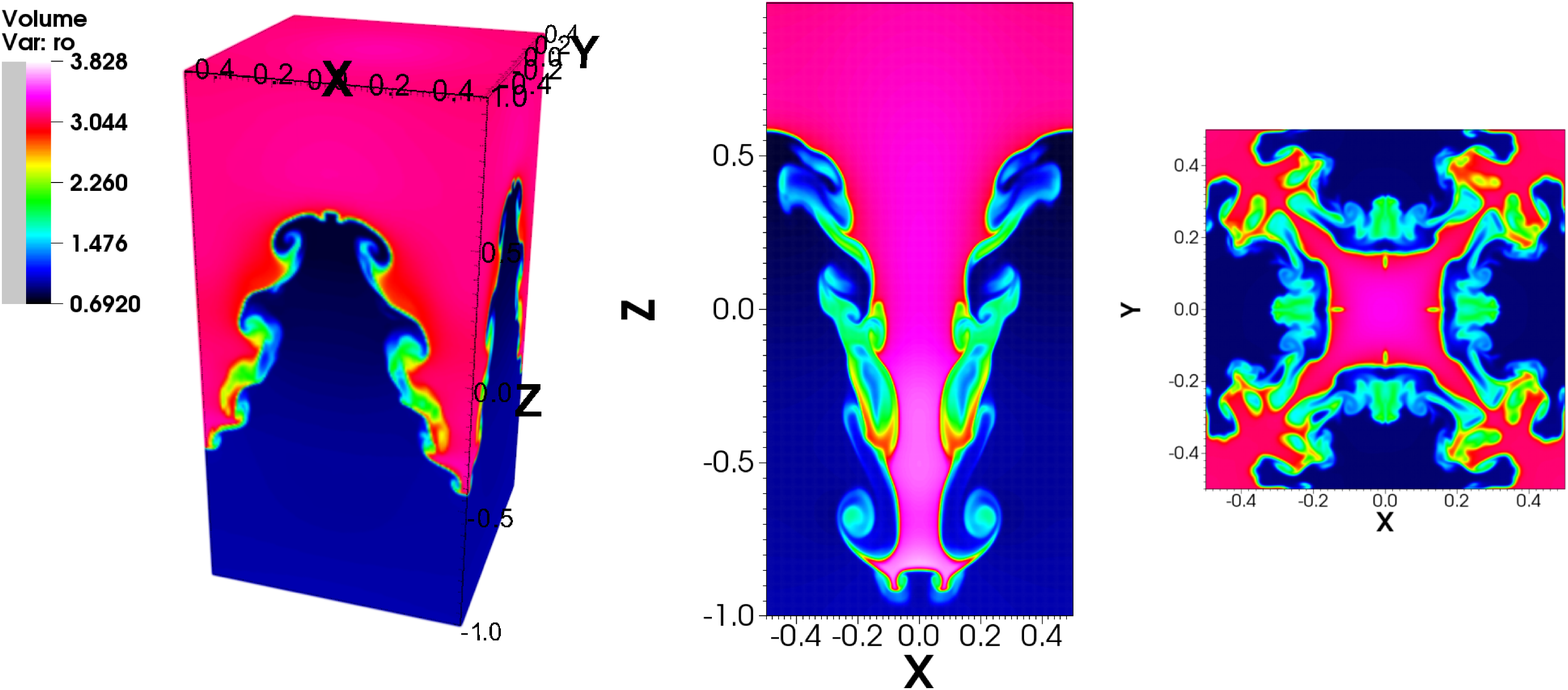}
\caption{Simulation result of the 3D rayleigh taylor instability at the time 
3.0, 3.5 and 4.0. Left panels: density distribution displayed by the volume rendering, 
which is visualized by the tool $VisIt$ developed by
the Department of Energy (DOE) Advanced Simulation and Computing
Initiative (ASCI). Middle panels: the density distribution on the $x$-$z$ plane at the 
position $y = 0$. Right panels: the density distribution on the $x$-$y$ plane at the 
position $z = 0$. The test is based on the WENO scheme with the HLLC approximate Riemann
solver.} \label{fig16}
\end{figure}

In the last application we check the performance of our MAP code in the 3D
domain $[-0.5,0.5] \times [-0.5,0.5] \times [-1,1]$ with a gravity
field ($g_x = 0$, $g_y = 0$, $g_z = -1$). The heavy gas ($\rho = 3$)
fills the upper half of the computational box ($z>0$) while the light
gas ($\rho = 1$) fills the lower half ($z<0$). A velocity
perturbation is introduced in the $z$-direction to trigger instability.
The mathematical form of the perturbation is $v_x = 0$, $v_y = 0$,
$v_z = 0.01 (1 + \cos(\pi x)) (1 + \cos(\pi y)) (1 + \cos(\pi z))$
in the cubic region $[-0.5,0.5] \times [-0.5,0.5] \times
[-0.5,0.5]$. The pressure ($p$) is changed from $1$ at the bottom
boundary to $5$ at the top boundary to get the hydrostatic balance. 
There is no magnetic field and
the adiabatic index $\gamma = 7 / 5$. The resuls are shown in
Fig.~\ref{fig16}. The statistics of the time consumption is given in
Table~\ref{list_time} for a 128-processor run and AMR level $L=4$.

This simulation can test the performance of the source term and the symmetry of 
our code. As we can observe from Fig. \ref{fig16}, the Rayleigh Taylor 
instability is successfully formed and the Kelvin-Helmholtz 
instability is also generated at the interface. The WENO scheme maintains 
a good symmetrical property in this problem. However, 
as we know, the final structure of this Rayleigh Taylor instability test 
is sensitive to the numerical diffusion of the scheme we used. If we 
use the Lax-Fridrichs flux instead of HLLC soler, the final shape of
the interface between the light and heavy fluids will differ wildly from
the result shown in Fig. \ref{fig16}. Thus, we do not give the results got by
other methods.

\begin{table}[htbp]
\caption{Statistics of the time consumption for the 3D rayleigh Taylor
instability. \label{list_time}}
\begin{tabular}{cc}
\hline
\hline
$N_p$                  & $128$      \\
Levels                 & $4$       \\
Effective resolution   & $256 \times 256 \times 512$ \\
Total elapsed time     & 5h 55m 11.15s       \\
\hline
Regridding, Load Balance   & 0.46\%         \\
Integration            & 83.64\%        \\
Data output            & 0.05\%         \\
Boundary condition     & 8.590\%       \\
\begin{tabular}{c}
Other (initial condition, data check, \\
calculate minimum $\Delta t$)       \\
\end{tabular}
                       & 6.96\%      \\
\hline
\end{tabular}
\end{table}

\section{Summary}
\label{summary}
We have presented an MHD code with adaptive mesh
refinement and parallelization for astrophysics, named MAP. We use
three kind of schemes, namely modified Mac Cormack Scheme (MMC),
Lax-Fridrichs scheme (LF) and weighted essentially non-oscillatory
(WENO) scheme, combined with TVD limiters and approximate Riemann
solvers (HLLC, HLLD, and Roe solvers). The divergence free condition
is guaranteed by the EGLM-MHD equations. As for the AMR parallelization
strategy, the conception of $framework$ and $gene$ may be discussed
with a different names in other codes. Nevertheless, there must be
many differences from other existing codes, for instance, the
treatments of the load balance, boundary condition, etc. It is
also useful for the readers to understand what is AMR or how to develop an
AMR code by themselves.

Our MAP code can be easily applied to the actual problems by simply
modifying the modules of initial and boundary conditions.
For some special cases, the resistivity model and the damping region can
also be changed. We are already using our MAP code for some solar
MHD applications, for example, magnetic reconnection between emerging flux
and the background canopy-type magnetic configuration which is aimed to
explain the microflares in the solar chromosphere and corona. These will be
discussed in detail in future papers.

This paper only describes the first version of our MAP code. The
code needs to be further improved in many aspects. Up to now, an
improvement should be done in the next job is the radiative
transfer. As well known, in the solar photosphere and chromosphere,
the optically-thick radiation is very important for the energy
transfer. How to implement it into MHD equations should be a heavy work and
needs a long time to code and test. Optically-thin case is relatively easy
to include, for instance, one can write a subroutine for adding a source
term to include the optically-thin transfer. The other implements like
the mesh geometries in cylindrical and spherical grids will be
added in the next version. It is emphasized here again that we did not take
the schemes with the accuracy higher than two orders into account as we
mentioned in Section~\ref{introduction}, since many operations in the AMR
algorithm and the treatment in source terms, for instance, the radiation
transfer mentioned above, can not reach such a higher accuracy. How to improve
the accuracy globally is another topic which is out of the scope of
this paper.





\section*{Acknowledgements}
The computations were done by using the IBM Blade
Center HS22 Cluster at high performance computing center of Nanjing
University of China. This work is supported by the National Natural
Science Foundation of China (NSFC) under the grant numbers 10221001,
10878002, 10403003, 10620150099, 10610099, 10933003, 11025314, and 10673004, as
well as the grant from the 973 project 2011CB811402.

\bibliographystyle{model1a-num-names}
\bibliography{map_v1}







\end{document}